\documentclass[journal,letter,10pt,twocolumn]{IEEEtran}
\pdfoutput=1

\renewenvironment{IEEEbiography}[1]
  {\IEEEbiographynophoto{#1}}
  {\endIEEEbiographynophoto}

\usepackage{amsmath, amssymb, amscd, amsthm, amsfonts}
\usepackage{bbm}
\usepackage[pdftex]{graphicx}
\usepackage[caption=false,font=footnotesize]{subfig}
\usepackage{dblfloatfix}    
\usepackage{color}
\usepackage{algorithm}
\usepackage[noend]{algpseudocode}
\usepackage{comment}

\usepackage{hyperref}

\usepackage{tikz,ifthen}
\usetikzlibrary{positioning,arrows,backgrounds}
\usetikzlibrary{fit}
\usetikzlibrary{calc}
 \usepackage{color}
 \usepackage{transparent}

\newtheorem{theorem}{Theorem}
\newtheorem{lemma}{Lemma}
\newtheorem{corollary}{Corollary}
\newtheorem{proposition}{Proposition}
\newtheorem{definition}{Definition}

\usepackage{pgfplots}
\pgfplotsset{width=10cm,compat=1.9}

\usepackage{subfiles}
\usepackage[noadjust]{cite}

\usepackage[english]{babel}

\addto\captionsenglish{}

\graphicspath{{../Figures/}}



\begin{document}

\title{On the Capacity-Achieving Input of Channels with Phase Quantization}

\author{
   \IEEEauthorblockN{Neil Irwin Bernardo, \textit{Graduate Student Member, IEEE}, Jingge Zhu, \textit{Member, IEEE}, and Jamie Evans, \textit{Senior Member, IEEE}
   }
   \thanks{Manuscript received Jun 21, 2021; revised Apr 14, 2022; accepted May 5, 2022. The work was supported in part by Australian Research Council under project DE210101497. N.I. Bernardo is a faculty (on study leave) of the University of the Philippines Diliman and his doctoral studies is supported by the Melbourne Research Scholarship of the University of Melbourne and the DOST-ERDT Faculty Development Fund of the Republic of the Philippines. Section \ref{section-awgn_case} is part of our preliminary work \cite{bernardo2021phase} which appeared in ISIT 2021.  (\emph{Corresponding author: Neil Irwin Bernardo}.) }
    \thanks{N.I. Bernardo, J. Zhu, and J. Evans are with the Department of Electrical and Electronic Engineering, The University of Melbourne, Parkville, VIC 3010, Australia (e-mail: bernardon@student.unimelb.edu.au, jingge.zhu@unimelb.edu.au;
jse@unimelb.edu.au).}
}

\maketitle
\begin{abstract}
Several information-theoretic studies on channels with output quantization have identified the capacity-achieving input distributions for different fading channels with 1-bit in-phase and quadrature (I/Q) output quantization. However, an exact characterization of the capacity-achieving input distribution for channels with multi-bit phase quantization has not been provided. In this paper, we consider four different channel models with multi-bit phase quantization at the output and identify the optimal input distribution for each channel model. We first consider a complex Gaussian channel with $b$-bit phase-quantized output and prove that the capacity-achieving distribution is a rotated $2^b$-phase shift keying (PSK). The analysis is then extended to multiple fading scenarios. We show that the optimality of rotated $2^b$-PSK continues to hold under noncoherent fast fading Rician channels with $b$-bit phase quantization when line-of-sight (LoS) is present. When channel state information (CSI) is available at the receiver, we identify $\frac{2\pi}{2^b}$-symmetry and constant amplitude as the necessary and sufficient conditions for the ergodic capacity-achieving input distribution; which a $2^b$-PSK satisfies. Finally, an optimum power control scheme is presented which achieves ergodic capacity when CSI is also available at the transmitter.
\end{abstract}


\IEEEpeerreviewmaketitle

\section{Introduction}\label{section-intro}

\IEEEPARstart{T}{he} use of low-resolution analog-to-digital converters (ADCs) has recently gained significant research interest because it addresses practical problems and scalability issues in 5G core technologies such as massive data processing, high power consumption, and cost \cite{Liu:2019}. Most studies on low-resolution ADCs have been more focused on investigating the fundamental limits and practical detection strategies in the context of Multiple-input Multiple-output (MIMO) and millimeter wave systems \cite{Jacobsson:2015,Bjornson:2015,Orhan:2015,Mezghani:2012}. However, these studies did not properly address the structure of the capacity-achieving input and only analyzed performance via capacity bounds using simplified analytical models. Low-resolution receiver design requires a shift in signal/code construction since Gaussian signaling is no longer optimal in channels with quantized output \cite{Vu:2019}.

Some research efforts have been invested in analyzing the capacity limits of channels with low-resolution quantization and finding the optimal signaling schemes for such channels. One of the first studies on this topic showed that binary antipodal signaling is optimal for real additive white Gaussian noise (AWGN) channels with 1-bit quantized output \cite{Singh:2007,Singh:2009}. Extension of capacity analysis to other wireless channels with 1-bit in-phase and quadrature (I/Q) ADCs revealed that Quadrature Phase Shift Keying (QPSK) is optimal for complex-valued AWGN channel \cite{Krone:2010}, noncoherent Rician channel \cite{Vu:2019}, and zero-mean Gaussian mixture channel \cite{Rahman:2020,Rahman2:2020}. Considering a block noncoherent Rayleigh channel with 1-bit ADC output, an on-off QPSK scheme with numerically-optimized duty cycle is the capacity-achieving input distribution \cite{Mezghani2:2008}. When channel state information (CSI) is granted at the receiver, it has been identified that any $\frac{\pi}{2}$-symmetric input distribution with constant amplitude is capacity-achieving in coherent Rayleigh channels with 1-bit ADC \cite{Krone:2010} and sum-capacity-achieving for multiple access Rayleigh channels with 1-bit ADC \cite{Ranjbar:2019,Ranjbar:2020}. For a multi-input single-output (MISO) channel with 1-bit ADC output, the capacity can be achieved using maximal ratio transmission (MRT) beamforming and QPSK signaling when CSI is granted to both transmitter and receiver\cite{Mo:2014}. However, identifying the structure of capacity-achieving input analytically for fading (and even static) channels with multi-bit I/Q quantization still remains as an open problem \cite{Vu:2019}. There are several algorithms available in the literature \cite{Abou-Faycal:2001,Huang:2005,Blahut:1972} to numerically construct the optimal input distribution for a specific channel. Even so, these algorithms lack explicit error bounds that may be useful in the analysis or may suffer from numerical instability and slow convergence under certain scenarios.

Motivated by the above discussion, we aim to extend the capacity results of 1-bit I/Q quantization to multi-bit quantization. However, we shall investigate multi-bit phase quantization instead of the conventional I/Q quantization. Phase quantization ignores the amplitude component thus eliminating the necessity for automatic gain control \cite{Singh:2013}. Furthermore, phase quantizers can be easily implemented in practice using analog phase detectors and 1-bit comparators which consume negligible power (in the order of mW) \cite{Gayan:2020}. Implementations based on time-to-digital converters (TDCs) can also be adopted to further reduce the area and power consumption of the phase quantizer \cite{Nazari:2014}.  Error rate analysis of low-resolution phase-modulated communication has been done for the single-input single-output (SISO) fading channel \cite{Gayan:2019,Gayan:2020}, relay channel \cite{Souryal:2008}, and multiuser MIMO channel \cite{Lopes:2020} but only investigated uncoded transmissions. Information rates of phase-quantized block noncoherent receiver and low signal-to-noise ratio (SNR) rates of $M$-PSK with hard decision detector have been studied before in \cite{Singh:2013} and \cite{Gursoy:2007}, respectively. However, the proponents of these studies did not establish the optimality of PSK and treated PSK signaling as a given in their problem setup. In fact, \cite{Singh:2013} only used QPSK signaling for a channel with 8-sector and 12-sector phase quantization even though there are more than four possible outputs per channel use.

An intuitive communications engineer would probably pick PSK modulation to transmit over channels with phase-quantized output due to the apparent rotational symmetry in the quantizer and the modulation format. Still, similar to its special case of 1-bit I/Q quantization, optimality conditions for the capacity-achieving input of channels with phase quantization at the output should be established. This paper provides a rigorous proof that $2^b$-PSK is the capacity-achieving input under different fading scenarios. We show that the analytical tractability of deriving the optimal input distribution of various 1-bit ADC channels \cite{Mezghani2:2008,Singh:2007,Singh:2009,Vu:2019,Krone:2010, Rahman:2020, Rahman2:2020} stems from the more general multi-bit phase quantization. Our main contributions are summarized as follows.

\begin{itemize}
    \item We show that a rotated $2^b$-PSK is the optimal modulation scheme for a complex Gaussian channel with $b$-bit phase quantization and fixed channel gain. We provide an expression for the capacity in terms of the SNR and phase quantizer resolution (Theorem \ref{theorem:AWGN_case}).
    \item We prove that $2^b$-PSK is still optimal for noncoherent fast fading Rician channel with $b$-bit phase quantization (Theorem \ref{theorem:noncoherent_rician_case}).
    \item We identify properties of the optimal modulation scheme for a Rayleigh fading channel with $b$-bit phase quantization when CSI is granted at the receiver. That is, the optimal input should satisfy a specific symmetry condition and should have a constant amplitude (Theorem \ref{theorem:Rayleigh_CSIR_case}). Moreover, we derive an ergodic capacity-achieving power control scheme when CSI is also available at the transmitter (Theorem \ref{theorem:Rayleigh_CSIT_case}).
\end{itemize}
Lastly, only symmetric phase quantization is considered in this study and is a given in the problem setup presented in the next section. Symmetric quantization strategy, however, is not necessarily optimal for all SNR regime in the 1-bit case as pointed out in \cite{Koch:2013}. Nonetheless, symmetric TDC-based phase quantizers are easier to construct since the logic delay buffers have identical designs.

\textit{Notation}: All $\log()$ terms in this paper are in base 2 unless specified otherwise. When it is clear from the context, we use $F_X$ and $f_X$ to denote the cumulative distribution function (CDF) $F_X(x)$ and the probability density function (PDF) $f_X(x)$, respectively. Equivalent notation can be used for the conditional and joint distributions. We write the probability mass function (PMF) of a random variable $Y$ as $p_Y(y;F_{X})$ if the distribution is induced by a choice of another distribution $F_{X}$. For instance, $p_{Y}(y;F_{X}) = \int p_{Y|X}\;dF_{X}$ for some conditional PMF $p_{Y|X}$. Lastly, the mutual information between two variables $X$ and $Y$ is typically denoted as $I(X;Y)$ or $I(F_{X};p_{Y|X})$. When the channel law $p_{Y|X}$ is clear from the context, we simply write $I(F_{X})$.

\section{Problem Formulation and Main Results}\label{section-sys_models}

\begin{figure*}[t]
  \centering
  \hspace*{-0.25cm}
  \subfloat[]{
\includegraphics[width = .5\textwidth]{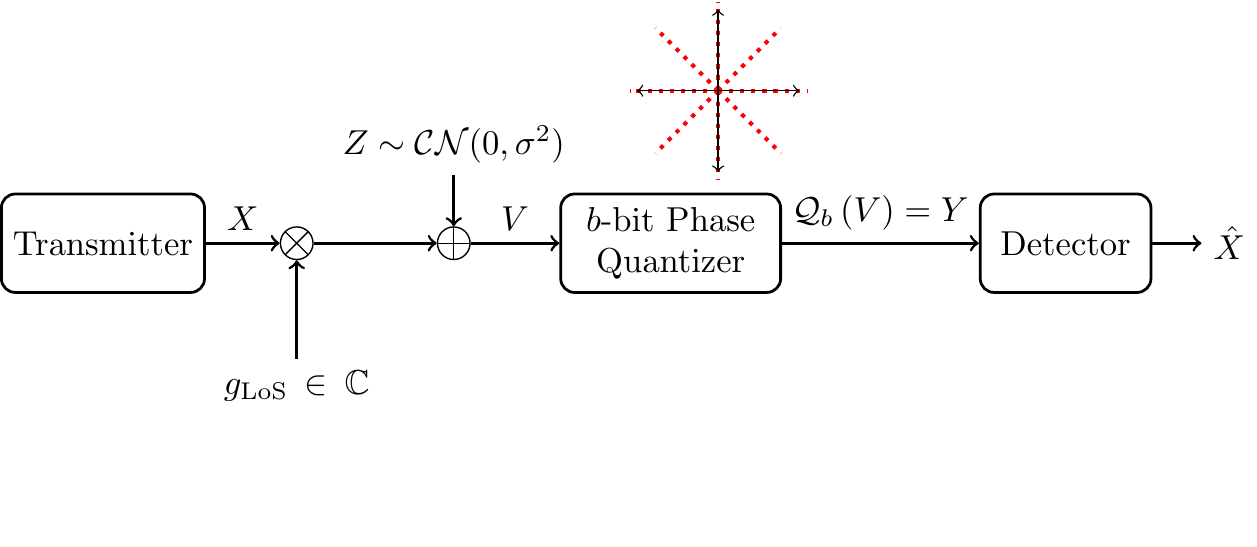}
\label{fig:sys_model_AWGN}
} \hspace*{0cm}%
 \subfloat[]{
\includegraphics[width = .5\textwidth]{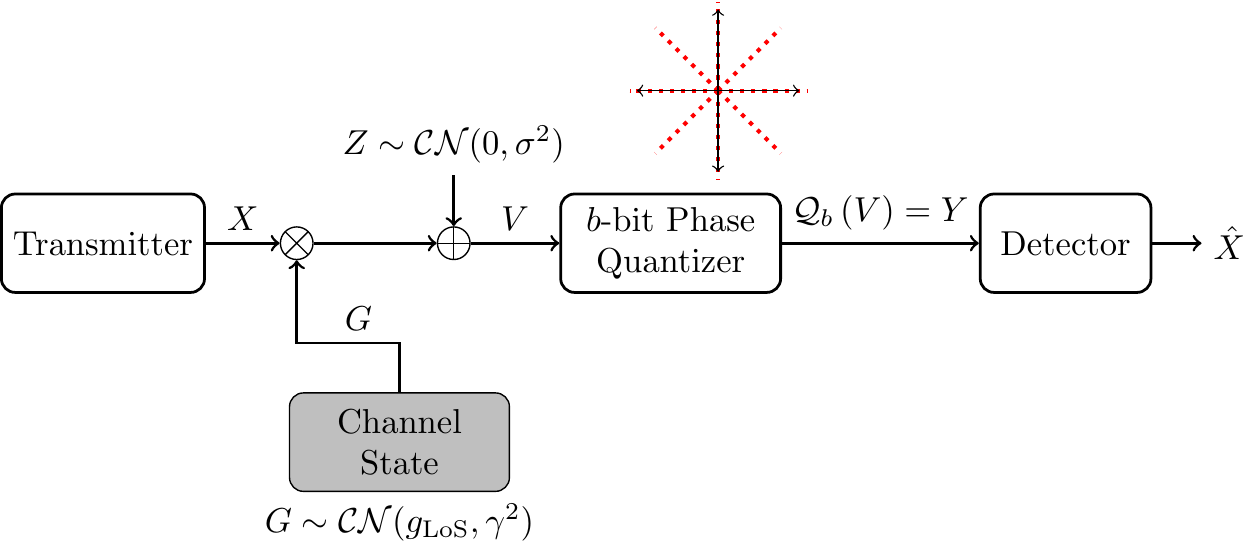}
    \label{fig:sys_model_NoCSI}
} \\
 \hspace*{-0.25cm}
 \subfloat[]{
\includegraphics[width = .5\textwidth]{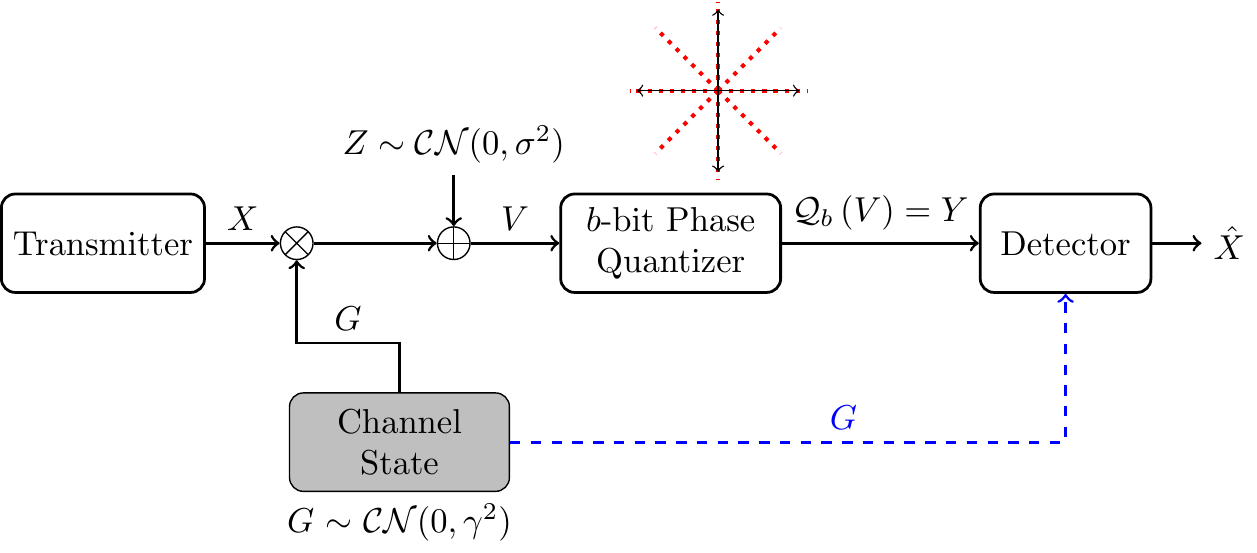}
    \label{fig:sys_model_CSIR}
} \hspace*{0cm}%
 \subfloat[ ]{
\includegraphics[width = .5\textwidth]{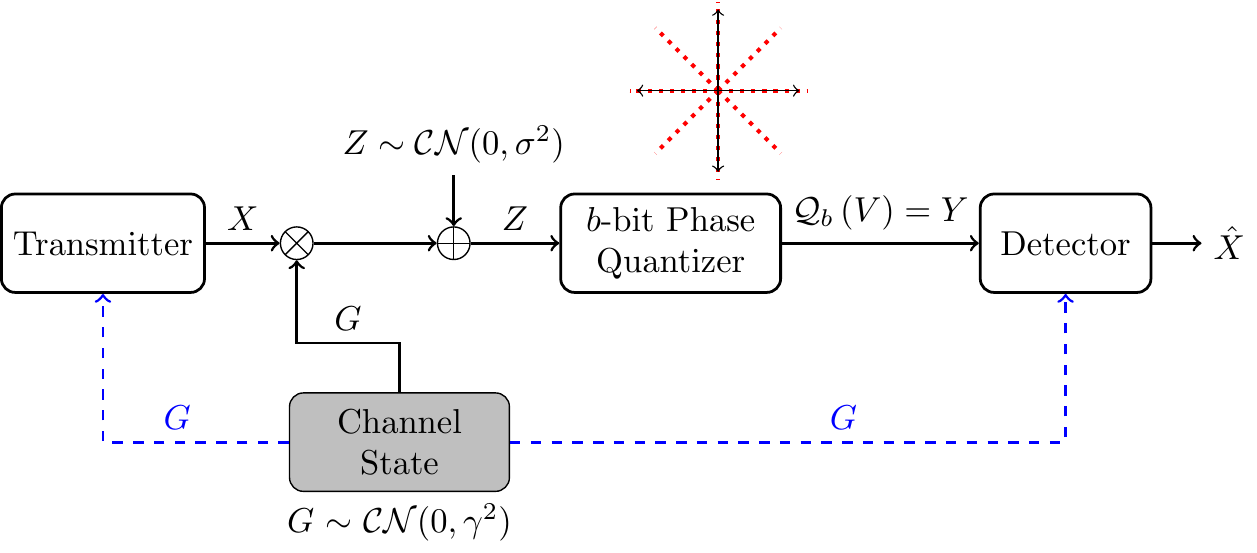}
    \label{fig:sys_model_CSIT}
} 
   
     \caption{System models with phase quantization at the output: (a) Gaussian channel with fixed channel gain, (b) noncoherent Rician Fading, (c) Rayleigh Fading with CSI at Receiver only (CSIR), and (d) Rayleigh Fading with CSI at both transmitter and receiver (CSIT)}
     \label{fig:sys_model_gen}
\end{figure*}
We consider four different discrete-time memoryless channel models shown in Figure \ref{fig:sys_model_gen}. In all cases, the input-output relationship between the transmitted signal $X$ and the unquantized received signal $V$ at each time instant can be expressed as
\begin{equation}
    V = GX + Z,
\end{equation}
where $X$ is the complex input with power constraint $\mathbb{E}[|X|^2]\leq P$, $Z$ is the zero-mean complex Gaussian noise with variance $\sigma^2$, and $G$ is the channel gain which is modeled differently for each case:

\begin{itemize}
    \item \textbf{Model A} \emph{(Gaussian Channel with Fixed Gain and Phase-Quantized Output)}: In this model, $G = g_{\mathrm{LoS}}$ is a complex constant representing the gain and direction of the LoS component and the transmitter and receiver have knowledge of this parameter (see Figure \ref{fig:sys_model_AWGN}).
    \item \textbf{Model B} \emph{(Noncoherent Fast Fading Rician Channel with Phase-Quantized Output)}: In this model, $G$ is the sum of a fixed (and known) LoS component $g_{\mathrm{LoS}} \in \mathbb{C}$ and a random non-LoS (nLoS) component $G_{\text{nLoS}}\sim\mathcal{CN}(0,\gamma^2)$. Equivalently, $G\sim \mathcal{CN}(g_{\mathrm{LoS}},\gamma^2)$. The Rician factor $\kappa$ is defined as $\kappa = \frac{|g_{\mathrm{LoS}}|^2}{\gamma^2}$ (see Figure \ref{fig:sys_model_NoCSI}).
    \item \textbf{Model C} \emph{(Rayleigh Fading Channel with Phase-Quantized Output and CSI at Receiver only)}: In this model, $G\sim\mathcal{CN}(0,\gamma^2)$ and the channel state is known only at the receiver  (see Figure \ref{fig:sys_model_CSIR}).
    \item \textbf{Model D} \emph{(Rayleigh Fading Channel with Phase-Quantized Output and CSI at the Transmitter and Receiver)}: The fading model is the same as Model C but the channel state is also known at the transmitter (see Figure \ref{fig:sys_model_CSIT}).
    
\end{itemize}
The unquantized received signal $V$ is sampled\footnote{Synchronized sampling at symbol rate is assumed. Here, each received sample corresponds to only one transmitted symbol.}
and then fed to a $b$-bit phase quantizer $\mathcal{Q}_b(\cdot)$ to produce an integer-valued output $Y \in \{0,\cdots,2^b - 1\}$. To be more precise, the output of the phase quantizer is $Y=y$ if $\angle V \in \mathcal{R}^{\text{PH}}_{y}$, where $\mathcal{R}^{\text{PH}}_{y}$ is given by
\begin{equation}
    \mathcal{R}^{\text{PH}}_{y} = \left\{\phi\in [-\pi,\pi]\;\Big|\;\frac{2\pi}{2^{b}}y  \leq \phi + \pi < \frac{2\pi}{2^{b}}(y+1)  \right\}.
\end{equation}
Due to the circular structure of the phase quantizer, the addition operation $Y+k$ for some $k\in\mathbb{Z}$ constitutes a modulo $2^b$ addition. In this quantization model, only a coarse phase information of the received signal is retained. The goal of the receiver is to reliably recover the message encoded in $X$ using the phase quantizer output, $Y$. From this problem description we are led to the following question: \emph{What is the capacity-achieving input distribution, denoted as $F_X(x)$, that maximizes the rate at which reliable communication is achievable?}

To answer this, we first define two important quantities that will appear frequently in the paper. 
\begin{definition}\label{definition:phase_quantization_func}
The phase quantization probability function ,$W_{y}^{(b)}(\nu,\theta)$, is defined as
\begin{equation}\label{eq:phase_quantization_prob}
    W_y^{(b)}(\nu,\theta) = \int_{\frac{2\pi}{2^b}y-\pi-\theta}^{\frac{2\pi}{2^b}(y+1)-\pi-\theta}f_{\Phi|N}(\phi|\nu)\;d\phi,
\end{equation}
where
\begin{align}\label{eq:prob_phi_given_nu}
     f_{\Phi|N}(\phi|\nu)
 =& \frac{e^{-\nu}}{2\pi}+\frac{\sqrt{\nu}\cos\left(\phi\right)e^{-\nu\sin^2\left(\phi\right)}}{\sqrt{\pi}}\nonumber\\
 &\cdot\left[1-Q\left(\sqrt{2\nu}\cos\left(\phi\right)\right)\right],\end{align}
$Q(x)$ is the Gaussian Q-function, $\theta\in[-\pi,\pi]$, and $\nu\geq 0$.
\end{definition}

\begin{definition}\label{definition:phase_quantization_entrop}
The phase quantization entropy, $w(\nu,\theta,b)$, is defined as
\begin{equation}\label{eq:phase_quantization_entrop}
    w(\nu,\theta,b) = -\sum_{y=0}^{2^{b-1}}W_y^{(b)}(\nu,\theta)\log W_y^{(b)}(\nu,\theta)
\end{equation}
for $\theta\in[-\pi,\pi]$, $\nu\geq 0$, and $b\geq 1$.
\end{definition}
Note that these are purely mathematical definitions and have no engineering significance so far. In essence, $W_{y}^{(b)}(\nu,\theta)$ and $w(\nu,\theta,b)$ describe the channel law and the conditional entropy of Model A, respectively. We defer giving the precise operational meaning for $W_{y}^{(b)}(\nu,\theta)$ and $w(\nu,\theta,b)$ to the next section. These expressions cannot be simplified further. However, $W_{y}^{(b)}(\nu,\theta)$ and $w(\nu,\theta,b)$ can still be used to identify the optimal input distribution and numerically compute the capacity of channels with $b$-bit phase quantization. 

We also define a class of complex-valued symmetric input distributions that will be relevant in our analysis. To be more precise, we will show that the capacity-achieving input distributions for Models A to D belong to this class of input distributions.

\begin{definition}\label{definition:2pi_2b-symmetry}
A distribution $F_X(x)$ is a $\frac{2\pi}{2^b}$-symmetric distribution if $F_X(x) \sim F_X(xe^{j\frac{2\pi k}{2^b}}) \;\forall k \in \mathbb{Z}$. To put it simply, applying an integer multiple rotation of $\frac{2\pi}{2^b}$ to $F_{X}(x)$ does not change its distribution.
\end{definition}

We now formally state the main results of this paper.
\begin{theorem}\label{theorem:AWGN_case} The capacity of a complex Gaussian channel with fixed channel gain and $b$-bit phase-quantized output (Model A) is
\begingroup
\allowdisplaybreaks
\begin{align}\label{eq:cap_AWGN}
    C = b - w\left(\frac{|g_{\mathrm{LoS}}|^2P}{\sigma^2},\frac{\pi}{2^b},b\right)\;\mathrm{bits/cu},
\end{align}
\endgroup
where $w(\cdot,\cdot,\cdot)$ is given in Definition \ref{definition:phase_quantization_entrop}. The capacity is achieved by a discrete input distribution with probability mass function (PMF) given by
\begin{align}
    f_X^*(x) =& 
    \bigg\{\frac{1}{2^b}\Big|x = \sqrt{P}e^{j\left(\frac{2\pi(k+0.5)}{2^b}-\angle g_{\mathrm{LoS}}\right)},\nonumber\\
    &\quad\forall k\in\{0,...,2^{b}-1\}\bigg\}.
\end{align}
In other words, the optimal input distribution is a rotated $2^b$-PSK with equiprobable symbols.
\end{theorem}
\noindent \textbf{Remark on Theorem \ref{theorem:AWGN_case}:} The proof of Theorem \ref{theorem:AWGN_case} establishes the properties of the capacity-achieving input for Model A such as rotational symmetry, constant amplitude per phase, boundedness, and discreteness. An optimality condition is derived to identify the positions of these discrete mass points (which collectively form a rotated $2^b$-PSK constellation).

\begin{theorem}\label{theorem:noncoherent_rician_case} The capacity of a noncoherent fast fading Rician channel with $b$-bit phase-quantized output (Model B) is
\begingroup
\allowdisplaybreaks
\begin{align}
    C = b - w\left(\frac{|g_{\mathrm{LoS}}|^2P}{\gamma^2P+\sigma^2},\frac{\pi}{2^b},b\right)\;\mathrm{bits/cu},
\end{align}
\endgroup
where $w(\cdot,\cdot,\cdot)$ is given in Definition \ref{definition:phase_quantization_entrop}. The capacity is achieved by a discrete input distribution with PMF given by
\begin{align}
    f_X^*(x) =&
    \bigg\{\frac{1}{2^b}\Big|x = \sqrt{P}e^{j\left(\frac{2\pi(k+0.5)}{2^b}-\angle g_{\mathrm{LoS}}\right)},\nonumber\\
    &\quad\forall k\in\{0,...,2^{b}-1\}\bigg\}.
\end{align}
In other words, the optimal input distribution is a rotated $2^b$-PSK with equiprobable symbols.
\end{theorem}
\noindent \textbf{Remark on Theorem \ref{theorem:noncoherent_rician_case}:} To prove Theorem \ref{theorem:noncoherent_rician_case}, we show that the mutual information of $X$ and $Y$ in Model B closely resembles that of Model A. A corollary is presented that extends the properties of the capacity-achieving input in Model A to Model B. Consequently, the capacity-achieving input is the same for Model A and Model B. We also note that Model A is a special case of Model B with $\gamma^2 = 0$. As such, the capacity expression in Theorem \ref{theorem:noncoherent_rician_case} simplifies to that of Theorem \ref{theorem:AWGN_case} when the noncoherent fading component is removed.


\begin{theorem} \label{theorem:Rayleigh_CSIR_case} When CSI is only available at the receiver, the ergodic capacity of a Rayleigh fading channel with $b$-bit phase-quantized output (Model C) is
\begingroup
\allowdisplaybreaks
\begin{align}
    C_{\mathrm{ergodic}} = b - \mathbb{E}_{|G|,\angle G}\left[w\left(\frac{|g|^2 P}{\sigma^2},\angle g,b\right)\right]\;\mathrm{bits/cu},
\end{align}
\endgroup
where $w(\cdot,\cdot,\cdot)$ is given in Definition \ref{definition:phase_quantization_entrop}. Moreover, any $\frac{2\pi}{2^b}$-symmetric input distribution with a single amplitude level $\sqrt{P}$ achieves the ergodic capacity of this channel.
\end{theorem}

\noindent \textbf{Remark on Theorem \ref{theorem:Rayleigh_CSIR_case}:} Unlike Model A and Model B, the input distribution that achieves ergodic capacity in Model C is not unique. Moreover, the ergodic capacity can be achieved by a continuous input distribution such as a circle with radius $\sqrt{P}$ and so the capacity-achieving distribution does not have to be discrete. Nonetheless, there is no loss of optimality when a $2^b$-PSK scheme is used.

\begin{theorem}\label{theorem:Rayleigh_CSIT_case} When CSI is available at both the transmitter and receiver, the ergodic capacity of a Rayleigh fading channel with $b$-bit phase-quantized output (Model D) is
\begingroup
\allowdisplaybreaks
\begin{align}
     C_{\mathrm{ergodic}} = b - \mathbb{E}_{|G|}\left[w\left(\frac{|g|^2 P}{\sigma^2},\frac{\pi}{2^b},b\right)\right]\;\mathrm{bits/cu},
\end{align}
\endgroup
where 
\begin{align}P^*(|g|^2) = \begin{cases}\frac{\left[\frac{\partial w}{\partial \nu}\right]^{-1}\left(-\frac{\eta}{|g|^2}\right)}{|g|^2/\sigma^2},\;\mathrm{ if }\;\;|g|^2>\frac{\eta}{w'_{\min}}\\
\qquad0\qquad\;\;,\;\mathrm{ otherwise}
\end{cases},
\end{align}
$w'_{\min}$ is given by
\begin{align}
    w'_{\min} =- \frac{2^{2b-1}}{2\pi}\sin^2\left(\frac{\pi}{2^b}\right),
\end{align}
and $\eta$ is a power control parameter that satisfies
\begin{align}
    \int_{\frac{\eta}{w'_{\min}}}^{\infty}\left[\frac{\partial w}{\partial \nu}\right]^{-1}\left(-\frac{\eta}{|g|^2}\right) \frac{f_{|G|}(|g|)}{|g|^2}\;d|g| = \frac{P}{\sigma^2}.
\end{align}
The function $\left[\frac{\partial w}{\partial \nu}\right]^{-1}\left(\cdot\right)$ is the inverse of the first-order derivative of $w\left(\nu,\frac{\pi}{2^b},b\right)$ with respect to $\nu$ and $f_{|G|}(|g|)$ is the probability density function (PDF) of a Rayleigh distribution. The capacity-achieving input distribution is a rotated $2^b$-PSK with equiprobable symbols and optimal power control given by
\begin{align}
    f_{X|G}^*(x|g) =& \bigg\{\frac{1}{2^b}\Big|x = \sqrt{P^*(|g|^2)}e^{j\left(\frac{2\pi(k+0.5)}{2^b}-\angle g\right)},\nonumber\\
    &\quad\forall k\in\{0,...,2^{b}-1\}\bigg\}.
\end{align}
\end{theorem}

\noindent \textbf{Remark on Theorem \ref{theorem:Rayleigh_CSIT_case}:} The power control scheme described in Theorem \ref{theorem:Rayleigh_CSIT_case} is fundamentally different from the ``water-pouring in time" power control \cite{Goldsmith:1997} and mercury/waterfilling algorithm \cite{Lozano:2006}. This distinction is discussed in Section \ref{subsection:coherent_CSIT}. It is observed that implementing the optimal power control when $|g|^2$ is also known at the transmitter provides additional boost, albeit a small amount, to the ergodic capacity (See Figure \ref{fig:ergodic_cap}).

It is worth mentioning at this point that the optimal distributions in Theorems \ref{theorem:AWGN_case}-\ref{theorem:Rayleigh_CSIT_case} have a fixed structure at all SNR regimes. This is in stark constrast to the numerical results for multi-bit I/Q quantization presented in \cite{Singh:2009,Vu:2019}. In particular, the capacity-achieving distribution constructed using numerical methods starts with a 2-mass point (4-mass point) distribution for real channels (complex channels) at vanishing SNR. Then, the number of mass points in the distribution is increased whenever the SNR exceeds a certain value. On the other hand, there is no loss of optimality in channels with phase quantization at the output if we still use a $2^b$-mass point distribution at vanishing SNR. This property is not obvious when only 1-bit I/Q ADC is considered.

The proofs of these channel capacity theorems are presented in the subsequent sections. We start with the simplest case of complex Gaussian channel with fixed gain in Section \ref{section-awgn_case} and then extend the results to the other three system models depicted in Figure \ref{fig:sys_model_gen}. The noncoherent fading case, coherent fading case with CSI at receiver only, and coherent fading case with CSI at both transmitter and receiver are discussed in Sections \ref{section-noncoherent}, \ref{subsection:coherent_CSIR}, and \ref{subsection:coherent_CSIT}, respectively. 


\section{Capacity of Gaussian Channels with Fixed Gain and Phase-Quantized Output}\label{section-awgn_case}

We derive the relevant quantities for the analysis of Model A. The relationship between the quantizer output and the channel input can be written as

\[Y = \mathcal{Q}_{b}(V) = \mathcal{Q}_{b}(g_{\text{LoS}}X+Z).
\]

The conditional PDF $f_{V|X}(v|x)$ is given by
\begin{equation}
       f_{V|X}(v|x) = \frac{1}{\pi \sigma^2}\exp\left(-\frac{|v- g_{\text{LoS}}x|^2}{\sigma^2}\right).
\end{equation}
Note that in a receiver with phase-quantized output, we discard the magnitude component. Suppose we represent $X$ and $V$ in polar form (i.e. $v = re^{j\phi}$ and $x = \sqrt{\alpha} e^{j\beta}$) and let $\beta' = \angle g_{\text{LoS}} + \beta$, $\alpha' = |g_{\text{LoS}}|^2\alpha$. The conditional PDF $f_{\Phi|X}(\phi|x)$ (or $f_{\Phi|A,B}(\phi|\alpha,\beta)$) can be written as
\begin{align}
    f_{\Phi|X}(\phi|x) =& \int_{0}^{\infty}r f_{V|X}\left(v=re^{j\phi}\Big|x = \sqrt{\alpha} e^{j\beta}\right)\;dr\nonumber\\\label{eq:prob_phi_given_x}
     =& \frac{e^{-\frac{\alpha'}{\sigma^2}}}{2\pi}+\frac{\sqrt{\alpha'}\cos\left(\phi - \beta'\right)e^{-\frac{\alpha'}{\sigma^2}\sin^2\left(\phi - \beta'\right)}}{\sqrt{\pi}\sigma}\nonumber\\
     &\quad\cdot\left[1-Q\left(\sqrt{2\frac{\alpha'}{\sigma^2}}\cos\left(\phi - \beta'\right)\right)\right],
\end{align}
where the first line follows from marginalizing $R$ and the last line is obtained from \cite[equation (10)]{Fu:2008}. Note the similarities between the structure of (\ref{eq:prob_phi_given_x}) and (\ref{eq:prob_phi_given_nu}). This implies that the conditional PMF $p_{Y|X}(y|x)$ (or $p_{Y|A,B}(y|\alpha,\beta)$) can be expressed as
\begingroup
\allowdisplaybreaks
\begin{align}\label{eq:prob_y_given_x}
        p_{Y|X}(y|x) =& \int_{ \mathcal{R}^{\text{PH}}_{y}} f_{\Phi|A,B}\left(\phi|\alpha,\beta\right)\;d\phi\nonumber\\
        =& \int_{\frac{2\pi}{2^b}y-\pi-\beta'}^{\frac{2\pi}{2^b}(y+1)-\pi-\beta'} f_{\bar{\Phi}|A}\left(\bar{\phi}|\alpha\right)\;d\bar{\phi} \quad(\bar{\phi} = \phi - \beta')\nonumber\\
        =& W_y^{(b)}\left(\frac{|g_{\text{LoS}}|^2\alpha}{\sigma^2},\beta'\right),
\end{align}
\endgroup
where the second line follows from the change of variable $\bar{\phi} = \phi - \beta'$. This moves $\beta'$ to the integral bounds. The third line follows from Definition \ref{definition:phase_quantization_func}. Equation (\ref{eq:prob_y_given_x}) provides an operational meaning to $W_y^{(b)}\left(\nu,\theta\right)$ as the conditional PMF of the $b$-bit phase quantizer output $Y$ when $x = \sqrt{\nu}e^{j\theta}$, $Z\sim\mathcal{CN}(0,1)$ and $g_{\text{LoS}} = 1$.

Now, consider a complex input distribution $F_X(x)$ with density function $f_X(x)$. For a given $F_X$, the PMF of $Y$ is
\begin{equation}\label{eq:pmf_y}
        \begin{split}
             p(y;F_X) =& \int_{\mathbb{C}} W_y^{(b)}\left(\frac{|g_{\text{LoS}}|^2\alpha}{\sigma^2},\beta'\right)\;dF_X
        \end{split}
\end{equation}
for all $y\in\{0\cdots,2^{b}-1\}$. Given the probability quantities, we can now express the mutual information between $X$ and $Y$ as follows:
\begingroup
\allowdisplaybreaks
\begin{align}\label{eq:MI_X_Y}
        I(X;Y) =& I(F_X;p_{Y|X}) = I(F_X) \nonumber\\
               =& H\left(Y\right) - H\left(Y|X\right),
\end{align}
where
\begin{align*}
H(Y)=& -\int_{\mathbb{C}}\sum_{y=0}^{2^b-1}W_y^{(b)}\left(\frac{|g_{\text{LoS}}|^2\alpha}{\sigma^2},\beta'\right)\log p(y;F_X)\;dF_X\nonumber\\
=&-\sum_{y=0}^{2^b-1}p(y;F_X)\log p(y;F_X)\;\nonumber
\end{align*}
and
\begin{align*}
        H(Y|X)=& -\int_{\mathbb{C}}\sum_{y=0}^{2^b-1}W_y^{(b)}\left(\frac{|g_{\text{LoS}}|^2\alpha}{\sigma^2},\beta'\right)\nonumber\\
        &\qquad\cdot\log W_y^{(b)}\left(\frac{|g_{\text{LoS}}|^2\alpha}{\sigma^2},\beta'\right)\;dF_X\nonumber\\
        =& \int_{\mathbb{C}}w\left(\frac{|g_{\text{LoS}}|^2\alpha}{\sigma^2},\beta',b\right)\;dF_X.\nonumber
\end{align*}
\endgroup
The above expression gives an operational meaning of $w(\nu,\theta,b)$ as the conditional entropy of the $b$-bit phase quantizer output $Y=y$ given $x = \sqrt{\nu}e^{j\theta}$, $Z\sim\mathcal{CN}(0,1)$ and $g_{\text{LoS}} = 1$. Moreoever, since it is clear from the context what the channel law $p_{Y|X}$ is, we use the notation $I(F_X)$.

The capacity for a given power constraint is the supremum of mutual information between $X$ and $Y$ over the set of all input distributions $F_X$ satisfying the power constraint $\mathbb{E}[|X|^2] \leq P$. In other words,
\begin{equation}\label{eq:cap_def}
    C = \sup_{F_X \in \Omega} I(F_X) = I(F_X^*),
\end{equation}  
where $\Omega$ is the set of all input distributions which have average power less than or equal to $P$ and $F_X^*$ is the optimal input distribution. The mutual information is concave with respect to $F_X$ \cite[Theorem 2.7.4]{Cover:2006_IT} and the power constraint ensures that $\Omega$ is convex and compact with respect to weak* topology\footnote{This is the coarsest topology in which all linear functionals of $dF_X$ of the form $\int g(x)dF_X$, where $g(x)$ is a continuous function, are continuous.} \cite{Abou-Faycal:2001}. The existence of $F_X^*$ is equivalent to showing that $I(F_X)$ is continuous over $F_X$. This is straightforward to prove  because of the finite cardinality of the phase quantizer output and the proof closely follows the method of \cite[Appendix A]{Singh:2009_techreport} and \cite[Lemma 1]{Vu:2019}.


\subsection{Properties of $f_{\Phi|N}(\phi|\nu)$, $W_y^{(b)}(\nu,\theta)$, and $w(\nu,\theta,b)$}
As primer to the derivation of the optimal input distribution, key lemmas and propositions are presented about the symmetry of $f_{\Phi|N}(\phi|\nu)$ and $W_y^{(b)}(\nu,\theta)$ and about the convexity and monotonicity of $w(\nu,\theta,b)$. 

\begin{lemma}\label{lemma:symmetry_p_phi_given_nu}
The function $f_{\Phi|N}(\phi|\nu)$ is even-symmetric. Moreover, if $\nu > 0$, $f_{\Phi|N}(\phi|\nu)$ is an increasing function of $\phi$ for $\phi \in (-\pi,0)$ and a decreasing function of $\phi$ for $\phi \in (0,\pi)$.
\end{lemma}
\begin{proof}
See Appendix \ref{appendix_A}.
\end{proof}
\begin{lemma}\label{lemma:symmetry_Wy}
The function $W_y^{(b)}(\nu,\theta)$ satisfies the following properties:
\begingroup\allowdisplaybreaks
\begin{align*}
    (i) && \; W_y^{(b)}\left(\nu,\theta + \frac{2\pi k}{2^b}\right) =& W_{y-k}^{(b)}\left(\nu,\theta\right)&&,\forall k\in\mathbb{Z}\\
    (ii) && \;W_{2^{b-1}-y}^{(b)}\left(\nu,\frac{\pi}{2^b}\right) =& W_{2^{b-1}+y}^{(b)}\left(\nu,\frac{\pi}{2^b}\right)  \\
    (iii) && \;W_{2^{b-1}-y}^{(b)}\left(\nu,0\right) =& W_{2^{b-1}-1+y}^{(b)}\left(\nu,0\right).
\end{align*}
\endgroup
\end{lemma}
\begin{proof}
See Appendix \ref{appendix_B}.
\end{proof}
Lemma \ref{lemma:symmetry_p_phi_given_nu} shows that $f_{\Phi|N}(\phi|\nu)$ is an even function of $\phi$ for arbitrary $\nu$. Lemma \ref{lemma:symmetry_Wy}.i states that shifting the input by $\frac{2\pi k}{2^b}$ corresponds to a shift in the phase quantizer output by $-k$. Meanwhile, Lemma \ref{lemma:symmetry_Wy}.ii and \ref{lemma:symmetry_Wy}.iii identify some symmetry properties of $W_y^{(b)}(\nu,\theta)$ when $\theta = 0$ and $\theta = \frac{\pi}{2^b}$. We show in the next two propositions that $w(\nu,\theta,b)$ in  (\ref{eq:phase_quantization_entrop}) is a decreasing convex function of $\nu$.

\begin{proposition}\label{proposition:monotonicity_w}
The function $w(\nu,\theta,b)$ is strictly decreasing on $\nu$ for all $\theta \in \left[-\pi,\pi\right)\text{ and } b\geq 1$.
\end{proposition}
\begin{proof}
See Appendix \ref{appendix_C}.
\end{proof}
\begin{proposition}\label{proposition:convexity_w}
The function $w(\nu,\theta,b)$ is strictly convex on $\nu$ for  all $\theta \in \left[-\pi,\pi\right)\text{ and } b\geq 1$.
\end{proposition}
\begin{proof}
See Appendix \ref{appendix_E}.
\end{proof}
Propositions \ref{proposition:monotonicity_w} and \ref{proposition:convexity_w} are used later to show mathematically that the optimal input distribution has a constant amplitude and uses full transmit power.

\subsection{Structural Properties of the Optimal Input: Symmetry and the Kuhn-Tucker Condition}
Given that an optimal input distribution exists in the set $\Omega$, this subsection focuses on finding the optimum $F_X$ and establishing the capacity of Gaussian channel with phase-quantized output. We first show that the capacity-achieving input distribution is $\frac{2\pi}{2^b}$-symmetric.
\begin{proposition}\label{proposition:symmetry_distribution}
For any input distribution $F_X$, we define another input distribution as
\begin{equation}
    F_X^{s} = \frac{1}{2^b}\sum_{i = 0}^{2^b-1}F_X(xe^{j\frac{2\pi i}{2^b}}),
\end{equation}
which is a $\frac{2\pi}{2^b}$-symmetric distribution. Then, $I(F_X^{s}) \geq I(F_X)$. Under this input distribution, $H(Y)$ is maximized and is equal to $b$.
\end{proposition}
\begin{proof}
See Appendix \ref{appendix_G}.
\end{proof}

Using Proposition \ref{proposition:symmetry_distribution}, we can simply search for $F_X^*$ in the set $\Omega_s$ defined as
\begin{align}\label{eq:Omega_s}
    \Omega_s = \left\{F_X\in \Omega \;\;\Big| \;\;F_X\sim F_X(xe^{j\frac{2\pi k}{2^b}}),k\in\mathbb{Z}\right\}.
\end{align}
Consequently, the capacity in (\ref{eq:cap_def}) can be simplified to
\begin{align}\label{eq:capacity_symmetric_in}
     C =& \sup_{F_X\in\Omega_s} I(F_X) \nonumber\\
     =& \sup_{F_X\in\Omega_s} \left\{b - \int_{\mathbb{C}}w\left(\frac{|g_{\mathrm{LoS}}|^2\alpha}{\sigma^2},\beta',b\right)\;dF_X\right\}\nonumber\\
     =& b - \inf_{F_X\in\Omega_s} \left\{\int_{\mathbb{C}}w\left(\frac{|g_{\mathrm{LoS}}|^2\alpha}{\sigma^2},\beta',b\right)\;dF_X\right\}
\end{align}


We next establish necessary and sufficient conditions on the optimal input distribution via the Kuhn-Tucker Conditions (KTC). We first establish that the Lagrange Multiplier Theorem can be applied to our problem. The following lemma shows that $I(F_{X})$ is weakly differentiable. That is, for a given $F_{X}^0\in\Omega_{s}$ and $\lambda\in [0,1]$, the weak derivative \begin{equation}\label{eq:weak_differentiability}
    I'_{F_X^0}(F_X) = \lim_{\lambda \rightarrow 0 }\frac{I\left((1-\lambda)F_X^0 + \lambda F_X\right) - I(F_X^0)}{\lambda}
\end{equation}
exists $\forall F_X \in \Omega_s$.
\begin{lemma}\label{lemma:weak_differentiability}
The functional 
\[I(F_{X}) = b - \int_{\mathbb{C}}w\left(\frac{|g_{\mathrm{LoS}}|^2\alpha}{\sigma^2},\beta',b\right)\;dF_X\]
is weakly differentiable with respect to $F_X$ and its weak derivative with respect to a point $F_{X}^0$ is 
\begin{align}
 I'_{F_X^0}(F_X) = I(F_X) - I(F_X^0).
\end{align}
\end{lemma}
\begin{proof}
See Appendix \ref{appendix_I}
\end{proof}
$I(F_X)$ is now a sum of a constant and a linear functional of $F_X$ when the input distribution is drawn from $\Omega_s$. Combining this with the weak differentiability of $I(F_X)$ and convexity and compactness of $\Omega_s$ implies the existence of a non-negative Lagrange multiplier $\mu$ such that
\begin{align*}
    C = \sup_{F_X\in \Omega_{s}} I(F_X) = \sup_{F_X\in \Omega^0_{s}} I(F_X) - \mu \phi(F_X),
\end{align*}
where $\phi(F_X) = \int |x|^2dF_X - P$ and $\Omega^0_{s}$ is the set of all $\frac{2\pi}{2^b}$-symmetric distributions. It is easy to show that $\phi(F_X)$ is also weakly differentiable over $F_X$ (i.e. $\phi_{F_X^0}'(F_X) = \phi(F_X) - \phi(F_X^0)$) and so is $I(F_X)-\mu\phi(F_X)$. Moreover, $I(F_X)-\mu\phi(F_X)$ can also be written as a sum of a constant and a linear term in $F_X$. Thus, a distribution $F_X^*$ is optimal if for all $F_X$, we have
\begin{align*}
        I'_{F_X^*}(F_X) - \mu\phi'_{F_X^*}(F_X) \leq& 0\\
        I(F_X)  - \mu\int_{\mathbb{C}}|x|^2\;dF_{X}\leq& I(F_{X}^*)  - \mu\int_{\mathbb{C}}|x|^2\;dF_{X}^*.
\end{align*}
The above inequality simplifies to
\begin{align*}
    b - \int_{\mathbb{C}}w\left(\frac{|g_{\mathrm{LoS}}|^2\alpha}{\sigma^2},\beta',b\right)\;dF_X - \mu\int_{\mathbb{C}}|x|^2\;dF_X\leq C - \mu P,
\end{align*}
where we used (\ref{eq:capacity_symmetric_in}), (\ref{eq:cap_def}), and the complementary slackness of the constraint (having $\int_{\mathbb{C}}|x|^2dF_X^*$ strictly less than $P$ makes $\mu = 0$ and the expression still holds). Finally, using the same contradiction argument in \cite[Theorem 4]{Abou-Faycal:2001}, noting that $|x| = \sqrt{\alpha}$, and after some algebraic manipulation, the KTC can be established as
\begin{equation}\label{eq:KTC}
  C - b + \mu(\alpha- P) +w\left(\frac{|g_\mathrm{LoS}|^2\alpha}{\sigma^2},\beta',b\right) \geq 0,
\end{equation}
and equality is achieved when $x = \sqrt{\alpha}e^{j\beta}$ is a mass point of $F_X^*$. The KTC will be used to prove some additional properties of $F_X^*$ as well as identify which mass points belong to $F_X^*$.

\subsection{Structural Properties of the Optimal Input: Boundedness and Discreteness}
Using the KTC, we prove that the optimal input distribution of a Gaussian channel with phase-quantization at the output must have a bounded and discrete support. The boundedness property of the optimal input is proven using the KTC and the asymptotic behavior of $H(Y|X=\sqrt{\alpha}e^{j\beta})$ as $\alpha\rightarrow \infty$. The key idea is to consider two cases of the Lagrange multiplier (i.e. $\mu = 0$ and $\mu > 0$) and show that in either case, equality cannot be achieved in (\ref{eq:KTC}) for an unbounded $\alpha$.
\begin{lemma}\label{lemma:boundedness}
The optimal input distribution $F_X^*$ has a bounded support.
\end{lemma}
\begin{proof}
See Appendix \ref{appendix_H}
\end{proof}
We use this boundedness property in Lemma \ref{lemma: discreteness} to
show that $F_X$ is discrete and identify an upper bound on the number of mass points. The proof closely follows the example application of Dubin's Theorem \cite{Dubins:1962} presented in \cite[Section II-C]{Witsenhausen:1980} but with consideration of the average power constraint.
\begin{lemma}\label{lemma: discreteness}
The support set of $F_X^*$ is discrete and contains at most $2^b+1$ mass points.
\end{lemma}
\begin{proof}
See Appendix \ref{appendix_K}.
\end{proof}
\subsection{Structural Properties of the Optimal Input: Location and Amplitude of Optimal Mass Points}
At this point, we have gained some insights about the symmetry, boundedness, and discreteness of the optimal distribution $F_X^*$. Since $F_X^*$ should be discrete with at most $2^b+1$ mass points (Lemma \ref{lemma: discreteness}), there are only two possible general input structures that satisfy the $\frac{2\pi}{2^b}$-symmetry (Proposition \ref{proposition:symmetry_distribution}): (1) A $2^b$-phase shift keying input distribution and (2) an on-off $2^b$ phase shift keying input distribution. The next proposition narrows down the choices for $F_{X}^*$ to input structure (1).
\begin{proposition}\label{proposition:constant_amplitude}
The optimal input distribution should have a single amplitude level per phase, denoted as $\sqrt{\alpha_{\beta}}$. 
\end{proposition}
\begin{proof}
The capacity in (\ref{eq:cap_def}) can be expressed as
\begingroup
\allowdisplaybreaks
\begin{align}\label{eq:capacity_bayes_rule}
    C =& b - \inf_{F_X\in\Omega_s} \mathbb{E}_{B}\left[\mathbb{E}_{A|B}\left[w\left(\frac{|g_{\mathrm{LoS}}|^2\alpha}{\sigma^2},\beta',b\right)\right]\right],
\end{align}
\endgroup
where we used Bayes' rule to express the complex PDF $f_X(x) = f_{A,B}(\alpha,\beta)$ as $f_{A|B}(\alpha|\beta)f_{B}(\beta)$ and perform the complex expectation as two real-valued expectations over $\alpha|\beta$ and $\beta$. Due to Proposition \ref{proposition:convexity_w}, Jensen's inequality can be applied to (\ref{eq:capacity_bayes_rule}). That is,
\begin{align*}
    &\mathbb{E}_{B}\left[\mathbb{E}_{A|B}\left[w\left(\frac{|g_{\mathrm{LoS}}|^2\alpha}{\sigma^2},\beta',b\right)\right]\right] \\&\qquad\qquad\qquad
    \geq \mathbb{E}_{B}\left[w\left(\frac{|g_{\mathrm{LoS}}|^2\mathbb{E}_{A|B}\left[\alpha\right]}{\sigma^2},\beta',b\right)\right],
\end{align*}
with equality if $\alpha$ is a deterministic function of $\beta'$ (effectively, this also means $\alpha$ is a deterministic function of $\beta$). Letting $\alpha_{\beta} = \mathbb{E}_{A|B}[\alpha]$ completes the proof.
\end{proof}
Due to Proposition \ref{proposition:constant_amplitude}, a mass point at the origin cannot be included in an optimal input distribution since it is an additional amplitude level for all $\beta \in [-\pi,\pi]$. Thus, the only candidate input structure is a $2^b$-PSK. Moreover, due to Proposition \ref{proposition:monotonicity_w}, for any PSK amplitude $\sqrt{\alpha^{(a)}} < \sqrt{P}$, we can find another PSK amplitude $\sqrt{\alpha^{(b)}}\in (\sqrt{\alpha^{(a)}},\sqrt{P}]$ such that
\[\mathbb{E}_{B}\left[w\left(\frac{|g_{\text{LoS}}|^2\alpha^{(a)}}{\sigma^2},\beta',b\right)\right] > \mathbb{E}_{B}\left[w\left(\frac{|g_{\text{LoS}}|^2\alpha^{(b)}}{\sigma^2},\beta',b\right)\right].\]
Thus, the amplitude of the $2^b$-PSK constellation should be $\sqrt{P}$ to attain capacity.

Using (\ref{eq:KTC}), we identify in Proposition \ref{proposition:optimal_loc} the location of the optimal mass points. By Proposition \ref{proposition:symmetry_distribution}, we can limit our search of $\beta^*$ in $[0,\frac{2\pi}{2^b})$ since if $\beta^* \in [0,\frac{2\pi}{2^b})$ is optimal, so are $\beta^* + \frac{2\pi k }{2^b}$ for $k\in\{1,\cdots,2^{b}-1\}$. Moreover, the optimal input distribution has a single mass point inside $[0,\frac{2\pi}{2^b})$ as a consequence of Lemma \ref{lemma: discreteness} and Proposition \ref{proposition:symmetry_distribution}.

\begin{proposition}\label{proposition:optimal_loc} The set containing the angles of the optimum mass points $x^*\in F_X^*$ is given by
\begin{align}
    \beta^* = \left\{\frac{2\pi(k+0.5)}{2^b} - \angle g_{\mathrm{LoS}}\right\}_{k = 0}^{2^b-1}.
\end{align}
\end{proposition}
\begin{proof}
See Appendix \ref{appendix_J}.
\end{proof}

The proof of the capacity-achieving input in Theorem \ref{theorem:AWGN_case} is completed by combining Lemma \ref{lemma: discreteness} and Propositions \ref{proposition:symmetry_distribution}, \ref{proposition:constant_amplitude}, and \ref{proposition:optimal_loc}. The use of the capacity-achieving input results in the capacity expression given in (\ref{eq:cap_AWGN}). Theorem \ref{theorem:AWGN_case} generalizes the previous results for $b = 1$ \cite[Theorem 2]{Singh:2009} and for $b = 2$ \cite[Lemma 1]{Mo:2014} to $b \geq 3$.
\begin{figure*}
    \centering
    \includegraphics[scale = .75,draft=false]{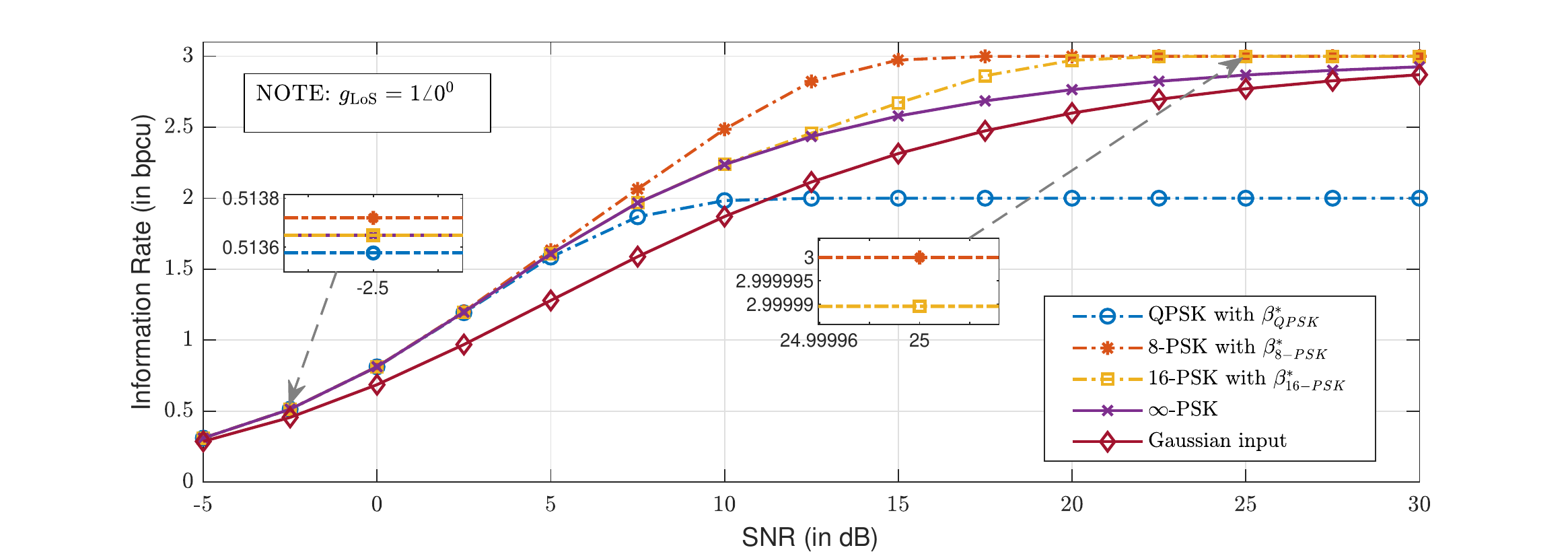}
     \caption{Information rates achieved by different modulation schemes when $g_{\text{LoS}} = 1 \angle 0^0$ and $b = 3$. Note that 8-PSK with optimal $\beta$ is capacity-achieving.}
    \label{fig:rate_vs_snr_AWGN}
\end{figure*}

To demonstrate the optimality of the signaling scheme, Figure \ref{fig:rate_vs_snr_AWGN} compares the rates achieved by using 4,8,16, and $\infty$-PSK (a circle) with equiprobable mass points on a Gaussian channel with 3-bit phase-quantized output. Each PSK constellation is rotated by a $\beta^*$ that maximizes the mutual information between $X$ and $Y$. The information rate of Gaussian input is also included. It can be observed that 8-PSK with optimal $\beta$ achieves the highest rate among all modulation orders considered.

\section{Capacity of Noncoherent Rician Fast Fading Channels with Phase-Quantized Output}\label{section-noncoherent}

We now extend the results of Section \ref{section-awgn_case} to noncoherent fading channels. We consider a small-scale Rician fading model with phase-quantized output (see Figure \ref{fig:sys_model_NoCSI}) to represent a wide range of communication channels including the static complex Gaussian channel and Rayleigh channel. The input-output relationship of the unquantized noncoherent Rician fading channel can be expressed as \cite{Gursoy:2005}:
\begin{align}
    V =& GX + Z\nonumber\\
    =& g_{\text{LoS}}X + \bar{G}X + Z,
\end{align}
where we decomposed $G$ as sum of a complex constant $g_{\text{LoS}}$, which represents the gain and direction of LoS component, and $\bar{G} \sim \mathcal{CN}(0,\gamma^2)$ , which accounts for the zero-mean nLoS component. Consequently, the conditional PDF $f_{V|X}(v|x)$ is given by
\begin{equation}
        f_{V|X}(v|x) = \frac{1}{\pi (\gamma^2|x|^2+\sigma^2)}\exp\left(-\frac{|v- g_{\text{LoS}}x|^2}{\gamma^2|v|^2+\sigma^2}\right).
\end{equation}
Suppose we represent $X$ and $V$ in the polar form (i.e. $v = re^{j\phi}$ and $x = \sqrt{\alpha} e^{j\beta}$) and define a function $\rho(\alpha)$ and a variable $\beta'$ as
\begin{equation}\label{eq:noncoherent_SINR}
     \rho(\alpha) = \frac{|g_{\text{LoS}}|^2\alpha}{\gamma^2\alpha + \sigma^2}\;\text{ and }\;\beta' = \beta + \angle g_{\text{LoS}}.
\end{equation}
Then, following the analysis in Section \ref{section-awgn_case} gives us the conditional probability mass function (PMF) $p_{Y|X}(y|x)$ (or $p_{Y|A,B}(y|\alpha,\beta)$)
\begin{equation}\label{eq:prob_y_given_x_noncoherent}
    \begin{split}
        p_{Y|X}(y|x) =& W_y^{(b)}\left(\rho(\alpha),\beta'\right).
    \end{split}
\end{equation}
This implies that the mutual information can be expressed as
\begin{align}\label{eq:MI_X_Y_noncoherent}
        I(X;Y) =& I(F_X)\nonumber \\
        =& H\left(Y\right) - H\left(Y|X\right),
\end{align}
where
\begingroup
\allowdisplaybreaks
\begin{align*}
    H(Y)=& -\int_{\mathbb{C}}\sum_{y=0}^{2^b-1}W_y^{(b)}\left(\rho(\alpha),\beta'\right)\log p(y;F_X)\;dF_X
\end{align*}
and
\begin{align*}
    H(Y|X) =&\int_{\mathbb{C}}w\left(\rho(\alpha),\beta'\right)\;dF_X.\nonumber
\end{align*}
\endgroup
The difference between (\ref{eq:MI_X_Y}) and (\ref{eq:MI_X_Y_noncoherent}) is only on the amplitude component (i.e. we used $\rho(\alpha)$ instead of $\frac{|g_{\text{LoS}}|^2\alpha}{\sigma^2}$). Thus, Lemmas \ref{lemma:symmetry_p_phi_given_nu}, \ref{lemma:symmetry_Wy}, \ref{lemma:weak_differentiability} and Propositions \ref{proposition:symmetry_distribution}, \ref{proposition:optimal_loc} continue to hold in this case as these properties are invariant of $\nu$ in $W_{y}^{(b)}(\nu,\theta)$ and $w(\nu,\theta,b)$. The following corollary of Propositions \ref{proposition:monotonicity_w} and \ref{proposition:convexity_w} shows that we can also apply these two propositions in the noncoherent setting.
\begin{corollary}\label{corollary:convex_monotonic} The function $w(\rho(\nu),\theta,b)$ where $\rho(\nu)$ is some strictly concave and strictly increasing function of $\nu$, is a strictly convex decreasing function of $\nu$ for all $\theta\in [0,\frac{2\pi}{2^b})$, $b\geq 1$.
\end{corollary}
\begin{proof}
Since $w(\nu,\theta,b)$ is strictly decreasing (Proposition \ref{proposition:monotonicity_w}), the function composition $w(\rho(\nu),\theta,b)$ is strictly decreasing. Moreover, since $w(\nu,\theta,b)$ is also strictly convex (Proposition \ref{proposition:convexity_w}), the function composition $w(\rho(\nu),\theta,b)$ is strictly convex \cite[Section 3.2.4]{Boyd:2004_convex-opt}.
\end{proof}
Since $\rho(\alpha)$ is a strictly concave increasing function, Propositions \ref{proposition:monotonicity_w} and \ref{proposition:convexity_w} also hold. It then follows that Proposition \ref{proposition:constant_amplitude} is a necessary condition of the optimal input distribution. 
The boundedness of the optimal input in the noncoherent case can be proven in a similar manner as Lemma \ref{lemma:boundedness} but with 
\begin{align*}
\lim_{\alpha \rightarrow \infty} W_y^{(b)}\left(\rho(\alpha),\beta\right) =  W_y^{(b)}\left(\kappa,\beta\right) 
\end{align*}
and
\begin{align*}
    \lim_{\alpha\rightarrow\infty}\;w\left(\rho(\alpha),\beta,b\right) = w\left(\kappa,\beta,b\right).
\end{align*}
Furthermore, we also use the KTC
\begin{equation}\label{eq:KTC_noncoherent}
  C - b + \mu(\alpha- P) +w\left(\rho(\alpha),\beta',b\right) \geq 0
\end{equation}
instead of (\ref{eq:KTC}). The proof of discreteness of the optimal input follows similarly to the proof of Lemma \ref{lemma: discreteness}. Applying similar reasoning as in the analysis of the capacity-achieving input distribution of Model A completes the proof of Theorem \ref{theorem:noncoherent_rician_case}. It can be seen that for sufficiently high SNR, the capacity approaches
\begingroup
\allowdisplaybreaks
\begin{align}
    C \approx\;& b - w\left(\frac{|g_{\text{LoS}}|^2}{\gamma^2},\frac{\pi}{2^b},b\right)\nonumber\\
    =\;&b - w\left(\kappa,\frac{\pi}{2^b},b\right)\;\text{bits/cu}.
\end{align}
\endgroup
The limits for the special case of complex Gaussian with fixed gain ($\kappa = \infty$) and noncoherent fast fading Rayleigh channel ($\kappa = 0$) are $b$ and 0, respectively. In fact, the capacity of noncoherent fast fading Rayleigh channel with phase quantization at the output is 0 in all SNR regimes. This is consistent with the results of \cite[Equation (64)]{Vu:2019} for $b = 2$.

\section{Ergodic Capacity of Fading Channels with with Phase-Quantized Output and Channel State Information}\label{section-coherent}

\subsection{Rayleigh Fading with CSI at the receiver only (CSIR)}\label{subsection:coherent_CSIR}
We now consider Model C in Figure \ref{fig:sys_model_CSIR} where the transmitted complex-valued signal $X$ ($x= \sqrt{\alpha}e^{j\beta}$) is also subjected to a random fading gain $G\sim \mathcal{CN}(0,\gamma^2)$ and CSI is only granted at the receiver. We assume that the fading process is ergodic. Since the receiver knows $G$, we treat the random variable pair $(Y,G)$ as the channel output. Equivalently, the input-output mutual information can be expressed as
\begingroup
\allowdisplaybreaks
\begin{align*}
    I(X;(Y,G)) =& I(X;G) + I(X;Y|G)\\
    =& I(X;Y|G)\\
    =& H(Y|G) - H(Y|X,G),
\end{align*}
where
\begin{align*}
    H(Y|G)=& \mathbb{E}_{G}\bigg\{-\int_{\mathbb{C}}\sum_{y=0}^{2^b-1}W_y^{(b)}\left(\frac{|g|^2\alpha}{\sigma^2},\beta + \angle g\right)\\
    &\qquad\qquad\qquad\cdot\log p(y;F_X)\;dF_X\bigg\}
\end{align*}
and
\begin{align*}
    H(Y|X,G)=&\mathbb{E}_{G}\left\{\int_{\mathbb{C}}w\left(\frac{|g|^2\alpha}{\sigma^2},\beta+\angle g,b\right)\;dF_X\right\}.
\end{align*}
\endgroup
The first line follows from the chain rule of mutual information and the second line follows from the independence between the channel gain and the transmitted signal. We used the notation $I(F_X|G)$ in the third line since the conditional mutual information $I(X;Y|G)$ is a result of choosing a specific input distribution $F_X$. Our goal is to identify the optimal input distribution $F_X^*$ that achieves the ergodic capacity
\begin{equation}\label{eq:ergodi_cap_def}
    C_{\text{ergodic}} = \sup_{F_X\in\Omega}I(F_X|G) = I(F_X^*|G).
\end{equation}
We now prove a corollary of Proposition \ref{proposition:symmetry_distribution} showing that the optimal input distribution in any fading channels with phase-quantized output should be $\frac{2\pi}{2^b}$-symmetric.
\begin{corollary}\label{corollary:symmetry_fading}
For any input distribution $F_X$, we define another input distribution as
\begin{equation*}
    F_X^{s} = \frac{1}{2^b}\sum_{i = 0}^{2^b-1}F_X(xe^{j\frac{2\pi i}{2^b}}),
\end{equation*}
which is a $\frac{2\pi}{2^b}$-symmetric distribution. Then, $I(F_X^{s}|G) \geq I(F_X|G)$. Under this input distribution, $H(Y|G)$ is maximized and is equal to $b$.
\end{corollary}
\begin{proof}
The corollary is proven by showing that
\begingroup
\allowdisplaybreaks
\begin{align*}
    I(F_X^{s}|G) \geq&\; I(F_X|G)
\end{align*}
or equivalently,
\begin{align*}
    \mathbb{E}_{G}\left[I(F_X^{s}|G = g)\right] \geq&\; \mathbb{E}_{G}\left[I(F_X|G = g)\right].
\end{align*}
\endgroup
Applying the result of Proposition \ref{proposition:symmetry_distribution} on the mutual information terms inside the expectation gives us
\begingroup
\allowdisplaybreaks
\begin{align*}
    &\mathbb{E}_{G}\left[b-H(Y|X,G=g)\right] \\
    &\qquad\qquad\qquad\geq\; \mathbb{E}_{G}\left[H(Y|G=g) - H(Y|X,G=g)\right],
\end{align*}
which simplifies to
\begin{align*}
    b \geq&\; \mathbb{E}_{G}\left[H(Y|G=g)\right].
\end{align*}
\endgroup
Noting that $b$ is the highest achievable output entropy concludes the proof.
\end{proof}
Similar to the analysis of Model A, Corollary \ref{corollary:symmetry_fading} allows us to reduce the search space of the optimal input distribution to the set $\Omega_s$ defined in (\ref{eq:Omega_s}). It then follows that the ergodic capacity is
\begingroup
\allowdisplaybreaks
\begin{align*}
    C_{\text{ergodic}} =& b - \inf_{F_X\in\Omega_s} \mathbb{E}_{G}\left\{\int_{\mathbb{C}}w\left(\frac{|g|^2\alpha}{\sigma^2},\beta+\angle g,b\right)\;dF_X\right\},
\end{align*}
which can be written as
\begin{align} \label{eq:C_csir_part1}
    =& b - \inf_{F_X\in\Omega_s} \mathbb{E}_{G}\bigg\{\mathbb{E}_{B}\bigg[\mathbb{E}_{A|B}\bigg[ w\left(\frac{|g|^2\alpha}{\sigma^2},\beta+\angle g,b\right)\bigg]\bigg]\bigg\}\nonumber\\
    =& b - \inf_{F_X\in\Omega_s} \mathbb{E}_{|G|}\bigg\{\mathbb{E}_{B,\angle G}\bigg\{\nonumber\\
    &\quad\qquad\qquad\mathbb{E}_{A\big|B,\angle G}\bigg[w\left(\frac{|g|^2\alpha}{\sigma^2},\beta+\angle g,b\right)\bigg]\bigg\}\bigg\}.
\end{align}
\endgroup
Here, we used Bayes' rule to express the complex PDF $f_X(x) = f_{A,B}(\alpha,\beta)$ as $f_{A|B}(\alpha|\beta)f_{B}(\beta)$ and perform the complex expectation as two real-valued expectations over $\alpha|\beta$ and $\beta$. Since the Rayleigh fading $G$ is independent of $X$ and its components, $|G|$ and $\angle G$, are independent as well, we can rewrite the joint distribution $f_{A,B,G}(\alpha,\beta,g)$ as
\begin{align*}
    f_{A,B,G}(\alpha,\beta,g) =& f_{A|B,\angle G,|G|}\cdot f_{B,\angle G\big| |G|}\cdot f_{|G|}\\
    =&f_{A|B,\angle G}\cdot f_{B,\angle G}\cdot f_{|G|}
\end{align*}
which implies the expectation in (\ref{eq:C_csir_part1}). For any arbitrary but fixed $\beta+\angle g$, Jensen's inequality can be applied to the first argument of $w\left(\frac{|g|^2\alpha}{\sigma^2},\beta+\angle g,b\right)$ due to Proposition \ref{proposition:convexity_w}. That is,
\begin{align*}
     &\mathbb{E}_{B,\angle G}\left\{\mathbb{E}_{A\big|B,\angle G}\left[w\left(\frac{|g|^2\alpha}{\sigma^2},\beta+\angle g,b\right)\right]\right\} \\
     &\quad\qquad\qquad\geq \mathbb{E}_{B,\angle G}\left\{w\left(\frac{|g|^2\mathbb{E}_{A\big|B,\angle G}\left[\alpha\right]}{\sigma^2},\beta+\angle g,b\right)\right\}
\end{align*}
with equality if $\alpha$ is a deterministic function of $\beta+\angle g \mod 2\pi$. Suppose we define $\tau = \beta + \angle g \mod 2\pi$ and let $\alpha_{\tau}$ be the value of $\alpha$ at a specific angle $\tau$. Then, the ergodic capacity can be expressed as
\begin{align}\label{eq:C_csir_part2}
    C_{\text{ergodic}} = & b - \inf_{\substack{F_{X}\in\Omega_{s}\\ \;\alpha_{\tau}}} \mathbb{E}_{|G|}\left\{\mathbb{E}_{T}\left\{w\left(\frac{|g|^2\alpha_{\tau}}{\sigma^2},\tau,b\right)\right\}\right\},
\end{align}
where $T \sim \mathrm{Unif}(-\pi,\pi)$. The simplification comes from the property that a mod $2\pi$ addition of a circular uniform distribution and any arbitrary circular distribution is a circular uniform distribution (See Appendix \ref{appendix_L}). Note, however, that a solution to the minimization problem in (\ref{eq:C_csir_part2}) exists if we can choose a distribution $F_X$ such that $\alpha$ is a function of $\tau$. We show in the following proposition that $\alpha$ is a deterministic function of $\tau = \beta+\angle g$ if and only if $F_{X}^*$ has a constant amplitude. In addition, full transmit power should be used so this amplitude is $\sqrt{P}$.
\begin{proposition}\label{proposition:fading_constant_amplitude}
$\alpha_{\tau}$ is a deterministic function of $\tau$, with $\tau = \beta + \angle g \;\mathrm{mod} \; 2\pi$, if and only if $F_{X}$ has a constant amplitude. Moreover, $\alpha_{\tau} = P$ to minimize (\ref{eq:C_csir_part2}).
\end{proposition}
\begin{proof}
We first prove necessity of the first statement. Let $x_1 = \sqrt{\alpha^{(1)}}e^{j\beta^{(1)}}$ and $x_2 = \sqrt{\alpha^{(2)}}e^{j\beta^{(2)}}$ be points in the support set of $F_{X}^*$. We introduce a `ghost sample' $\angle G'$ which is an independent copy of $\angle G$ and also follows the same distribution as $\angle G$. Then, $x_1e^{j\angle g} = \sqrt{\alpha^{(1)}}e^{j\tau_1}$ and $x_2e^{j\angle g'} = \sqrt{\alpha^{(2)}}e^{j\tau_2}$, where $\tau_1 = \beta^{(1)} + \angle g\;\mathrm{mod}\;2\pi$ and $\tau_2 = \beta^{(2)} + \angle g'\;\mathrm{mod}\;2\pi$ are uniform circular distributions. Since $\alpha_{\tau_1}$ should be equal to $ \alpha_{\tau_2}$ when $\tau_1 = \tau_2$, then $\alpha^{(1)} = \alpha^{(2)}$.

Next, we prove sufficiency of the first statement.
Pick $\beta^{(1)},\beta^{(2)}$, and $K$ such that $x_1 = \sqrt{K}e^{j\beta^{(1)}}$ and $x_2 = \sqrt{K}e^{j\beta^{(2)}}$ are points in the support set of $F_{X}^*$. We introduce again a `ghost sample' $\angle G'$. Let $x_1e^{j\angle g} = \sqrt{K}e^{j\tau_1}$ and $x_2e^{j\angle g'} = \sqrt{K}e^{j\tau_2}$, where $\tau_1 = \beta^{(1)} + \angle g\;\mathrm{mod}\;2\pi$ and $\tau_2 = \beta^{(2)} + \angle g'\;\mathrm{mod}\;2\pi$ are uniform circular distributions. Then, whenever $\tau_1$ is equal to $ \tau_2$, we have $\alpha_{\tau_1} = \alpha_{\tau_2} = K$.

Finally, the second statement follows from the fact that $w(\nu,\theta,b)$ is a decreasing function of $\nu$ (Proposition \ref{proposition:monotonicity_w}).
\end{proof}

By applying Proposition \ref{proposition:fading_constant_amplitude} to (\ref{eq:C_csir_part2}), we obtain
\begin{align}\label{eq:C_csir_part3}
    C_{\text{ergodic}} = & b -  \mathbb{E}_{|G|}\left\{\mathbb{E}_{T}\left\{w\left(\frac{|g|^2P}{\sigma^2},\tau,b\right)\right\}\right\}\nonumber\\
    = &b -  \mathbb{E}_{|G|}\left\{\mathbb{E}_{\angle G}\left\{w\left(\frac{|g|^2P}{\sigma^2},\angle g,b\right)\right\}\right\}.
\end{align}
Here, we note that $T$ in the first line has the same distribution as $\angle G$ regardless of the distribution of $B$. Since $\angle G$ is uniformly distributed, the expectation attains the same value independent of $\beta$. Thus, without loss of generality, we can set $\beta = 0$ in the second line. Since (\ref{eq:C_csir_part3}) is invariant of the input distribution (provided that the input distribution is $\frac{2\pi}{2^b}$-symmetric and has a single amplitude level $\sqrt{\alpha} = \sqrt{P}$), then the two necessary conditions established are also sufficient. This completes the proof of Theorem \ref{theorem:Rayleigh_CSIR_case}. Theorem \ref{theorem:Rayleigh_CSIR_case} generalizes the result in \cite[Theorem 3]{Krone:2010} to $b \geq 3$.

\subsection{Rayleigh Fading with CSI at the receiver and transmitter (CSIT)}\label{subsection:coherent_CSIT}
The situation changes when the transmitter also has CSI knowledge since the transmitted signal can be adapted depending on the instantaneous realization of $G$. The ergodic capacity becomes
\begin{equation}\label{eq:ergodi_cap_def_CSIT}
    C_{\text{ergodic}} = \sup_{F_{X|G}\in\Omega}I(F_X|G).
\end{equation}
Here, the transmitter optimizes the input distribution depending on the instantaneous CSI. Moreover, the transmitter can use different transmit power for different fading realizations, provided that the scheme still satisfies the average power requirement $\mathbb{E}[|X|^2] \leq P$. In this case, the ergodic capacity expression can be written as 
\begingroup
\allowdisplaybreaks
\begin{align}
    \label{eq:ergodic_cap_long_term_P}
        C_{\text{ergodic}} =&\sup_{F_{X|G}:\;\mathbb{E}\left[|X|^2\right]\leq P}I(F_X|G)\nonumber\\
    =&b - \inf_{\mathbb{E}[P\left(|g|^2\right)]\leq P} \mathbb{E}_{|G|}\left[w\left(\frac{|g|^2 P\left(|g|^2\right)}{\sigma^2},\frac{\pi}{2^b},b\right)\right]\nonumber\\
    =& b - \mathbb{E}_{|G|}\left[w\left(\frac{|g|^2 P^*\left(|g|^2\right)}{\sigma^2},\frac{\pi}{2^b},b\right)\right],
\end{align}
\endgroup
where the second line is obtained from the fact that a rotated $2^b$-PSK is capacity-achieving for any $G$. With the knowledge of channel phase, the transmitter can rotate the $2^b$-PSK accordingly such that the symbols will be placed at optimal angles. $P(|g|^2)$ denotes an optimal power control strategy that depends on the CSI $|g|^2$. In the last line, we introduced $P^*(|g|^2)$ to denote the optimal power control strategy that minimizes the second term of the second line of (\ref{eq:ergodic_cap_long_term_P}). To derive $P^*\left(|g|^2\right)$, we consider the functional
\begingroup
\allowdisplaybreaks
\noindent
\begin{align}\label{eq:power_alloc_functional}
    \mathcal{J}\left(P\left(|g|^2\right)\right) =&\mathbb{E}_{|G|}\left[w\left(\frac{|g|^2 P\left(|g|^2\right)}{\sigma^2},\frac{\pi}{2^b},b\right)\right] \nonumber\\
    &+ \eta\left[\int_{0}^{\infty}P\left(|g|^2\right)f_{|G|}(|g|)d|g| - P\right].
\end{align}
\endgroup
This functional is the Lagrangian functional of the convex optimization problem term in (\ref{eq:ergodic_cap_long_term_P}) and $\eta$ is the Lagrange multiplier. We also define a quantity $w_{\min}'$ to be
\begingroup
\allowdisplaybreaks
\begin{align}
    w_{\min}' =& \lim_{\nu\rightarrow 0 }\frac{\partial w\left(\nu,\frac{\pi}{2^b},b\right)}{\partial \nu}\nonumber\\
    =& -\frac{2^{2b-1}}{2\pi}\sin^2\left(\frac{\pi}{2^b}\right),
\end{align}
\endgroup
which states that $w_{\min}'$ is the conditional entropy per unit energy at $\nu = 0$. The second equality is its closed-form expression for $2^b$-PSK (the expression is based on \cite[Theorem 3]{Gursoy:2007}). An optimal power allocation function should satisfy the stationary condition
\begin{align}\frac{\partial \mathcal{J}\left(P(|g|^2)\right)}{\partial P(|g|^2)} = 0.
\end{align}
Differentiating (\ref{eq:power_alloc_functional}) with respect to $P(|g|^2)$ gives us
\begingroup
\allowdisplaybreaks
\begin{align*}
    \frac{|g|^2}{\sigma^2}\cdot \frac{\partial w\left(|g|^2 \frac{P(|g|^2)}{\sigma^2},\frac{\pi}{2^b},b\right)}{\partial \nu} + \eta =& 0,
\end{align*}
\endgroup
which yields the optimal power allocation strategy
\begingroup
\allowdisplaybreaks
\begin{align*}
    P^*(|g|^2) =\begin{cases} \frac{\sigma^2}{|g|^2}\left\{\left[\frac{\partial w}{\partial \nu}\right]^{-1}\left(-\frac{\sigma^2\eta}{|g|^2}\right)\right\},\quad |g|^2 >\frac{\eta}{w_{\min}'}\\
    \qquad\;\qquad 0\qquad\;\qquad,\quad \text{otherwise}\end{cases}
\end{align*}
\endgroup
such that $\eta$ satisfies 
\[\int_{\frac{\eta}{w_{\min}'}}^{\infty}\left\{\left[\frac{\partial w}{\partial \nu}\right]^{-1}\left(-\frac{\sigma^2\eta}{|g|^2}\right)\right\}\frac{f_{|G|}(|g|)}{|g|^2}\;d|g| = \frac{P}{\sigma^2}.\]
This completes the proof of Theorem \ref{theorem:Rayleigh_CSIT_case}. The dependence of $P^*(|g|^2)$ on the fading distribution is only through the parameter $\eta$. Thus, this power allocation strategy can be used to other fading distributions by replacing $f_{|G|}(|g|)$. The quantity $\frac{\eta}{w_{\min}'}$ is called the cut-off value. This is the lowest value of the fading gain $|g|^2$ for which the power control scheme allocates a non-zero power.  

\begin{figure}[t]
\centering
\hspace*{-.55cm}
    \includegraphics[width=.525\textwidth]{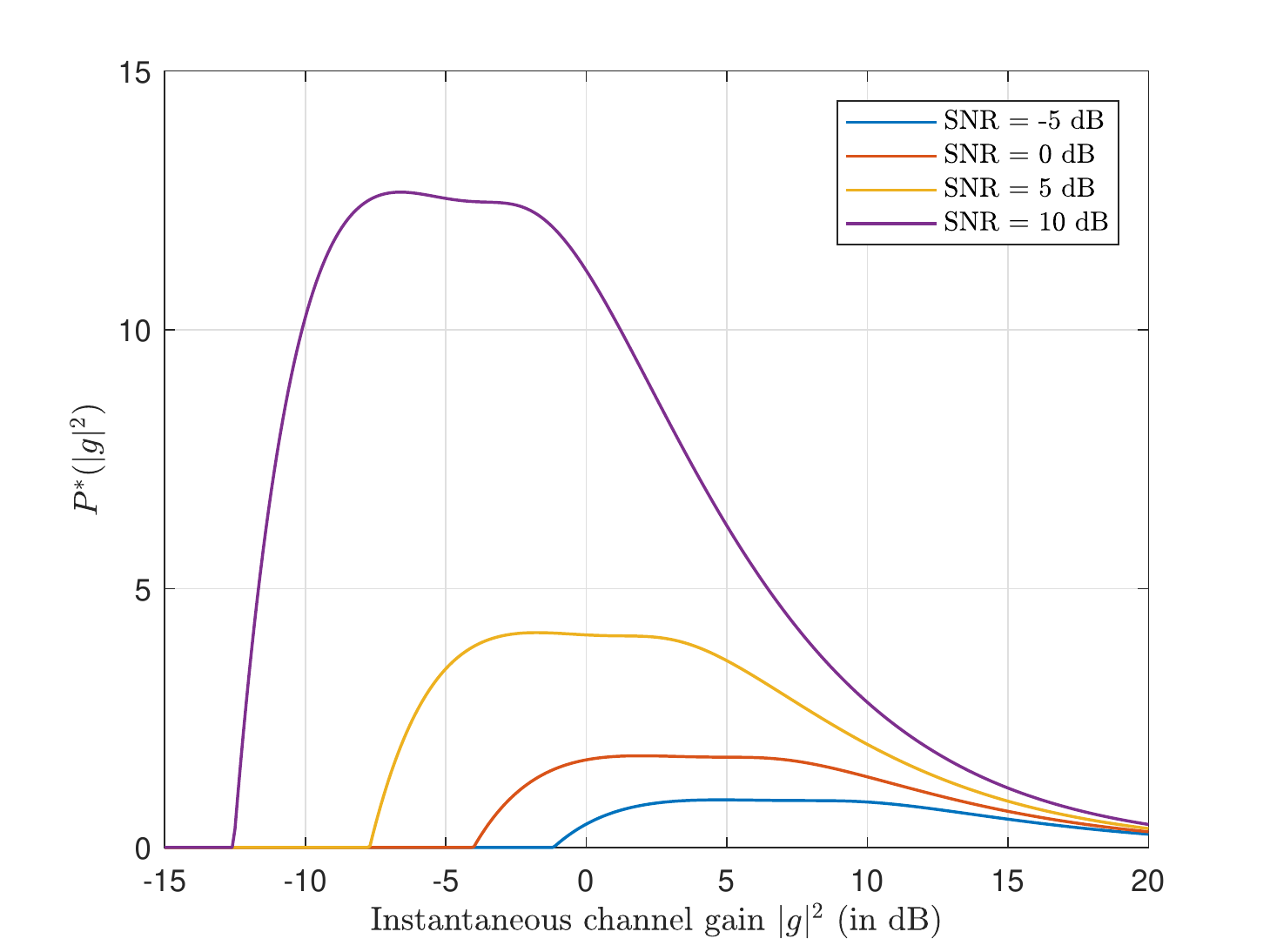}
    \caption{Allocated power using optimal power control policy vs. instantaneous channel gain $|g|^2$ ($b = 3$)}    \label{fig:PC_control}
\end{figure}

The plots of the optimal power control scheme for a Rayleigh fading channel with 3-bit phase quantization at the output as a function of $|g|^2$ are shown in Figure \ref{fig:PC_control} for different SNR (i.e. $\frac{P}{\sigma^2}$) and $\gamma^2 = 1$. To vary the SNR, noise power is fixed at $\sigma^2 = 1$ and the average power $P$ is varied. Note that unlike the ``water-pouring in time" approach \cite{Goldsmith:1997}, our optimal power control does not allocate a lot of power on fading states with extremely high channel gain. This is because the capacity is bounded above by $b$ bits/cu at high SNR regime so there is little or no benefit in allocating power on the channel realizations with very high gain. There are also subtle differences between our optimal power allocation scheme and the mercury/waterfilling algorithm in \cite{Lozano:2006}. Mercury/waterfilling is based on the I-MMSE relationship \cite{Guo:2005} which only holds for Gaussian channels. Hence, the quantity $\frac{\partial w(\nu,\frac{\pi}{2^b},b)}{\partial \nu}$ is not necessarily related to the minimum mean square error. Phase quantization also impacts $w_{\min}'$ so the cut-off values of the two power allocation strategies are different. Ultimately, mercury/waterfilling power allocation maximizes the achievable rate of a suboptimal input transmitted over an AWGN channel while our scheme is the optimal strategy for fading channel with phase quantization at the output.

\begin{figure}[t]
\centering
\hspace*{-.5cm}
    \includegraphics[angle =90 ,width=.525\textwidth]{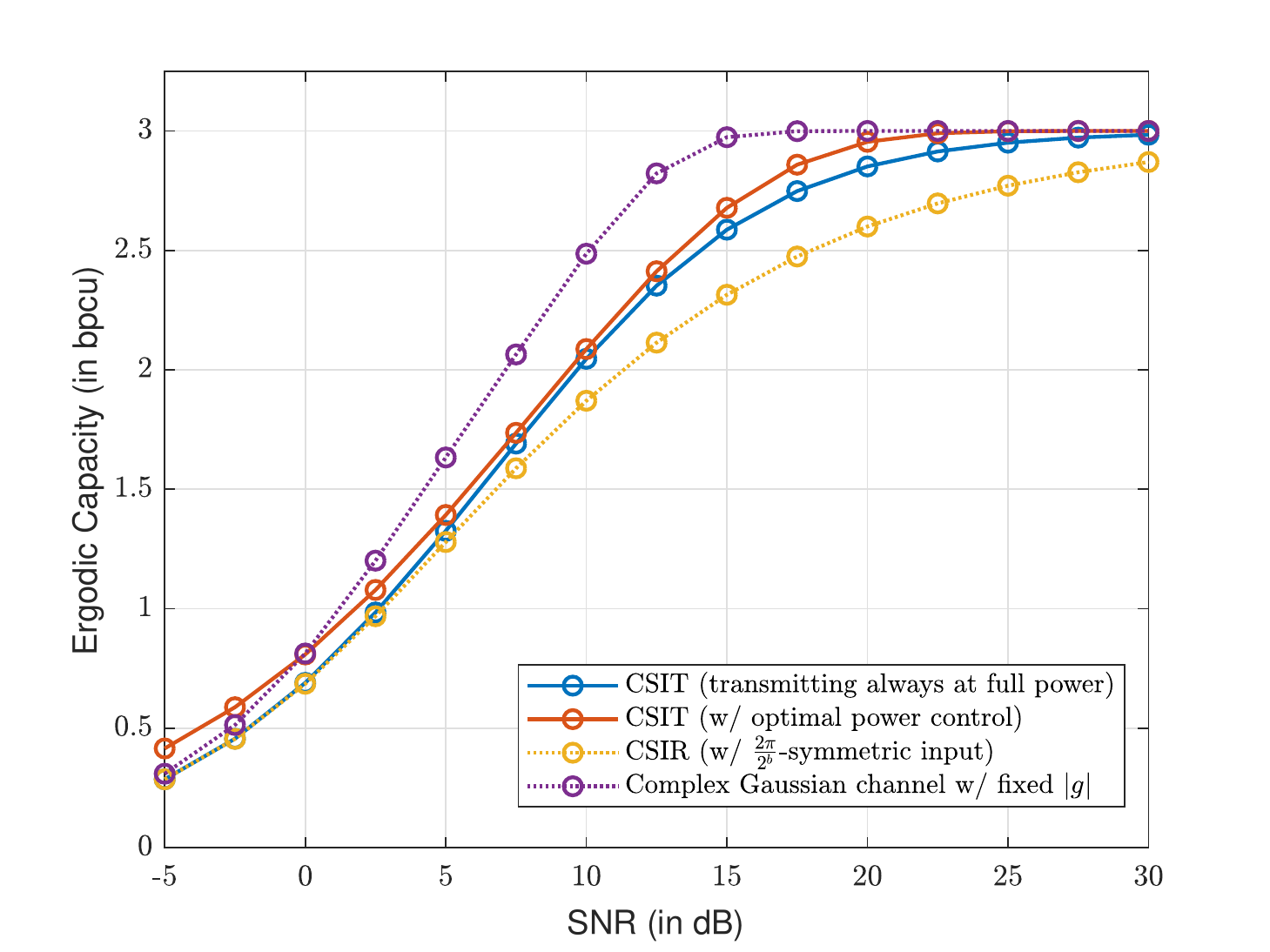}
    \caption{Ergodic Capacity of Rayleigh Fading with 3-bit Phase Quantizer under different CSI availability. Here, we assume that $\gamma^2 = 1$ and $|g_{\text{LoS}}| = 1$ for the AWGN channel so that the average SNR is defined as $\frac{P}{\sigma^2}$ in all cases.}    \label{fig:ergodic_cap}
\end{figure}

The ergodic capacity of Rayleigh fading with phase quantization at the output is depicted in Figure \ref{fig:ergodic_cap} under different CSI availability. Here, we assume $\gamma^2 = 1$ for the fading channels and $|g_{\text{LoS}}| = 1$ for the AWGN channel so that the average SNR in all cases can be defined as $SNR = \frac{P}{\sigma^2}$. The capacity of Gaussian channel with phase quantization (purple) is also superimposed in this plot for reference. The CSIR capacity (yellow) is obtained using an 8-PSK constellation although any $\frac{2\pi}{2^b}$-symmetric distribution with constant amplitude will produce the same ergodic capacity. It can be observed that most of the improvements in granting CSI at the transmitter comes from compensating channel phase rotation (blue) but there is still some additional benefit in doing power control when channel magnitude is known (red), especially in the low SNR regime. In fact, the optimal power control can take advantage of fading in this regime to obtain an ergodic capacity higher than the fixed channel gain case.

\section{Conclusion}\label{section-conclusion}

In this work, we generalized the capacity results of channels with 1-bit I and Q ADC from several past studies to multi-bit phase quantization. Our first contribution is a rigorous proof that a rotated $2^b$-PSK is the capacity-achieving input to complex Gaussian channels with fixed channel gain $b$-bit phase quantization at the output. We then showed that the optimality of this input distribution still holds under different fading scenarios such as noncoherent Rician fading, Rayleigh fading with CSIR only, and Rayleigh fading with CSIT. Capacity expressions for different system models are established using the derived capacity-achieving input distribution. Unlike the multi-bit I and Q quantization case which relies on numerical computation to identify the capacity-achieving input at every SNR value, PSK is capacity-achieving for channels with phase-quantization at the output in all SNR regimes. This also gives merit to multi-bit phase quantization from a practical viewpoint since a conventional modulation scheme is capacity-achieving. 
In contrast, the information rates of conventional modulation schemes are far from the capacity of channels with multi-bit I/Q ADC at the output \cite{Vu:2019}. Nonetheless, multi-bit I/Q ADC may achieve higher capacity than phase quantization since it takes into account the amplitude of the signal.

One notable future research direction is to extend the analysis to low-resolution polar receivers\footnote{In contrast to a I/Q receiver which recovers the in-phase and quadrature components, a polar receiver recovers the magnitude and phase of the signal \cite{Nazari:2014}}. Wireless receivers equipped with polar quantizers are shown to have advantage over its Cartesian counterpart in terms of energy-efficiency \cite{Nazari:2014,Wang:2020}. However, information-theoretic limits of quantized polar receivers should be established. We conjecture that the capacity-achieving signaling scheme for such channel would be amplitude-phase shift keying (APSK) but further research needs to be conducted to prove or disprove this conjecture. While only point-to-point channels are considered in this work, the extension of this study to multi-user and multi-antenna setting is very interesting and is currently being explored.

\section*{Acknowledgement}\label{section-acknowledgement}

The authors would like to thank the Associate Editor and the anonymous reviewers for their valuable feedback. Their feedback helped improve the quality and presentation of our paper.

\begin{appendices}
 \section{Proof of Lemma \ref{lemma:symmetry_p_phi_given_nu}}\label{appendix_A}
 
From the definition of $f_{\Phi|N}(\phi|\nu)$ in (\ref{eq:prob_phi_given_nu}), the claim about the symmetry of $f_{\Phi|N}(\phi|\nu)$ can be proven by showing that
\[f_{\Phi|N}(\phi|\nu) = \frac{1}{2}\left[ f_{\Phi|N}(\phi|\nu) +  f_{\Phi|N}(-\phi|\nu)\right].
\]
This is true because $\cos(\cdot)$ is an even-symmetric function. Next, we obtain its derivative with respect to $\phi$. 
\begingroup
\allowdisplaybreaks
\begin{align*}
  &\frac{\partial f_{\Phi|N}(\phi|\nu)}{\partial \phi} \\
  &\qquad= -\sqrt{\frac{\nu}{\pi}}\sin\phi e^{-v\sin^2\phi}[2\nu\cos^2\phi + 1]Q(-\sqrt{2\nu}\cos\phi) \\
  &\qquad\quad-\sqrt{\frac{\nu}{\pi}}\cos\phi e^{-\nu\sin^2\phi}\left[\frac{e^{-v\cos^2\phi}}{\sqrt{2\pi}}\sqrt{2\nu}\sin\phi\right]\\
  &\qquad= -\sqrt{\frac{\nu}{\pi}}\sin\phi e^{-\nu\sin^2\phi}\bigg[[2\nu\cos^2\phi +1]Q(-\sqrt{2\nu}\cos\phi) \\
  &\qquad\quad+ \frac{\sqrt{\nu}\cos\phi e^{-\nu\cos^2\phi}}{\sqrt{\pi}}\bigg]
\end{align*}%
\endgroup
Since $-\sqrt{\nu}\sin(\phi)$ is positive for $\phi \in (-\pi,0)$ and negative for $\phi\in (0,\pi)$, we simply need to prove that
\[[2\nu\cos^2\phi +1]Q(-\sqrt{2\nu}\cos\phi) + \frac{\sqrt{\nu}\cos\phi e^{-\nu\cos^2\phi}}{\sqrt{\pi}} > 0.
\]
This is true for $\phi \in (-\pi/2,\pi/2)$ because $\cos\phi > 0$. For $\phi \notin (-\pi/2,\pi/2)$, we have
\begingroup
\allowdisplaybreaks
\begin{align*}
 &[2\nu\cos^2\phi' +1]Q(\sqrt{2\nu}\cos\phi') - \frac{\sqrt{\nu}\cos\phi' e^{-\nu\cos^2\phi'}}{\sqrt{\pi}} \\
 &\qquad\quad>
  [2\nu\cos^2\phi' +1]\underbrace{\frac{e^{-\nu\cos^2\phi'}}{\sqrt{2\pi}\sqrt{2\nu}\cos\phi'}\cdot\frac{2\nu\cos^2\phi'}{1+2\nu\cos^2\phi'}}_{< Q(\sqrt{2\nu}\cos\phi')} \\
  &\qquad\qquad\quad- \frac{\sqrt{\nu}\cos\phi' e^{-\nu\cos^2\phi'}}{\sqrt{\pi}}\\
  &\qquad\quad=\underbrace{[1-1]}_{=0}\frac{\sqrt{\nu}\cos\phi' e^{-\nu\cos^2\phi'}}{\sqrt{\pi}}.
\end{align*}%
\endgroup
The first line follows from a change of variable ($\phi' = -\phi$) and by using the following lower bound of the Q-function \cite[Proposition A.8]{Moser_LecNotes}: 
\[Q(x) > \frac{e^{-\frac{x^2}{2}}}{\sqrt{2\pi}x}\left(\frac{x^2}{1+x^2}\right)\qquad \forall x\in\mathbb{R}.\]
Since the LHS is positive, the claim is proven.

\section{Proof of Lemma \ref{lemma:symmetry_Wy}}\label{appendix_B}

From (\ref{eq:phase_quantization_prob}), we can express $W_y^{(b)}\left(\nu,\theta + \frac{2\pi k}{2^b}\right)$ as
\begingroup
\allowdisplaybreaks
\begin{align*}
    W_y^{(b)}\left(\nu,\theta + \frac{2\pi k}{2^b}\right) =& \int_{\frac{2\pi}{2^b}y-\pi-\theta-\frac{2\pi k}{2^b}}^{\frac{2\pi}{2^b}(y+1)-\pi-\theta-\frac{2\pi k}{2^b}}f_{\Phi|N}(\phi|\nu)\;d\phi \\
    =&  \int_{\frac{2\pi}{2^b}(y-k)-\pi-\theta}^{\frac{2\pi}{2^b}(y+1-k)-\pi-\theta}f_{\Phi|N}(\phi|\nu)\;d\phi \\
    =& W_{y-k}^{(b)}\left(\nu,\theta\right)
\end{align*}%
\endgroup
which proves Lemma \ref{lemma:symmetry_Wy}.i. For Lemma \ref{lemma:symmetry_Wy}.ii, we have
\begingroup
\allowdisplaybreaks
\begin{align*}
    W_{2^{b-1}-y}^{(b)}\left(\nu,\frac{\pi}{2^b}\right) =&\int_{-\frac{2\pi y}{2^b}-\frac{\pi}{2^b}}^{-\frac{2\pi y}{2^b}+\frac{\pi}{2^b}}f_{\Phi|N}\left(\phi|\nu\right)d\phi\\ =& \int_{\frac{2\pi y}{2^b}-\frac{\pi}{2^b}}^{\frac{2\pi y}{2^b}+\frac{\pi}{2^b}}f_{\Phi|N}\left(\phi'|\nu\right)d\phi'\;\text{(Let $\phi'=-\phi$)}\\
    =& W_{2^{b-1}+y}^{(b)}\left(\nu,\frac{\pi}{2^b}\right),
\end{align*}%
\endgroup
where we obtain the last line by noting that the change of variable did not affect $f_{\Phi|N}\left(\phi|\nu\right)$ since it is even-symmetric about $\phi = 0$ (Lemma \ref{lemma:symmetry_p_phi_given_nu}). Finally, following the same arguments in the proof of Lemma \ref{lemma:symmetry_Wy}.ii, Lemma \ref{lemma:symmetry_Wy}.iii is proven as follows:
\begingroup
\allowdisplaybreaks
\begin{align*}
    W_{2^{b-1}-y}^{(b)}\left(\nu,0\right) =&\int_{-\frac{2\pi y}{2^{b}}}^{-\frac{2\pi y}{2^{b}} + \frac{2\pi}{2^b} }\;f_{\Phi|N}\left(\phi|\nu\right)\;d\phi\\
         =&\int_{\frac{2\pi y}{2^{b}}- \frac{2\pi}{2^b}}^{\frac{2\pi y}{2^{b}}  }\;f_{\Phi|N}\left(\phi'|\nu\right)\;d\phi'\;\; \text{(Let $\phi' = -\phi$)}\\
         =& W_{2^{b-1}-1+y}^{(b)}\left(\nu,0\right).
\end{align*}%
\endgroup

\section{Proof of Proposition \ref{proposition:monotonicity_w}}\label{appendix_C}
 
Note that the special case of $b = 1$ is already proven in \cite[Appendix D]{Singh:2009_techreport} and the proof can be readily extended to $b = 2$ by treating this case as two 1-bit quantizers for I and Q. Thus, we only need to consider $b\geq 3$. Consider first the case $\theta\in[0,\frac{2\pi}{2^b})$ (or equivalently, $\theta\in\mathcal{R}_{2^{b-1}}^{\text{PH}}$). To show that $w(\nu,\theta,b)$ is strictly decreasing function of $\nu$ for $\theta\in[0,\frac{2\pi}{2^b})$, we first prove a lemma about $f_{\Phi|N}(\phi|\nu)$.

\begin{lemma}\label{lemma:decreasing_p_phi_nu}
The conditional pdf $f_{\Phi|N}(\phi|\nu)$ is a decreasing function of $\nu$ for the following cases:
\begin{align*}
    (A): & \;\phi \in (\pi/2,\pi] \cup [-\pi,-\pi/2) \\
    (B): & \; \phi \in [-\pi/2,\pi/2]\backslash\{0\}\cap \nu \geq \nu_0,\;\text{for some }\nu_0 \leq\frac{1}{2\sin^2\phi}
\end{align*}
\end{lemma}
\begin{proof}
The first-order derivative of $f_{\Phi|N}(\phi|\nu)$ with respect to $\nu$ is
\begin{align}\label{eq:first_diff_p_phi_given_nu}
        &\frac{\partial f_{\Phi|N}(\phi|\nu)}{\partial \nu} \nonumber\\
        &\qquad\quad=\;\frac{e^{-\nu\sin^2\phi}Q\left(-\sqrt{2\nu}\cos\phi\right)\cos\phi (1-2\nu\sin^2\phi)}{2\sqrt{\nu}\sqrt{\pi}}\nonumber\\
        &\qquad\qquad\quad\;-\frac{e^{-\nu}\sin^2\phi}{2\pi}.
\end{align}
This derivative is analyzed under two scenarios.\\
\noindent\textbf{Scenario A ($\phi \in (\pi/2,\pi] \cup [-\pi,-\pi/2) $):}\\
In this case, $\cos(\phi) < 0$ and we have
\begin{align*}
        &\frac{\partial f_{\Phi|N}(\phi|\nu)}{\partial \nu} \\
        &\qquad=-\underbrace{\frac{e^{-\nu\sin^2\phi}Q\left(-\sqrt{2\nu}\cos\phi\right)(-\cos\phi)}{2\sqrt{\nu}\sqrt{\pi}}}_{> 0}(1-2\nu\sin^2\phi)\\
        &\qquad\qquad\;-\underbrace{\frac{e^{-\nu}\sin^2\phi}{2\pi}}_{>0}
\end{align*}
which is negative for all $\nu \leq \frac{1}{2\sin^2\phi}$. For $\nu >   \frac{1}{2\sin^2\phi}$, $\frac{\partial f_{\Phi|N}(\phi|\nu)}{\partial \nu}$ can be bounded from above as follows:
\begingroup
\allowdisplaybreaks
\begin{align*}
    \frac{\partial f_{\Phi|N}(\phi|\nu)}{\partial \nu}
        <&\;\frac{e^{-\nu\sin^2\phi}\cos\phi}{2\sqrt{\nu}\sqrt{\pi}}\cdot \frac{e^{-\nu\cos^2\phi} (2\nu\sin^2\phi-1)}{\sqrt{2\pi}(\sqrt{2\nu}\cos\phi)}\\
        &\;\qquad-\frac{e^{-\nu}\sin^2\phi}{2\pi}\\
        =&\;-\frac{e^{-\nu}\sin^2\phi}{2\pi}+\frac{e^{-\nu}}{4\pi\nu}(2\nu\sin^2\phi-1)\\
        =&-\frac{e^{-\nu}}{4\pi\nu}.
\end{align*}%
\endgroup
The inequality comes from the Q-function upper bound $Q(x) < \frac{e^{-\frac{x^2}{2}}}{\sqrt{2\pi}x}$ and the second and third lines are obtained through simplification. This shows that (\ref{eq:first_diff_p_phi_given_nu}) is negative in Scenario A.\\
\noindent\textbf{Scenario B ($\phi \in [-\pi/2,\pi/2]$):}\\
In this case, $\cos(\phi) > 0$ and $\frac{\partial f_{\Phi|N}(\phi|\nu)}{\partial \nu}$ becomes
\begin{align*}
&\frac{\partial f_{\Phi|N}(\phi|\nu)}{\partial \nu}\\
        &\quad=\;-\frac{e^{-\nu}\sin^2\phi}{2\pi}\\
        &\qquad\;\;+\underbrace{\frac{e^{-\nu\sin^2\phi}\left[1-Q\left(\sqrt{2\nu}\cos\phi\right)\right]\cos\phi}{2\sqrt{\nu}\sqrt{\pi}}}_{> 0}(1-2\nu\sin^2\phi).
\end{align*}
The second term is negative if $\nu \geq \frac{1}{2\sin^2\phi}$. Note also that $\frac{\partial f_{\Phi|N}(\phi|\nu)}{\partial \nu} > 0$ as $\nu \rightarrow 0^+$ and $\frac{\partial f_{\Phi|N}(\phi|\nu)}{\partial \nu} < 0$ at $\nu = \frac{1}{2\sin^2\phi}$. Moreover, for a fixed $\phi \ne 0$, the first term is monotonic increasing on $\nu$ while the second term is monotonic decreasing on $\nu$. Thus, by Intermediate Value Theorem, there exists a $\nu_0 \leq \frac{1}{2\sin^2\phi}$ such that $f_{\Phi|N}(\phi|\nu)$ is non-increasing for $\nu \geq \nu_0$ and increasing for $\nu < \nu_0$. 

Combining the results of both scenarios completes the proof.
\end{proof}
To prove the proposition, we consider the first-order derivative of $w(\nu,\theta,b)$ with respect to $\nu$, which can be written as
\begingroup
\allowdisplaybreaks
\begin{align}
\label{eq:first_diff_w_nu_theta}
     \frac{\partial w(\nu,\theta,b)}{\partial \nu} =&
     -\sum_{y=0}^{2^b-1}\frac{1+\ln W_y^{(b)}\left(\nu,\theta\right)}{\ln2}\cdot\frac{\partial W_y^{(b)}\left(\nu,\theta\right)}{\partial\nu}\nonumber\\
     =&-\sum_{y\ne2^{b-1}}\frac{1+\ln W_y^{(b)}\left(\nu,\theta\right)}{\ln2}\frac{\partial W_y^{(b)}\left(\nu,\theta\right)}{\partial\nu} \nonumber\\
     &\qquad- \frac{1+\ln W_{2^{b-1}}^{(b)}\left(\nu,\theta\right)}{\ln2}\frac{\partial W_{2^{b-1}}^{(b)}\left(\nu,\theta\right)}{\partial\nu}\nonumber\\
     =&\sum_{y\ne2^{b-1}}\frac{\partial W_y^{(b)}\left(\nu,\theta\right)}{\partial\nu}\underbrace{\left[\log\frac{W_{2^{b-1}}^{(b)}\left(\nu,\theta\right)}{W_{y}^{(b)}\left(\nu,\theta\right)}\right]}_{\geq 0}.
\end{align}%
\endgroup
The first line follows from the chain rule of differentiation. The third line follows from some algebraic manipulation and the fact that $\sum_{\text{all }y}W_{y}^{(b)}(\nu,\theta) = 1$ so the following equalities hold:
\begin{equation}\label{eq:phase_quantization_prob_relationship}
\begin{cases}
   W_{2^{b-1}}^{(b)}(\nu,\theta) = 1 - \sum_{y\ne 2^{b-1}}W_{y}^{(b)}(\nu,\theta) \\\frac{\partial W_{2^{b-1}}^{(b)}(\nu,\theta)}{\partial\nu} = -\sum_{y\ne 2^{b-1}}\frac{\partial W_{y}^{(b)}(\nu,\theta)}{\partial\nu}.
   \end{cases}
\end{equation}
Moreover, $W_{2^{b-1}}(\nu,\theta) \geq W_{y}(\nu,\theta)\;\forall y\ne2^{b-1}$ since $\theta \in \mathcal{R}_{2^{b-1}}^{\text{PH}}$. We also drop the $\ln 2$ since this will not affect the sign of the quantity. Because of Lemma \ref{lemma:decreasing_p_phi_nu} and Lemma \ref{lemma:symmetry_p_phi_given_nu}, there exists two integers $y^{(1)}_0 \leq 2^{b-1}$ and $y^{(1)}_1 \geq 2^{b-1}$ such that we can define the integer sets $\mathcal{Y}_1 = \{y_0^{(1)}+1,\cdots,y_1^{(1)}\}$ and $\mathcal{Y}_1^c = \{0,\cdots,y_0^{(1)}\}\cup\{y_1^{(1)}+1,\cdots,2^{b}-1\}$ satisfying
\begingroup
\allowdisplaybreaks
\begin{align*}
     &\text{(i) }\frac{\partial W_{y}^{(b)}(\nu,\theta)}{\partial\nu} \geq 0, y\in\mathcal{Y}_1,\\
     &\text{(ii) }\frac{\partial W_{y}^{(b)}(\nu,\theta)}{\partial\nu} < 0, y\in\mathcal{Y}_1^c,\text{ and}\\
    & \text{(ii) }\min_{k\in\mathcal{Y}_1}\{W_{k}^{(b)}(\nu,\theta)\} \geq \max_{j\in\mathcal{Y}_1^c}\{W_{j}^{(b)}(\nu,\theta)\}.
\end{align*}%
\endgroup
Note that $\mathcal{Y}_1$ can be an empty set if $y_0^{(1)} = y_1^{(1)}$. Given these, $\frac{\partial w(\nu,\theta,b)}{\partial\nu}$ can be upper bounded as
\begingroup
\allowdisplaybreaks
\begin{align*}
     \frac{\partial w(\nu,\theta,b)}{\partial \nu} 
     =&\sum_{y\in \mathcal{Y}_1}\underbrace{\frac{\partial W_y^{(b)}\left(\nu,\theta\right)}{\partial\nu}}_{\geq 0}\left[\log\frac{W_{2^{b-1}}^{(b)}\left(\nu,\theta\right)}{W_{y}^{(b)}\left(\nu,\theta\right)}\right] \\
     &\qquad+ \sum_{y\in \mathcal{Y}_1^c}\underbrace{\frac{\partial W_y^{(b)}\left(\nu,\theta\right)}{\partial\nu}}_{< 0}\left[\log\frac{W_{2^{b-1}}^{(b)}\left(\nu,\theta\right)}{W_{y}^{(b)}\left(\nu,\theta\right)}\right]\\
     \leq &\; K_1\sum_{y\in \mathcal{Y}_1\backslash\{2^{b-1}\}}\frac{\partial W_y^{(b)}\left(\nu,\theta\right)}{\partial\nu} \\
     &\qquad+  K_2\sum_{y\in \mathcal{Y}_1^c\backslash\{2^{b-1}\}}\frac{\partial W_y^{(b)}\left(\nu,\theta\right)}{\partial\nu}\\
     \leq &\; K_2\sum_{y\ne 2^{b-1}}\frac{\partial W_y^{(b)}\left(\nu,\theta\right)}{\partial\nu}\\
     = &\; -K_2\frac{\partial W_{2^{b-1}}^{(b)}\left(\nu,\theta\right)}{\partial\nu},
\end{align*}%
\endgroup
where $K_1$ and $K_2$ are nonnegative constants given by
\begingroup
\allowdisplaybreaks
\begin{align}
    K_1 = \log\frac{W_{2^{b-1}}^{(b)}\left(\nu,\theta\right)}{\min_{y\in\mathcal{Y}_1\backslash\{2^{b-1}\}}\{ W_{y}^{(b)}\left(\nu,\theta\right)\}}
\end{align}%
and
\begin{align}
    K_2 = \log\frac{W_{2^{b-1}}^{(b)}\left(\nu,\theta\right)}{\max_{y\in\mathcal{Y}_1^c\backslash\{2^{b-1}\}}\{ W_{y}^{(b)}\left(\nu,\theta\right)\}}.
\end{align}
\endgroup
The first line is obtained by placing all the positive terms in the first summation and all the negative terms in the second summation. The upper bound in the second line is obtained from using $K_1$ and $K_2$ to replace the $\log(\cdot)$ terms of the first and second summations, respectively. The third line follows from the fact that $K_1 \leq K_2$ so replacing $K_1$ by $K_2$ increases the positive terms. The last line follows from (\ref{eq:phase_quantization_prob_relationship}). The last line is negative since $W_{2^{b-1}}^{(b)}(\nu,\theta)$ is a strictly increasing function of $\nu$ for $b\geq 3$ (see Appendix \ref{appendix_D}).

Next, suppose $\theta\notin [0,\frac{2\pi}{2^b})$. Then there exists a $\theta'\in[0,\frac{2\pi}{2^b})$ and $j\in\mathbb{Z}$ such that $\theta' = \theta - \frac{2\pi j}{2^b}$. The function $w(\nu,\theta,b)$ can be expressed as
\begingroup
\allowdisplaybreaks
\begin{align*}
 w(\nu,\theta,b)
 =&-\sum_{y=0}^{2^{b-1}}W_y^{(b)}(\nu,\theta'+\frac{2\pi j}{2^b})\log W_y^{(b)}(\nu,\theta'+\frac{2\pi j}{2^b}) \\
 =&-\sum_{y=0}^{2^{b-1}}W_{y-j}^{(b)}(\nu,\theta')\log W_{y-j}^{(b)}(\nu,\theta') \\
 =& w(\nu,\theta',b).
\end{align*}
\endgroup
The second line follows from the relationship between $\theta'$ and $\theta$. The third line follows from Lemma \ref{lemma:symmetry_Wy}.i and the circular structure of the phase quantizer. This implies that $w(\nu,\theta,b)$ and $w(\nu,\theta',b)$ share the same properties so the claim holds for all $\theta\in[-\pi,\pi)$.

\section{Proof of Proposition \ref{proposition:convexity_w}}\label{appendix_E}

Note that the special case of $b = 1$ is already proven in \cite[Appendix D]{Singh:2009_techreport} and the proof can be readily extended to $b = 2$ by treating this case as two 1-bit quantizers for I and Q (as in the method of \cite{Krone:2010}). Thus, we only need to consider $b\geq 3$. Consider first the case $\theta\in[0,\frac{2\pi}{2^b})$ (or equivalently, $\theta\in\mathcal{R}_{2^{b-1}}^{\text{PH}}$). To show that $w(\nu,\theta,b)$ is strictly convex function of $\nu$ for $\theta\in[0,\frac{2\pi}{2^b})$, we prove that its second-order derivative with respect to $\nu$ is positive. That is,
\begingroup
\allowdisplaybreaks
\begin{align}
    \label{eq:second_diff_w_nu_theta}
     \frac{\partial^2 w(\nu,\theta)}{\partial \nu^2} =& \sum_{y\ne 2^{b-1}}\frac{\partial^2 W_y^{(b)}\left(\nu, \theta\right)}{\partial \nu^2}\log_2 \left\{\frac{W_{2^{b-1}}^{(b)}\left(\nu, \theta\right)}{W_y^{(b)}\left(\nu, \theta\right)}\right\}\nonumber\\
     & - \sum_{y= 0}^{2^b-1}\frac{1}{W_y^{(b)}(\nu,\theta)}\frac{\partial W_y^{(b)}(\nu,\theta)}{\partial \nu},
\end{align}
\endgroup
which is obtained by applying chain rule of differentiation to (\ref{eq:first_diff_w_nu_theta}), is greater than or equal to zero. $ \frac{\partial^2 w(\nu,\theta)}{\partial \nu^2}$ can be lower bounded as follows:
\begingroup
\allowdisplaybreaks
\begin{align*}
     &\frac{\partial^2 w(\nu,\theta)}{\partial \nu^2}\\ 
     &\qquad=- \sum_{y\in\mathcal{Y}_1}\frac{1}{W_y^{(b)}(\nu,\theta)}\underbrace{\frac{\partial W_y^{(b)}(\nu,\theta)}{\partial \nu}}_{\geq 0} \\
     &\qquad\quad+ \sum_{y\in\mathcal{Y}_1^c}\frac{1}{W_y^{(b)}(\nu,\theta)}\underbrace{\left[-\frac{\partial W_y^{(b)}(\nu,\theta)}{\partial \nu}\right]}_{
    \geq 0} \\
     &\qquad\quad+ \sum_{y\ne 2^{b-1}}\frac{\partial^2 W_y^{(b)}\left(\nu, \theta\right)}{\partial \nu^2}\log_2 \left\{\frac{W_{2^{b-1}}^{(b)}\left(\nu, \theta\right)}{W_y^{(b)}\left(\nu, \theta\right)}\right\} \\
     &\qquad\geq \sum_{y\in\mathcal{Y}_1}K_1'\underbrace{\left[-\frac{\partial W_y^{(b)}(\nu,\theta)}{\partial \nu}\right]}_{\leq 0} + \sum_{y\in\mathcal{Y}_1^c}K_2'\underbrace{\left[-\frac{\partial W_y^{(b)}(\nu,\theta)}{\partial \nu}\right]}_{
    \geq 0} \\
     &\qquad\quad+ \sum_{y\ne 2^{b-1}}\frac{\partial^2 W_y^{(b)}\left(\nu, \theta\right)}{\partial \nu^2}\log_2 \left\{\frac{W_{2^{b-1}}^{(b)}\left(\nu, \theta\right)}{W_y^{(b)}\left(\nu, \theta\right)}\right\}\\
     &\qquad\geq K_2'\underbrace{\sum_{y=0}^{2^{b-1}-1}\left[-\frac{\partial W_y^{(b)}(\nu,\theta)}{\partial \nu}\right]}_{=0} \\
     &\qquad\quad+ \sum_{y\ne 2^{b-1}}\frac{\partial^2 W_y^{(b)}\left(\nu, \theta\right)}{\partial \nu^2}\log_2 \left\{\frac{W_{2^{b-1}}^{(b)}\left(\nu, \theta\right)}{W_y^{(b)}\left(\nu, \theta\right)}\right\},
\end{align*}
\endgroup
where $\mathcal{Y}_1$ and $\mathcal{Y}_1^c$ are integer sets defined in Appendix \ref{appendix_C} and $K_1'$,$K_2'$ are nonnegative constants given by
\begingroup
\allowdisplaybreaks
\begin{align}
    K_1' =\frac{1}{\min_{y\in\mathcal{Y}_1}\{ W_{y}^{(b)}\left(\nu,\theta\right)\}}
\end{align}%
and
\begin{align}
    K_2' = \frac{1}{\max_{y\in\mathcal{Y}_1^c}\{ W_{y}^{(b)}\left(\nu,\theta\right)\}}.
\end{align}
\endgroup
The first line follows from placing all the positive $\frac{\partial^2 W_y^{(b)}\left(\nu, \theta\right)}{\partial \nu^2}$ terms in the first summation and all the negative $\frac{\partial^2 W_y^{(b)}\left(\nu, \theta\right)}{\partial \nu^2}$ terms in the second summation. The second inequality follows from replacing all the $\log(\cdot)$ terms in the first and second inequality with $K_1'$ and $K_2'$, respectively. The third inequality follows from the fact that $K_1' \leq K_2'$ so replacing $K_1'$ by $K_2'$ increases the negative terms. Consequently, the first summation term in the last line becomes zero because of (\ref{eq:phase_quantization_prob_relationship}).


To proceed with the proof, we first present a key lemma about the convexity of $f_{\Phi|N}(\phi|\nu)$.
\begin{lemma}\label{lemma:convexity_p_phi_nu}
The conditional pdf $f_{\Phi|N}(\phi|\nu)$ is a convex function of $\nu$ for the following cases:
\begin{align*}
    (A): & \;\phi \in (\pi/2,\pi] \cup [-\pi,-\pi/2) \\
    (B): & \; \phi \in [-\pi/2,\pi/2]\backslash\{0\}\cap\nu \geq \nu_0\text{, for }\nu_0 \leq\frac{(1+\sqrt{2})}{2\sin^2\phi}
\end{align*}
\end{lemma}
\begin{proof}
The second-order derivative of $f_{\Phi|N}(\phi|\nu)$ with respect to $\nu$, denoted $\frac{\partial^2 f_{\Phi|N}(\phi|\nu)}{\partial \nu^2}$, is
\begingroup
\allowdisplaybreaks
\begin{align}\label{eq:second_diff_p_phi_given_nu}
      &\frac{\partial^2 f_{\Phi|N}(\phi|\nu)}{\partial \nu^2} \nonumber\\
      &\qquad=\;\frac{\partial}{\partial \nu}\Bigg\{-\frac{e^{-\nu}\sin^2\phi}{2\pi}\nonumber\\
      &\qquad\quad+\frac{e^{-\nu\sin^2\phi}Q\left(-\sqrt{2\nu}\cos\phi\right)\cos\phi}{2\sqrt{\nu}\sqrt{\pi}}(1-2\nu\sin^2\phi)\Bigg\}\nonumber\\
        &\qquad= \frac{\cos\phi e^{-\nu\sin^2\phi}Q\left(-\sqrt{2\nu}\cos\phi\right)\left[(2\nu\sin^2\phi-1)^2-2\right]}{4\sqrt{\pi}\nu^{\frac{3}{2}}}\nonumber\\
        &\qquad\quad+\frac{e^{-\nu}\cos^2\phi}{4\pi\nu}+\frac{e^{-\nu}\sin^4\phi}{2\pi}.
\end{align}
\endgroup
We now analyze the two scenarios mentioned in the lemma.\\
\noindent\textbf{Scenario A:}\\
In this case, $\cos(\phi) < 0$ and $\frac{\partial^2 f_{\Phi|N}(\phi|\nu)}{\partial \nu^2}$ becomes
\begingroup
\allowdisplaybreaks
\begin{align*}
        &\frac{\partial^2 f_{\Phi|N}(\phi|\nu)}{\partial \nu^2}\\
        &\qquad= \underbrace{\frac{\cos\phi e^{-\nu\sin^2\phi}Q\left(-\sqrt{2\nu}\cos\phi\right)}{4\sqrt{\pi}\nu^{\frac{3}{2}}}}_{< 0}\left[(2\nu\sin^2\phi-1)^2-2\right]\\
        &\qquad\quad+\frac{e^{-\nu}\cos^2\phi}{4\pi\nu}+\frac{e^{-\nu}\sin^4\phi}{2\pi},
\end{align*}
\endgroup
which is negative if $\nu \leq \frac{1+\sqrt{2}}{2\sin^2\phi}$. For $\nu > \frac{1+\sqrt{2}}{2\sin^2\phi}$, we have
\begingroup
\allowdisplaybreaks
\begin{align*}
        &\frac{\partial^2 f_{\Phi|N}(\phi|\nu)}{\partial \nu^2}\\
        &= \underbrace{\frac{(-\cos\phi) e^{-\nu\sin^2\phi}\left[2-(2\nu\sin^2\phi-1)^2\right]}{4\sqrt{\pi}\nu^{\frac{3}{2}}}}_{< 0}Q\left(-\sqrt{2\nu}\cos\phi\right)\\
        &\qquad+\frac{e^{-\nu}\cos^2\phi}{4\pi\nu}+\frac{e^{-\nu}\sin^4\phi}{2\pi}\\
        &>\frac{(-\cos\phi) e^{-\nu\sin^2\phi}\left[2-(2\nu\sin^2\phi-1)^2\right]}{4\sqrt{\pi}\nu^{\frac{3}{2}}}\left[\frac{-e^{-\nu\cos^2\phi}}{\sqrt{2\pi}\sqrt{2\nu}\cos\phi}\right]\\
        &\qquad+\frac{e^{-\nu}\cos^2\phi}{4\pi\nu}+\frac{e^{-\nu}\sin^4\phi}{2\pi}\\
        &=\frac{ e^{-\nu}\left[2-(2\nu\sin^2\phi-1)^2\right]}{8\pi\nu^2}+\frac{e^{-\nu}\cos^2\phi}{4\pi\nu}+\frac{e^{-\nu}\sin^4\phi}{2\pi},
\end{align*}
\endgroup
where the inequality in the second line is obtained using the upper bound $Q(x) < \frac{e^{-\frac{x^2}{2}}}{\sqrt{2\pi}x}$. Further algebraic manipulation leads to the following lower bound:
\begingroup
\allowdisplaybreaks
\begin{align*}
       \frac{\partial^2 f_{\Phi|N}(\phi|\nu)}{\partial \nu^2}
        >& \frac{e^{-\nu}}{4\pi\nu}\left[\frac{1}{2\nu}+\cos^2\phi\right]  -\frac{e^{-\nu}\sin^4\phi}{2\pi}\\
        &+\frac{e^{-\nu}\sin^2\phi}{2\pi\nu}+\frac{e^{-\nu}\sin^4\phi}{2\pi}\\
        =&\frac{e^{-\nu}}{4\pi\nu}\left[\frac{1}{2\nu}+\cos^2\phi\right]  +\frac{e^{-\nu}\sin^2\phi}{2\pi\nu} > 0.
\end{align*}
\endgroup
Combining both regions of $\nu$ completes the proof that $f_{\Phi|N}(\phi|\nu)$ is convex $\nu$ for Scenario A.\\
\noindent\textbf{Scenario B:}\\
In this case, $\cos(\phi) > 0$ and $\frac{\partial^2 f_{\Phi|N}(\phi|\nu)}{\partial \nu^2}$ becomes
\begin{equation*}
    \begin{split}
        &\frac{\partial^2 f_{\Phi|N}(\phi|\nu)}{\partial \nu^2}\\
        &\qquad= \underbrace{\frac{\cos\phi e^{-\nu\sin^2\phi}Q\left(-\sqrt{2\nu}\cos\phi\right)}{4\sqrt{\pi}\nu^{\frac{3}{2}}}}_{> 0}\left[(2\nu\sin^2\phi-1)^2-2\right]\\
        &\qquad\quad+\frac{e^{-\nu}\cos^2\phi}{4\nu\pi}+\frac{e^{-\nu}\sin^4\phi}{2\pi}
    \end{split}
\end{equation*}
and it is guaranteed that $  f_{\Phi|N}(\phi|\nu)$ is convex on $\nu$ in this scenario if $\nu > \frac{1+\sqrt{2}}{2\sin^2\phi}$. Note also that $\frac{\partial^2 f_{\Phi|N}(\phi|\nu)}{\partial \nu^2} < 0$ as $\nu \rightarrow 0^+$ and $\frac{\partial^2 f_{\Phi|N}(\phi|\nu)}{\partial \nu^2} > 0$ at $\nu = \frac{1+\sqrt{2}}{2\sin^2\phi}$. Moreover, for a fixed $\phi \ne 0$, the first term is monotonic increasing on $\nu$ and the sum of second and third terms is monotonic decreasing on $\nu$. Thus, by Intermediate Value Theorem, there exists a $\nu_0 \leq \frac{1+\sqrt{2}}{2\sin^2\phi}$ such that $f_{\Phi|N}(\phi|\nu)$ is convex for $\nu \geq \nu_0$ and nonconvex for $\nu < \nu_0$. 
\end{proof}
Because of Lemma \ref{lemma:convexity_p_phi_nu} and Lemma \ref{lemma:symmetry_p_phi_given_nu}, there exists two integers $y^{(2)}_0 \leq 2^{b-1}$ and $y^{(2)}_1 \geq 2^{b-1}$ such that we can define the integer sets $\mathcal{Y}_2 = \{y_0^{(2)}+1,\cdots,y_1^{(2)}\}$ and $\mathcal{Y}_2^c = \{0,\cdots,y_0^{(2)}\}\cup\{y_1^{(2)}+1,\cdots,2^{b}-1\}$ satisfying
\begingroup
\allowdisplaybreaks
\begin{align*}
     &\text{(i) }\frac{\partial^2 W_{y}^{(b)}(\nu,\theta)}{\partial\nu^2} \leq 0, y\in\mathcal{Y}_2,\\
     &\text{(ii) }\frac{\partial^2 W_{y}^{(b)}(\nu,\theta)}{\partial\nu^2} > 0, y\in\mathcal{Y}_2^c,\text{ and}\\
    & \text{(ii) }\min_{k\in\mathcal{Y}_2}\{W_{k}^{(b)}(\nu,\theta)\} \geq \max_{j\in\mathcal{Y}_2^c}\{W_{j}^{(b)}(\nu,\theta)\}.
\end{align*}%
\endgroup
Note that $\mathcal{Y}_2$ can be an empty set if $y_0^{(2)} = y_1^{(2)}$. Using a similar approach in Appendix \ref{appendix_C}, we define two nonnegative constants, $K_1''$ and $K_2''$, which are given by
\begingroup
\allowdisplaybreaks
\begin{align}
    K_1'' = \log\frac{W_{2^{b-1}}^{(b)}\left(\nu,\theta\right)}{\min_{y\in\mathcal{Y}_2\backslash\{2^{b-1}\}}\{ W_{y}^{(b)}\left(\nu,\theta\right)\}}
\end{align}%
and
\begin{align}
    K_2'' = \log\frac{W_{2^{b-1}}^{(b)}\left(\nu,\theta\right)}{\max_{y\in\mathcal{Y}_2^c\backslash\{2^{b-1}\}}\{ W_{y}^{(b)}\left(\nu,\theta\right)\}}
\end{align}
\endgroup
and then get the following lower bound for $ \frac{\partial^2 w(\nu,\theta)}{\partial \nu^2}$:
\begingroup
\allowdisplaybreaks
\begin{align*}
     \frac{\partial^2 w(\nu,\theta)}{\partial \nu^2} \geq& \sum_{y\ne 2^{b-1}}\frac{\partial^2 W_y^{(b)}\left(\nu, \theta\right)}{\partial \nu^2}\log_2 \left\{\frac{W_{2^{b-1}}^{(b)}\left(\nu, \theta\right)}{W_y^{(b)}\left(\nu, \theta\right)}\right\}\\ \geq& K_2''\sum_{y\ne 2^{b-1}}\frac{\partial^2 W_y^{(b)}\left(\nu, \theta\right)}{\partial \nu^2}\\
     =&  -K_2''\cdot\frac{\partial^2 W_{2^{b-1}}^{(b)}\left(\nu, \theta\right)}{\partial \nu^2}.
\end{align*}
\endgroup
The last line follows from extending (\ref{eq:phase_quantization_prob_relationship}) to second-order derivatives. The last line is positive since $W_{2^{b-1}}^{(b)}(\nu,\theta)$ is a strictly concave function of $\nu$ for $b\geq 3$ (see Appendix \ref{appendix_F}).

Next, suppose $\theta\notin [0,\frac{2\pi}{2^b})$. Then there exists a $\theta'\in[0,\frac{2\pi}{2^b})$ and $j\in\mathbb{Z}$ such that $\theta' = \theta - \frac{2\pi j}{2^b}$. The function $w(\nu,\theta,b)$ can be expressed as
\begingroup
\allowdisplaybreaks
\begin{align*}
 w(\nu,\theta,b) 
 =&-\sum_{y=0}^{2^{b-1}}W_y^{(b)}(\nu,\theta'+\frac{2\pi j}{2^b})\log W_y^{(b)}(\nu,\theta'+\frac{2\pi j}{2^b}) \\
 =&-\sum_{y=0}^{2^{b-1}}W_{y-j}^{(b)}(\nu,\theta')\log W_{y-j}^{(b)}(\nu,\theta')\\
 =& w(\nu,\theta',b).
\end{align*}
\endgroup
The second line follows from the relationship between $\theta'$ and $\theta$. The third line follows from Lemma \ref{lemma:symmetry_Wy}.i and the circular structure of the phase quantizer. This implies that $w(\nu,\theta,b)$ and $w(\nu,\theta',b)$ share the same properties so the claim holds for all $\theta\in[-\pi,\pi)$.


 \section{Proof of Proposition \ref{proposition:symmetry_distribution}} \label{appendix_G}

We first define the notations
\begingroup
\allowdisplaybreaks
\begin{align*}
         H_{F_X}(Y) =& -\sum_{y=0}^{2^b-1} p(y;F_X)\log p(y;F_X)\\
         H_{F_X}(Y|X)=& \int_{\mathbb{C}}w\left(\frac{|g_{\text{LoS}}|^2\alpha}{\sigma^2},\beta',b\right)\;dF_X,
\end{align*}%
\endgroup
where we used the subscript $F_X$ to note that the entropy and conditional entropy are induced by the input distribution in the subscript. We want to show that
\[H_{F_X^{s}}(Y) - H_{F_X^{s}}(Y|X) \geq H_{F_X}(Y) - H_{F_X}(Y|X)
\]
for any input distribution $F_X$. The conditional output entropy using $F_X^s$ is
\begingroup
\allowdisplaybreaks
\begin{align*}
    H_{F_X^s}(Y|X) =& \int_{\mathbb{C}}w\left(\frac{|g_{\text{LoS}}|^2\alpha}{\sigma^2},\beta',b\right)d\left[\frac{1}{2^b}\sum_{i = 0}^{2^b-1}F_X(xe^{j\frac{2\pi i}{2^b}}) \right]\\
    =& \frac{1}{2^b}\int_{\mathbb{C}}\sum_{i = 0}^{2^b-1}w\left(\frac{|g_{\text{LoS}}|^2\alpha}{\sigma^2},\beta'-\frac{2\pi i}{2^b},b\right)dF_X(x)
\end{align*}
\endgroup
Using Lemma \ref{lemma:symmetry_Wy}.i and the circular structure of phase quantizer, it is easy to show that 
\begin{equation*}
    \sum_{i = 0}^{2^b-1}w\left(\frac{|g_{\text{LoS}}|^2\alpha}{\sigma^2},\beta'-\frac{2\pi i}{2^b},b\right) = \sum_{i = 0}^{2^b-1}w\left(\frac{|g_{\text{LoS}}|^2\alpha}{\sigma^2},\beta',b\right).
\end{equation*}
Consequently, $H_{F_X^s}(Y|X)$ can be simplified to
\begingroup
\allowdisplaybreaks
\begin{align*}
     H_{F_X^s}(Y|X) =&\frac{1}{2^b}\sum_{i = 0}^{2^b-1}\int_{\mathbb{C}}w\left(\frac{|g_{\text{LoS}}|^2\alpha}{\sigma^2},\beta',b\right)\;dF_X \\
     =& H_{F_X}(Y|X).
\end{align*}
\endgroup
For the output entropy, we examine the PMF $p(y;F_X^s)$:
\begingroup
\allowdisplaybreaks
\begin{align*}
     p(y;F_X^s) =& \int_{\mathbb{C}} W_y^{(b)}\left(\frac{|g_{\text{LoS}}|^2\alpha}{\sigma^2},\beta'\right)d\left[\frac{1}{2^b}\sum_{i = 0}^{2^b-1}F_X(xe^{j\frac{2\pi i}{2^b}})\right]\\
     =&\frac{1}{2^b}\int_{\mathbb{C}}\sum_{i = 0}^{2^b-1} W_y^{(b)}\left(\frac{|g_{\text{LoS}}|^2\alpha}{\sigma^2},\beta'-\frac{2\pi i}{2^b}\right)dF_X\\
     =&\frac{1}{2^b}\int_{\mathbb{C}}\underbrace{\sum_{i = 0}^{2^b-1} W_{y+i}^{(b)}\left(\frac{|g_{\text{LoS}}|^2\alpha}{\sigma^2},\beta'\right)}_{=1}dF_X\\
     =&\frac{1}{2^b}\int_{\mathbb{C}}\;dF_X = \frac{1}{2^b}.
\end{align*}
\endgroup
The second line comes from the rotation of the input distribution and the third line follows from Lemma \ref{lemma:symmetry_Wy}.i. The expression $\sum_{i = 0}^{2^b-1} W_{y+i}^{(b)}\left(\frac{|g_{\text{LoS}}|^2\alpha}{\sigma^2},\beta'\right)$ is essentially summing $p_{Y|X}(y|x = \frac{|g_{\mathrm{LoS}}|^2\alpha e^{j\beta'}}{\sigma^2})$ over all values of $y \in \{0,\cdots,2^{b}-1\}$. Consequently, it is equal to 1. Thus, the PMF of $Y$ is a uniform distribution and also maximizes the output entropy. More specifically, the output entropy is $H_{F_X^s}(Y) = b$. This also implies that $I(F_X^s) \geq I(F_X)$.

\section{Proof of Lemma \ref{lemma:weak_differentiability}} \label{appendix_I}

Suppose we let $F_X^\lambda = (1-\lambda)F_X^0 + \lambda F_X$. Then, $I(F_X^{\lambda}) - (F_X^{0})$ can be written as
\begingroup
\allowdisplaybreaks
\begin{align*}
&I(F_X^{\lambda}) - (F_X^{0})\\
      &\qquad= (1-\lambda)\left[b-\int_{\mathbb{C}}w\left(\frac{|g_{\mathrm{LoS}}|^2\alpha}{\sigma^2},\beta',b\right)\;dF_X^0\right] \\
      &\qquad\qquad+ \lambda\left[b-\int_{\mathbb{C}}w\left(\frac{|g_{\mathrm{LoS}}|^2\alpha}{\sigma^2},\beta',b\right)\;dF_X\right]\\
      &\qquad\qquad-\left[b-\int_{\mathbb{C}}w\left(\frac{|g_{\mathrm{LoS}}|^2\alpha}{\sigma^2},\beta',b\right)\;dF_X^0\right]\\
      &\qquad= \lambda\bigg[-b+\int_{\mathbb{C}}w\left(\frac{|g_{\mathrm{LoS}}|^2\alpha}{\sigma^2},\beta',b\right)\;dF_X^0 \\
      &\qquad\qquad+ b - \int_{\mathbb{C}}w\left(\frac{|g_{\mathrm{LoS}}|^2\alpha}{\sigma^2},\beta',b\right)\;dF_X\bigg]\\
      &\qquad=\lambda[I(F_X) - I(F_X^0)].
\end{align*}
\endgroup
Using the definition of weak differentiability in (\ref{eq:weak_differentiability}), we can express $I'_{F_X^0}(F_X)$ as
\begin{align*}
     I'_{F_X^0}(F_X) = \lim_{\lambda \rightarrow 0 }\frac{I\left(F_X^\lambda\right) - I(F_X^0)}{\lambda} = I(F_X) - I(F_X^0).
\end{align*}
The proof is concluded by noting that the weak derivative above exists because both terms are finite (also a consequence of discrete nature and finite cardinality of the output).

\section{Proof of Lemma \ref{lemma:boundedness}} \label{appendix_H}

Without loss of generality, we assume $g_{\mathrm{LoS}} = 1$. For all $\phi$ and $\nu$, $f_{\Phi|N}(\phi|\nu)$ can be bounded from below by
\begingroup
\allowdisplaybreaks
\begin{align*}
    f_{\Phi|N}(\phi|\nu)  
    \geq&  \underbrace{\begin{cases} \frac{\sqrt{\nu}\cos\left(\phi\right)e^{-\nu\sin^2\left(\phi\right)}}{\sqrt{\pi}},\quad \phi \in [-\pi/2,\pi/2]\\
          \qquad\;\;\;\;0 \qquad\;\;\;,\quad\mathrm{otherwise}
    \end{cases}}_{{f_\mathrm{\mathrm{LB}}(\nu,\phi)}}.
\end{align*}
\endgroup
The inequality comes from the Q-function upper bound $Q(x) < \frac{e^{-\frac{x^2}{2}}}{\sqrt{2\pi}x}$ for $x > 0$ and the fact that $\cos(\phi) < 0$ when $\phi\notin[-\pi/2,\pi/2]$. We obtain the RHS after some algebraic manipulation. Similarly, an upper bound for $f_{\Phi|N}(\phi|\nu)$ can be established as follows:
\begingroup
\allowdisplaybreaks
\begin{align*}
    f_{\Phi|N}(\phi|\nu) 
    \leq& \underbrace{\begin{cases}
         \frac{e^{-\nu}}{2\pi}+\frac{\sqrt{\nu}\cos\left(\phi\right)e^{-\nu\sin^2\left(\phi\right)}}{\sqrt{\pi}},\quad \phi \in [-\pi/2,\pi/2]\\
         \qquad\quad\;\;\;\frac{e^{-\nu}}{2\pi}\;\;\;\quad\qquad,\quad\mathrm{otherwise}
    \end{cases}}_{f_\mathrm{UB}(\nu,\phi) \;=\; f_\mathrm{\mathrm{LB}}(\nu,\phi) + \frac{e^{-\nu}}{2\pi}}.
\end{align*}
\endgroup
The inequality comes from the fact that $\cos(\phi) < 0$ when $\phi\notin[-\pi/2,\pi/2]$ and $Q(x)$ is nonnegative for all $x$. The negative terms are dropped to obtain the upper bound.

Let
\begin{align*}
    F_{\mathrm{LB}}(\nu,\beta) =&  \int_{\frac{2\pi}{2^b}y-\pi-\beta}^{\frac{2\pi}{2^b}(y+1)-\pi-\beta} f_{\mathrm{\mathrm{LB}}}\left(\nu,\phi\right)d\phi\\
    =&\Bigg[Q\Bigg(\frac{\sqrt{\alpha}\sin\left(\frac{2\pi(y)}{2^b}-\beta\right)}{\sigma}\Bigg)\\
    &\qquad\quad- Q\Bigg(\frac{\sqrt{\alpha}\sin\left(\frac{2\pi(y+1)}{2^b}-\beta\right)}{\sigma}\Bigg)\Bigg]_+.
\end{align*}
where $[\cdot]_+ = \max\{\cdot,0\}$. This expression coincides with the approximation given in \cite{Cahn:1959}. For an input $x = \sqrt{\alpha}e^{j\beta}$, the channel law $W_y^{(b)}\left(\frac{\alpha}{\sigma^2},\beta\right)$ can be bounded from above and from below as follows:
\begin{align*}
    F_{\mathrm{LB}}\left(\frac{\alpha}{\sigma^2},\beta\right)\; \leq\; W_y^{(b)}\left(\frac{\alpha}{\sigma^2},\beta\right)\;\leq\; \frac{e^{-\frac{\alpha}{\sigma^2}}}{2^b} + F_{\mathrm{LB}}\left(\frac{\alpha}{\sigma^2},\beta\right).
\end{align*}
By the Squeeze Theorem, we have 
\begin{equation}\label{eq:integ_limit_f_phi_nu}
    \begin{split}
       \lim_{\alpha \rightarrow \infty} W_y^{(b)}\left(\frac{\alpha}{\sigma^2},\beta\right) =&  \lim_{\alpha \rightarrow \infty} F_{\mathrm{LB}}\left(\frac{\alpha}{\sigma^2},\beta\right).
    \end{split} 
\end{equation}

To prove the boundedness of the support, we consider two cases of the KTC coefficient $\mu$.\\

\noindent \textbf{Case A ($\mu > 0$)}: \\
Suppose we define $y_\beta$ to be the phase quantizer output that satisfies $\beta \in \mathcal{R}_{y_\beta}^{\text{PH}}$. It can be shown using (\ref{eq:integ_limit_f_phi_nu}) that
\begin{align}\label{eq:limit_Wy_part1}
\lim_{\alpha \rightarrow \infty}W_y^{(b)}\left(\frac{\alpha}{\sigma^2},\beta\right) = \mathbbm{1}_{\{y_\beta\}}(y)
\end{align}
when $\beta \ne \frac{2\pi y_{\beta}}{2^{b}}$ (i.e. when $\beta$ does not fall exactly at the boundary of $\mathcal{R}_{y_{\beta}}^{\text{PH}}$), and 
\begin{align}\label{eq:limit_Wy_part2}
    \lim_{\alpha \rightarrow \infty}W_y^{(b)}\left(\frac{\alpha}{\sigma^2},\beta\right) = \frac{1}{2}\mathbbm{1}_{\{y_{\beta},y_{\beta}-1\}}(y)
\end{align}
when $\beta = \frac{2\pi y_{\beta}}{2^{b_1}}$ (i.e. when $\beta$ falls exactly at the boundary of $\mathcal{R}_{y_{\beta}}^{\text{PH}}$). The notation $\mathbbm{1}_{A}(\cdot)$ refers to the indicator function. Consequently, $\underset{\alpha\rightarrow \infty}{\lim}\; w\left(\frac{\alpha}{\sigma^2},\beta,b\right)$ is  given by
\begin{equation*}
= \begin{cases}
0,\quad\text{if $\beta \ne \frac{2\pi y_\beta}{2^b},y_\beta\in \{0,1,\cdots,2^{b}-1\}$}\\
1,\quad\text{if $\beta = \frac{2\pi y_\beta}{2^b},y_\beta\in \{0,1,\cdots,2^{b}-1\}$}
\end{cases}.
\end{equation*}
Since $C$ and $b$ are non-negative numbers with $C \leq b$, the LHS of (\ref{eq:KTC}) goes to $\infty$ as $\alpha \rightarrow \infty$ and equality is not achieved. Equivalently, $x = \sqrt{\alpha}e^{j\beta} \in F_X^*$ should have finite magnitude.\\

\noindent \textbf{Case B ($\mu = 0$)}: \\
In this case, the KTC in (\ref{eq:KTC}) becomes $C \geq b-w\left(\frac{\alpha}{\sigma^2},\beta,b\right)$. Similar to the approach in \cite{Rahman:2020} and \cite{Singh:2009}, we want to show that there exists a finite constant $\alpha_0$ such that for $\alpha > \alpha_0$, equality in (\ref{eq:KTC}) cannot be achieved with $\mu = 0$ and any $\beta \in \mathcal{R}_{y_{\beta}}^{\text{PH}}$. Mathematically,
\begin{align*}
\exists \alpha_0\in\mathbb{R^+}\;|\;\forall \alpha > \alpha_0: \;\;  w\left(\frac{\alpha}{\sigma^2},\beta;b\right) > \underset{\alpha'\rightarrow\infty}{\lim}\;  w\left(\frac{\alpha'}{\sigma^2},\beta;b\right).    
\end{align*}
The strictly decreasing property of $w\left(\frac{\alpha}{\sigma^2},\beta;b\right)$ established in Proposition \ref{proposition:monotonicity_w} implies the existence of such $\alpha_0$. Combining the results of both cases concludes the proof.


 \section{Proof of Lemma \ref{lemma: discreteness}} \label{appendix_K}
 
First, let $P_0 \leq P$ and $R(y) = p(y;F_X^*)$ be the power and output distribution corresponding to the optimal input. Also, let $\mathcal{B}(l)$ be a Borel set of $x\in\mathbb{C}$ with $|x|^2 \leq l$. Due to Lemma \ref{lemma:boundedness}, there exists a finite $T$ such that $\text{supp}(F_X^*) \subset \mathcal{B}(T)$. Define a convex and compact set $\mathcal{S}$ to be
\begin{align*}
    \mathcal{S} = \{F_X| \text{supp}(F_X) \subset \mathcal{B}(T)\},
\end{align*}
and the corresponding subset $\mathcal{M}$ of $\mathcal{S}$ as
\begin{align*}
    \mathcal{M} = \left\{F_X \in \mathcal{S}| p(y;F_X) = R(y),\mathbb{E}[|X|^2] = P_0\right\}.
\end{align*}
It is clear that $F_X^* \in \mathcal{M}$ for some finite $T$ since the output PMF should be $p(y;F_X) = 1/2^b\;\forall y\in\{0,\cdots,2^b-1\}$ and $F_X^*$ is bounded. Thus, we can rewrite the capacity formula as
\begin{align*}
        C =& \sup_{F_U\in \mathcal{S}} I(F_X) = \sup_{F_U\in \mathcal{M}} \left\{b  - \int_\mathbb{C} w\left(\frac{\alpha}{\sigma^2},\beta,b\right)\;dF_X\right\}
\end{align*}
for some non-negative multiplier $\mu$. Note that $I(F_X)$ is a sum of a constant term plus a linear functional term with respect to $F_X$. As such, it has a maximum at an extreme point in $\mathcal{M}$ and this extreme point is $F_X^*$. Moreover, we consider $\mathcal{M}$ as intersection of $\mathcal{S}$, the $2^{b}-1$ output probability hyperplanes given by
\begin{equation*}
    \mathcal{H}_{y} : \int_{B(T)}W_{y}^{(b)}\left(\frac{\alpha}{\sigma^2},\beta\right)\;dF_X = \frac{1}{2^b},
\end{equation*}
for all $y \in \{0,1,\cdots,2^{b}-2\}$ (defining $\mathcal{H}_{2^{b}-1}$ is redundant since the probability of all mass points should sum up to 1), and an additional hyperplane given by
\begin{equation*}
    \mathcal{H}_{P} : \int_{B(T)}|x|^2\;dF_X = P_0.
\end{equation*}
By applying Dubin's Theorem \cite{Dubins:1962} in a similar manner as in \cite{Witsenhausen:1980} and \cite{Singh:2009}, the optimal distribution $F_X^*$ is a convex combination of at most $2^b+1$ extreme points of $\Omega_s$. These extreme points are the set of at most $L$ unit masses $\left\{\delta(x_{i})\right\}_{i=1}^{i=L}$, where $x_{i} \in \mathcal{B}(T)$ and $L\leq 2^b +1$.

\section{Proof of Proposition \ref{proposition:optimal_loc}} \label{appendix_J}

Suppose we define a function $\mathcal{L}(\beta')$ as
\begin{equation*}
    \mathcal{L}(\beta') = C -b+w\left(\frac{|g_{\text{LoS}}|^2P}{\sigma^2},\beta',b\right)
\end{equation*}
(i.e. the LHS of equation (\ref{eq:KTC}) with $\alpha = P$). Note that $\mathcal{L}(\beta') = 0$ implies that $x = \sqrt{P}e^{j(\beta' - \angle g_{\mathrm{LoS}})} \in F_{X}^*$ while $\mathcal{L}(\beta') > 0$ implies otherwise. Then, a necessary condition for $\beta'$ to be a minimizer of $\mathcal{L}(\beta')$ is
\begin{equation}
    \frac{\partial\mathcal{L}(\beta')}{\partial \beta'} = \frac{\partial w\left(\nu^*,\beta',b\right)}{\partial \beta'} = 0,
\end{equation}
where we let $\nu^* = \frac{|g_{\text{LoS}}|^2P}{\sigma^2}$. This expression for $\frac{\partial w\left(\nu^*,\beta',b\right)}{\partial \beta'} $ can be explicitly written as
\begingroup
\allowdisplaybreaks
\begin{align}\label{eq:diff_L_beta}
     = \sum_{y \ne 2^{b-1}} \frac{\partial W_{y}^{(b)}\left(\nu^*,\beta'\right)}{\partial \beta'}\log\left[\frac{W_{2^{b-1}}^{(b)}\left(\nu^*,\beta'\right)}{W_{y}^{(b)}\left(\nu^*,\beta'\right)}\right],
\end{align}
\endgroup
which is obtained using chain rule of differentiation and the fact that $\sum_{y=0}^{2^{b-1}} \frac{\partial W_{y}^{(b)}\left(\nu^*,\beta'\right)}{\partial \beta'} = 0$ (i.e. adopting (\ref{eq:phase_quantization_prob_relationship}) for $\beta'$ instead of $\nu$). Note that we have dropped the factor $\frac{1}{\ln 2}$ since it does not affect the sign of the differential. We can rewrite (\ref{eq:diff_L_beta}) as
\begingroup
\allowdisplaybreaks
\begin{align*}
     =& \sum_{y=1}^{y = 2^{b-1}-1} \Bigg\{\frac{\partial W_{2^{b-1}-y}^{(b)}\left(\nu^*,\beta'\right)}{\partial \beta'}\log\left[\frac{W_{2^{b-1}}^{(b)}\left(\nu^*,\beta'\right)}{W_{2^{b-1}-y}^{(b)}\left(\nu^*,\beta'\right)}\right] \\
     &\qquad\qquad+ \frac{\partial W_{2^{b-1}+y}^{(b)}\left(\nu^*,\beta'\right)}{\partial \beta'}\log\left[\frac{W_{2^{b-1}}^{(b)}\left(\nu^*,\beta'\right)}{W_{2^{b-1}+y}^{(b)}\left(\nu^*,\beta'\right)}\right]\Bigg\}\\
     &\qquad\qquad+ \frac{\partial W_{0}^{(b)}\left(\nu^*,\beta'\right)}{\partial \beta'}\log\left[\frac{W_{2^{b-1}}^{(b)}\left(\nu^*,\beta'\right)}{W_{0}^{(b)}\left(\nu^*,\beta'\right)}\right].
\end{align*}
\endgroup
Suppose $\beta' = \frac{\pi}{2^b}$. We can apply Leibniz integral rule to get
\begingroup
\allowdisplaybreaks
\begin{align*}
    \frac{\partial W_{2^{b-1}-y}^{(b)}\left(\nu^*,\beta'\right)}{\partial \beta'}\Big|_{\beta' = \frac{\pi}{2^b}} =& f_{\Phi|N}\left(-\frac{2\pi(y+0.5)}{2^b}\Big|\nu^*\right) \\
    &- f_{\Phi|N}\left(-\frac{2\pi(y-0.5)}{2^b}\Big|\nu^*\right)\\
    \frac{\partial W_{2^{b-1}+y}^{(b)}\left(\nu^*,\beta'\right)}{\partial \beta'}\Big|_{\beta' = \frac{\pi}{2^b}} =& f_{\Phi|N}\left(\frac{2\pi(y-0.5)}{2^b}\Big|\nu^*\right) \\
    &- f_{\Phi|N}\left(\frac{2\pi(y+0.5)}{2^b}\Big|\nu^*\right)\\
    \frac{\partial W_{0}^{(b)}\left(\nu^*,\beta'\right)}{\partial \beta'}\Big|_{\beta' = \frac{\pi}{2^b}} =&f_{\Phi|N}\left(-\frac{\pi}{2^b}\Big|\nu^*\right) - f_{\Phi|N}\left(\frac{\pi}{2^b}\Big|\nu^*\right)
\end{align*}
\endgroup
Due to even-symmetry of $f_{\Phi|N}(\phi|\nu)$ (Lemma \ref{lemma:symmetry_p_phi_given_nu}), we have
\begingroup
\allowdisplaybreaks
\begin{align*}
     \frac{\partial W_{2^{b-1}-y}^{(b)}\left(\nu^*,\beta'\right)}{\partial \beta'}\Big|_{\beta' = \frac{\pi}{2^b}} =&  -\frac{\partial W_{2^{b-1}+y}^{(b)}\left(\nu^*,\beta'\right)}{\partial \beta'}\Big|_{\beta' = \frac{\pi}{2^b}},
\end{align*}
for all $y\in\{1,\cdots,2^{b-1}-1\}$, and 
\begin{align*}
     \frac{\partial W_{0}^{(b)}\left(\nu^*,\beta'\right)}{\partial \beta'}\Big|_{\beta' = \frac{\pi}{2^b}} =& 0.
\end{align*}
\endgroup
By combining this with Lemma \ref{lemma:symmetry_Wy}.ii, (\ref{eq:diff_L_beta}) becomes 0. Alternatively, we can write (\ref{eq:diff_L_beta}) as
\begingroup
\allowdisplaybreaks
\begin{align*}
     =& \sum_{y=1}^{y = 2^{b-1}-1} \Bigg\{\frac{\partial W_{2^{b-1}-y}^{(b)}\left(\nu^*,\beta'\right)}{\partial \beta'}\log\left[\frac{W_{2^{b-1}}^{(b)}\left(\nu^*,\beta'\right)}{W_{2^{b-1}-y}^{(b)}\left(\nu^*,\beta'\right)}\right] \nonumber\\
     &\quad\quad+ \frac{\partial W_{2^{b-1}-1+y}^{(b)}\left(\nu^*,\beta'\right)}{\partial \beta'}\log\left[\frac{W_{2^{b-1}}^{(b)}\left(\nu^*,\beta'\right)}{W_{2^{b-1}-1+y}^{(b)}\left(\nu^*,\beta'\right)}\right]\Bigg\}\nonumber\\
     &\quad\quad+ \frac{\partial W_{2^{b-1}-1}^{(b)}\left(\nu^*,\beta'\right)}{\partial \beta'}\log\left[\frac{W_{2^{b-1}}^{(b)}\left(\nu^*,\beta'\right)}{W_{2^{b-1}-1}^{(b)}\left(\nu^*,\beta'\right)}\right].
\end{align*}
\endgroup
Suppose we have $\beta' = 0$. The last term becomes zero due to Lemma \ref{lemma:symmetry_Wy}.iii. We apply Leibniz integral rule to get
\begingroup
\allowdisplaybreaks
\begin{align*}
    \frac{\partial W_{2^{b-1}-y}^{(b)}\left(\nu^*,\beta'\right)}{\partial \beta'}\Big|_{\beta' = 0} =& f_{\Phi|N}\left(-\frac{2\pi y}{2^b}\Big|\nu^*\right) \\
    &- f_{\Phi|N}\left(-\frac{2\pi(y-1)}{2^b}\Big|\nu^*\right)\\
    \frac{\partial W_{2^{b-1}-1+y}^{(b)}\left(\nu^*,\beta'\right)}{\partial \beta'}\Big|_{\beta' = 0} =& f_{\Phi|N}\left(\frac{2\pi (y-1)}{2^b}\Big|\nu^*\right) \\
    &- f_{\Phi|N}\left(\frac{2\pi y}{2^b}\Big|\nu^*\right)
\end{align*}
\endgroup
Due to even-symmetry of $f_{\Phi|N}(\phi|\nu)$ (Lemma \ref{lemma:symmetry_p_phi_given_nu}), we have
\begingroup
\allowdisplaybreaks
\begin{align*}
    \frac{\partial
    W_{2^{b-1}-y}^{(b)}\left(\nu^*,\beta'\right)}{\partial \beta'}\Big|_{\beta' = 0} =&  -\frac{\partial W_{2^{b-1}-1+y}^{(b)}\left(\nu^*,\beta'\right)}{\partial \beta'}\Big|_{\beta' = 0},
\end{align*}
\endgroup
for all $y\in\{1,\cdots,2^{b-1}\}$. By combining this with Lemma \ref{lemma:symmetry_Wy}.iii, equation (\ref{eq:diff_L_beta}) becomes 0. Thus, the two stationary points occur at $\beta' \in \{0,\frac{\pi}{2^b}\}$. To prove that $\beta' = 0$ is not a minimizer, it suffices to show that 
\begingroup
\allowdisplaybreaks
\begin{align*}
     \mathcal{L}(0) > & \mathcal{L}\left(\frac{\pi}{2^b}\right)
\end{align*}
or, when written explicitly,
\begin{align*}
    &2\sum_{y=0}^{2^{b-1}-1}W_{y}^{(b)}\left(\nu^*,0\right)\log \frac{1}{W_{y}^{(b)}\left(\nu^*,0\right)}\\
    &\qquad\qquad > \sum_{y=0}^{2^{b-1}-1}\Bigg\{W_{y+1}^{(b)}\left(\nu^*,\frac{\pi}{2^b}\right)\log\frac{1}{ W_{y+1}^{(b)}\left(\nu^*,\frac{\pi}{2^b}\right)}\\
    &\qquad\qquad\qquad+ W_{y}^{(b)}\left(\nu^*,\frac{\pi}{2^b}\right)\log\frac{1}{ W_{y}^{(b)}\left(\nu^*,\frac{\pi}{2^b}\right)}\Bigg\}.
\end{align*}
\endgroup
The LHS follows from Lemma \ref{lemma:symmetry_Wy}.iii and the RHS follows from Lemma \ref{lemma:symmetry_Wy}.ii and the circular structure of the phase quantizer. Since $f_{\Phi|N}(\phi|\nu)$ is non-decreasing function of $\phi$ for $\phi < 0$ and any $\nu > 0$ (Lemma \ref{lemma:symmetry_p_phi_given_nu}), it is easy to verify that
\[W^{(b)}_y(\nu^*,0) < W^{(b)}_{y+1}\left(\nu^*,\frac{\pi}{2^b}\right)
\]
and 
\[
W^{(b)}_y(\nu^*,0) > W^{(b)}_{y}\left(\nu^*,\frac{\pi}{2^b}\right),\quad\forall y\in\{0,\cdots,2^{b-1}-1\}
\]
from Definition \ref{definition:phase_quantization_func}. We prove the following lemma about the entropy of a discrete distribution. Essentially, this lemma tells us that if the discrete distribution has a form $\{p_i\}_{i=1}^{i=N}$ such that $p_i = p_{\frac{N}{2}+i}\;\forall i \in \{1,\cdots,N/2\}$, then breaking these ties in the probabilities reduces entropy.
\begin{lemma}\label{lemma:symmetric_entropy}
For all $a_y >  0$ such that $\sum_{y}a_y = 0.5$, the inequality
\begin{align}\label{eq:inequality_entopy}
    &-\sum_{y}\left[(a_y-b_y)\log (a_y-b_y) + (a_y+c_y)\log (a_y+c_y)\right] \nonumber \\
    &\qquad\qquad\qquad\leq -2\sum_{y}a_y\log a_y 
\end{align}
holds for all $c_y \geq 0$ and $a_{y} > b_y \geq 0$ that satisfy $\sum_{y}c_y = \sum_y b_y$.
\end{lemma}
\begin{proof}
 Let $f(\boldsymbol{b},\boldsymbol{c})$ be the RHS of (\ref{eq:inequality_entopy}). The second-order derivatives of $f(\boldsymbol{b},\boldsymbol{c})$ with respect to $b_y$ and $c_y$ are
\begin{align*}
&\frac{\partial^2 f(\boldsymbol{b},\boldsymbol{c})}{\partial b_y^2} = -\frac{1}{a_y-b_y}
,\qquad\frac{\partial^2 f(\boldsymbol{b},\boldsymbol{c})}{\partial c_y^2} = -\frac{1}{a_y+c_y},
\end{align*}
and
\begin{align*}
    \frac{\partial^2 f(\boldsymbol{b},\boldsymbol{c})}{\partial b_y b_{i\ne y}}=\frac{\partial^2 f(\boldsymbol{b},\boldsymbol{c})}{\partial c_yc_{i\ne y}} = 0, 
\end{align*}
which are all nonpositive in the domain of $f(\boldsymbol{b},\boldsymbol{c})$. Thus, $f(\boldsymbol{b},\boldsymbol{c})$ is concave. Next, we construct the Lagrangian as
\begin{align*}
f(\boldsymbol{b},\boldsymbol{c},\lambda) =& -\sum_{k}\bigg[(a_y-b_y)\log (a_y-b_y) \\&
\quad+ (a_y+c_y)\log (a_y+c_y)\bigg] -\lambda\left(\sum_y c_y-b_y\right)
\end{align*}
for $\lambda > 0$. The optimal solution $(\boldsymbol{b}^*,\boldsymbol{c}^*)$ should satisfy
\begin{align*}
    \frac{\partial f(\boldsymbol{b},\boldsymbol{c},\lambda)}{\partial b_y} =& [\log(a_y-b_y)+1]+\lambda = 0,\\
    \frac{\partial f(\boldsymbol{b},\boldsymbol{c},\lambda)}{\partial c_y} =& -[\log(a_y+c_y)+1]-\lambda = 0.
\end{align*}
Combining the two stationary conditions, we get
\begin{equation*}
    \begin{split}
        \log(a_y-b_y) =& \log(a_y+c_y),
    \end{split}
\end{equation*} 
which is only satisfied if $c_y = b_y = 0$. Thus, $\boldsymbol{b}^* = \boldsymbol{0}$ and $\boldsymbol{c}^* = \boldsymbol{0}$. Plugging $\boldsymbol{b}^*$ and $\boldsymbol{c}^*$ to the LHS of (\ref{eq:inequality_entopy}) gives the RHS of (\ref{eq:inequality_entopy}).
\end{proof}
If we let $a_y = W_y(\nu^*,0)$, $a_y - b_y = G_y(\nu^*,\pi/2^b)$, and $a_y+c_y = G_{y+1}(\nu^*,\pi/2^b)$, and apply Lemma \ref{lemma:symmetric_entropy}, then $\beta' = 0$ is not a minimizer. Noting that the input distribution should be $\frac{2\pi}{2^b}$-symmetric and that $\beta = \beta' - \angle g_{\mathrm{LoS}}$ concludes the proof.

\section{$W_{2^{b-1}}^{(b)}(\nu,\theta)$ is a strictly increasing function of $\nu$ for $b\geq 3$}\label{appendix_D}

To prove that $W_{2^{b-1}}^{(b)}(\nu,\theta)$ is an increasing function of $\nu$, we need to show that $\frac{\partial W_{2^{b-1}}^{(b)}(\nu,\theta)}{\partial\nu}$ can be written as
\begingroup
\allowdisplaybreaks
\begin{align}\label{eq:first_diff_W_2b-1}
    &\frac{\partial W_{2^{b-1}}^{(b)}(\nu,\theta)}{\partial\nu}\nonumber\\
    &= \int_{-\theta}^{\frac{2\pi}{2^b}-\theta}\;\frac{\partial f_{\Phi|N}\left(\phi|\nu\right)}{\partial\nu}\;d\phi \nonumber\\
    &=2\int_{0}^{\theta}\;\frac{\partial f_{\Phi|N}\left(\phi|\nu\right)}{\partial\nu}\;d\phi + \int_{\theta}^{\frac{2\pi}{2^b}-\theta}\;\frac{\partial f_{\Phi|N}\left(\phi|\nu\right)}{\partial\nu}\;d\phi,
    \end{align}%
\endgroup
where $\frac{\partial f_{\Phi|N}\left(\phi|\nu\right)}{\partial\nu}$ is given in (\ref{eq:first_diff_p_phi_given_nu}), is positive. The first integral term in the last equality of (\ref{eq:first_diff_W_2b-1}) follows from Lemma \ref{lemma:symmetry_p_phi_given_nu}. We first analyze a simple lower bound of the first term of (\ref{eq:first_diff_W_2b-1}) for arbitrary $\theta$ and show that it is positive for $b \geq 3$. We breakdown the problem into two regions of $\nu$.\\

\noindent\textbf{Region 1} ( $\nu \in \left[\frac{1}{2\sin^2\theta},+\infty\right)$ ): We first give a lower bound of $\frac{\partial f_{\Phi|N}\left(\phi|\nu\right)}{\partial\nu}$ for $\nu > \frac{1}{2\sin^2\phi}$.
\begingroup
\allowdisplaybreaks
\begin{align*}
    \frac{\partial f_{\Phi|N}\left(\phi|\nu\right)}{\partial\nu}
            >&-\frac{e^{-\nu}\sin^2\phi}{2\pi}-\frac{e^{-\nu\sin^2\phi}\cos\phi(2\nu\sin^2\phi-1)}{2\sqrt{\nu}\sqrt{\pi}}\\
            &\cdot\left[1-\frac{e^{-\nu\cos^2\phi}}{\sqrt{2\pi}\sqrt{2\nu}\cos\phi}\left(1-\frac{1}{2\nu\cos^2\phi}\right)\right]\\
            =&-\frac{e^{-\nu\sin^2\phi}\cos\phi(2\nu\sin^2\phi-1)}{2\sqrt{\nu}\sqrt{\pi}}\\
            &-\frac{e^{-\nu}}{4\pi\nu}\left(1-\frac{1}{2\nu\cos^2\phi}\right)-\frac{e^{-\nu}\tan^2\phi}{4\pi\nu}\\
            =&\frac{e^{-\nu\sin^2\phi}\cos\phi(1-2\nu\sin^2\phi)}{2\sqrt{\nu}\sqrt{\pi}}\\
            &-\frac{e^{-\nu}\sec^2\phi}{4\pi\nu}\left(1-\frac{1}{2\nu}\right).
\end{align*}%
\endgroup
The first line is obtained using the Q-function lower bound $Q(x) > \frac{\exp(-x^2/2)}{\sqrt{2\pi}x}\left(1-\frac{1}{x^2}\right)$. The second and third lines are obtained after some algebraic manipulation. Meanwhile, for $\nu \leq \frac{1}{2\sin^2\phi}$, the lower bound becomes
\begingroup
\allowdisplaybreaks
\begin{align*}
    \frac{\partial f_{\Phi|N}\left(\phi|\nu\right)}{\partial\nu}
        >&-\frac{e^{-\nu}\sin^2\phi}{2\pi}+\frac{e^{-\nu\sin^2\phi}\cos\phi(1-2\nu\sin^2\phi)}{2\sqrt{\nu}\sqrt{\pi}}\\
        &\qquad\cdot\left[1-\frac{e^{-\nu\cos^2\phi}}{\sqrt{2\pi}\sqrt{2\nu}\cos\phi}\right]\\
        =&\frac{e^{-\nu\sin^2\phi}\cos\phi(1-2\nu\sin^2\phi)}{2\sqrt{\nu}\sqrt{\pi}}-\frac{e^{-\nu}}{4\pi\nu}.
\end{align*}%
\endgroup
The inequality is obtained using the Q-function upper bound $Q(x) < \frac{\exp(-x^2/2)}{\sqrt{2\pi}x}$. Suppose we define $\theta_0 = \sin^{-1}\left(\frac{1}{\sqrt{2\nu}}\right) > 0$ for some fixed $\nu$. Then, $ \int_{0}^{\theta}\;\frac{\partial f_{\Phi|N}\left(\phi|\nu\right)}{\partial\nu}\;d\phi$ can be bounded by
\begingroup
\allowdisplaybreaks
\begin{align*}
&\int_{0}^{\theta}\;\frac{\partial f_{\Phi|N}\left(\phi|\nu\right)}{\partial\nu}\;d\phi\\
     &\qquad> \int_{\theta_0}^{\theta}\bigg[\frac{e^{-\nu\sin^2\phi}\cos\phi(1-2\nu\sin^2\phi)}{2\sqrt{\nu}\sqrt{\pi}}\\
     &\qquad\qquad\qquad-\frac{e^{-\nu}\sec^2\phi}{4\pi\nu}\left(1-\frac{1}{2\nu}\right)\bigg]d\phi\\
        &\qquad\quad+\int_{0}^{\theta_0}\left[\frac{e^{-\nu\sin^2\phi}\cos\phi(1-2\nu\sin^2\phi)}{2\sqrt{\nu}\sqrt{\pi}}-\frac{e^{-\nu}}{4\pi\nu}\right]d\phi\\ 
     &\qquad=\;\frac{e^{-\nu}\left[\tan(\theta_0)-\tan\theta\right]}{4\pi\nu}\left(1-\frac{1}{2\nu}\right)\\
     &\qquad\quad-\frac{e^{-\nu}\theta_0}{4\pi\nu}+\frac{\sin\theta e^{-\nu\sin^2\theta}}{2\sqrt{\nu}\sqrt{\pi}}\\
     &\qquad=\;\frac{e^{-\frac{1}{2}\csc^2\theta_0}\sin^2(\theta_0)\cos^2(\theta_0)\left[\tan(\theta_0)-\tan\left(\theta\right)\right]}{2\pi}\\
         &\qquad\quad\;-\frac{e^{-\frac{1}{2}\csc^2\theta_0}\theta_0\sin^2\theta_0}{2\pi}\\
         &\qquad\quad+\frac{\sqrt{2\pi}\sin(\theta_0)\sin\left(\theta\right)\exp\left(-\frac{\sin^2\left(\theta\right)}{2\sin^2(\theta_0)}\right)}{2\pi}.
    \end{align*}%
\endgroup
The second line is obtained by evaluating the integrals while the third line follows from expressing the $\nu$'s in terms of $\theta_0$. We want to show that the above expression is positive. Equivalently, the claim is proven for Region 1 by showing that
\begin{equation}\label{eq:check_first_diff_reg1}
    \begin{split}
        &\sqrt{2\pi}\sin\left(\theta\right)\exp\left(\frac{\cos^2\left(\theta\right)}{2\sin^2(\theta_0)}\right) \\
        &\qquad\qquad> \sin\theta_0\left[\theta_0-\cos^2(\theta_0)\left[\tan(\theta_0)-\tan\left(\theta\right)\right]\right]
    \end{split}
\end{equation}
holds for $\theta \in(0,\frac{\pi}{4}]$ and  $\theta_0\in[0,\theta]$. At $\theta_0 = \theta$, (\ref{eq:check_first_diff_reg1}) becomes
\begin{equation*}
    \sqrt{2\pi}\sin\theta\exp\left(\frac{\cot^2\theta}{2}\right) > \theta\sin\theta,
\end{equation*}
which holds for $\theta = (0,\frac{\pi}{4}]$. Moreover, for a fixed $\theta$, The LHS of (\ref{eq:check_first_diff_reg1}) increases as $\theta_0$ moves towards 0. The RHS of (\ref{eq:check_first_diff_reg1}) can be verified to be an increasing function of $\theta_0$ for $\theta \in [0,\frac{\pi}{4})$ by inspecting its derivative with respect to $\theta_0$. That is, 
\begin{equation*}
    \begin{split}
        \cos\theta_0[\tan(\theta_0)-\tan(\theta)][2\sin^2(\theta_0)-\cos^2(\theta)] +\theta_0\cos\theta_0\; 
    \end{split}
\end{equation*}
is positive $\forall \theta_0 \leq \theta$ and $\forall \theta\in(0,\frac{\pi}{4}]$. As such, the RHS of (\ref{eq:check_first_diff_reg1}) decreases as $\theta_0$ moves towards 0. The claim holds for region 1.\\

\noindent\textbf{Region 2} ( $\nu \in [0, \frac{1}{2\sin^2\theta})$ ): Similar to Region 1, we give a lower bound of $\frac{\partial f_{\Phi|N}\left(\phi|\nu\right)}{\partial\nu}$ for $\nu < \frac{1}{2\sin^2\phi}$. This lower bound is written as
\begingroup
\begin{align*}
        \frac{\partial f_{\Phi|N}\left(\phi|\nu\right)}{\partial\nu}
        >&-\frac{e^{-\nu}\sin^2\phi}{2\pi}+\frac{e^{-\nu\sin^2\phi}\cos\phi(1-2\nu\sin^2\phi)}{2\sqrt{\nu}\sqrt{\pi}}\\
        &\qquad\qquad\cdot\left[1-\frac{e^{-\nu\cos^2\phi}}{2}\right]\\
\end{align*}
\endgroup
where the inequality is obtained using the Q-function upper bound $Q(x) < \frac{\exp(-x^2/2)}{2}$. Consequently, 
\begingroup
\allowdisplaybreaks
\begin{align*}
    &\int_{0}^{\theta}\;\frac{\partial f_{\Phi|N}\left(\phi|\nu\right)}{\partial\nu}\;d\phi\\ &\qquad> \int_{0}^{\theta}\Bigg[\frac{e^{-\nu\sin^2\phi}\cos\phi(1-2\nu\sin^2\phi)}{2\sqrt{\nu}\sqrt{\pi}}\left[1-\frac{e^{-\nu\cos^2\phi}}{2}\right]\\
    &\qquad\qquad\qquad-\frac{e^{-\nu}\sin^2\phi}{2\pi}\Bigg]d\phi\\
        &\qquad= \frac{-e^{-\nu}}{4\pi\sqrt{\nu}}\Bigg[\sqrt{\nu}\theta - \frac{2\sqrt{\pi}\nu \sin^3\theta}{3} - \sqrt{\nu}\sin\theta\cos\theta \\
        &\qquad\qquad\qquad- 2\sqrt{\pi}\sin\theta e^{\nu\cos^2\theta}+\sqrt{\pi}\sin\theta\Bigg]
    \end{align*}%
\endgroup
We want to show that the above expression is positive. Equivalently, the claim is proven for Region 2 by showing that
\begin{equation}\label{eq:check_first_diff_reg2}
    \begin{split}
        2\sin\theta e^{\nu\cos^2\theta} + \frac{2\nu\sin^3\theta}{3}> \sqrt{\frac{\nu}{\pi}}\left[\theta-\sin\theta\cos\theta\right] + \sin\theta 
    \end{split}
\end{equation}
holds for all $\nu \in [0, \frac{1}{2\sin^2\theta})$. At $\nu = 0$, we have
\begin{equation*}
    2\sin\theta > \sin\theta
\end{equation*}
which is satisfied for all $\theta$ considered. Both sides of (\ref{eq:check_first_diff_reg2}) increase with $\nu$ but the LHS increases at a faster rate than the RHS. At the endpoint $\nu = \frac{1}{2\sin^2\theta}$, we have
\begin{equation*}
    2\sin\theta e^{\cot^2\theta}  > \sqrt{\frac{1}{2\pi}}\left[\frac{\theta}{\sin\theta} - \cos\theta\right] + \frac{2\sin\theta}{3} 
\end{equation*}
which still holds for $\theta\in(0,\frac{\pi}{4}]$. Combining the results for both Region 1 and Region 2 shows that the first integral term of (\ref{eq:first_diff_W_2b-1}) is positive. Going back to (\ref{eq:first_diff_W_2b-1}), $\frac{\partial W_{2^{b-1}}^{(b)}(\nu,\theta)}{\partial \nu}$ can be lower bounded by
\begingroup
\allowdisplaybreaks
\begin{equation*}
        \begin{split}
             &\frac{\partial W_{2^{b-1}}^{(b)}(\nu,\theta)}{\partial \nu}\\
         &\quad=\;2\underbrace{\int_{0}^{\theta }\; \frac{\partial f_{\Phi|N}(\phi|\nu)}{\partial \nu}\;d\phi}_{\geq 0} + \int_{ \theta}^{\frac{2\pi}{2^b}-\theta }\; \frac{\partial f_{\Phi|N}(\phi|\nu)}{\partial \nu}\;d\phi\\
             &\quad\geq\;\underbrace{\int_{0}^{\theta }\; \frac{\partial f_{\Phi|N}(\phi|\nu)}{\partial \nu}\;d\phi}_{\geq 0} + \int_{ \theta}^{\frac{2\pi}{2^b}-\theta }\; \frac{\partial f_{\Phi|N}(\phi|\nu)}{\partial \nu}\;d\phi\\
             &\quad=\;\int_{0}^{\frac{2\pi}{2^b}-\theta}\; \frac{\partial f_{\Phi|N}(\phi|\nu)}{\partial \nu}\;d\phi,
        \end{split}
\end{equation*}
\endgroup
which is positive when $ 0 < |\frac{2\pi}{2^b} - \theta| \leq \frac{\pi}{4}$. This is satisfied for $\theta\in[0,\frac{2\pi}{2^b})$ when $b > 3$. Note also that if $\theta = \frac{2\pi}{2^b}$, (\ref{eq:first_diff_W_2b-1}) becomes
\begingroup
\allowdisplaybreaks
\begin{align*}
             \frac{\partial W_{2^{b-1}}^{(b)}(\nu,\theta)}{\partial \nu}\Big|_{\theta = \frac{2\pi}{2^b}} =&\;2\int_{0}^{\frac{2\pi}{2^b}}\; \frac{\partial f_{\Phi|N}(\phi|\nu)}{\partial \nu}\;d\phi \\
             &\qquad+ \int_{\frac{2\pi}{2^b}}^{0}\; \frac{\partial f_{\Phi|N}(\phi|\nu)}{\partial \nu}\;d\phi\\
            =&\;\underbrace{\int_{0}^{\frac{2\pi}{2^b}}\; \frac{\partial f_{\Phi|N}(\phi|\nu)}{\partial \nu}\;d\phi}_{> 0}
\end{align*}
\endgroup
which is also positive for $b = 3$. Thus, the claim holds for $b \geq 3$.

\section{$W_{2^{b-1}}^{(b)}(\nu,\theta)$ is strictly concave on the parameter $\nu$ for $b\geq 3$} \label{appendix_F}

To prove that $W_{2^{b-1}}^{(b)}(\nu,\theta)$ is a strictly concave function of $\nu$, we need to show that
\begingroup
\allowdisplaybreaks
\begin{align}\label{eq:second_diff_W_2b-1}
&\frac{\partial^2 W_{2^{b-1}}^{(b)}(\nu,\theta)}{\partial\nu^2}\nonumber\\
     &\;\;= \int_{-\theta}^{\frac{2\pi}{2^b}-\theta}\;\frac{\partial^2 f_{\Phi|N}\left(\phi|\nu\right)}{\partial\nu^2}\;d\phi\nonumber \\
    &\;\;=2\int_{0}^{\theta}\frac{\partial^2 f_{\Phi|N}\left(\phi|\nu\right)}{\partial\nu^2}\;d\phi + \int_{\theta}^{\frac{2\pi}{2^b}-\theta}\frac{\partial^2 f_{\Phi|N}\left(\phi|\nu\right)}{\partial\nu^2}d\phi,
    \end{align}%
\endgroup
where $\frac{\partial^2 f_{\Phi|N}\left(\phi|\nu\right)}{\partial\nu^2}$ is given in (\ref{eq:second_diff_p_phi_given_nu}), is negative. The first integral term in the last equality of (\ref{eq:second_diff_W_2b-1}) follows from Lemma \ref{lemma:symmetry_p_phi_given_nu}. We first analyze a simple upper bound of the first term of (\ref{eq:second_diff_W_2b-1}) for arbitrary $\theta$ and show that it is negative for $b \geq 3$. We breakdown the problem into two regions of $\nu$.\\

\noindent\textbf{Region 1} ( $\nu \in \left[\frac{1+\sqrt{2}}{2\sin^2\theta},+\infty\right)$ ):  We first give an upper bound of $\frac{\partial^2 f_{\Phi|N}\left(\phi|\nu\right)}{\partial\nu^2}$ for $\nu > \frac{1+\sqrt{2}}{2\sin^2\phi}$.
\begingroup
\allowdisplaybreaks
\begin{align*}
     \frac{\partial^2 f_{\Phi|N}\left(\phi|\nu\right)}{\partial\nu^2}
        <& \frac{\cos\phi e^{-\nu\sin^2\phi}\left[(2\nu\sin^2\phi-1)^2-2\right]}{4\sqrt{\pi}\nu^{\frac{3}{2}}}\\
        &\quad\cdot\left[1-\frac{e^{-\nu\cos^2\phi}}{\sqrt{2\pi}\sqrt{2\nu}\cos\phi}\left(1-\frac{1}{2\nu\cos^2\phi}\right)\right]\\
        &+\frac{e^{-\nu}\cos^2\phi}{4\pi\nu}+\frac{e^{-\nu}\sin^4\phi}{2\pi}\\
        =& \frac{\cos\phi e^{-\nu\sin^2\phi}\left[(2\nu\sin^2\phi-1)^2-2\right]}{4\sqrt{\pi}\nu^{\frac{3}{2}}}\\
        &+\frac{e^{-\nu}[1+\sin^2\phi]}{4\pi\nu}+\frac{e^{-\nu}}{8\pi\nu^2}\\
        &+\frac{ e^{-\nu}\left[(2\nu\sin^2\phi-1)^2-2\right]}{16\pi\nu^{3}\cos^2\phi}
\end{align*}%
\endgroup
The first line follows from using the Q-function lower bound $Q(x) > \frac{\exp(-x^2/2)}{\sqrt{2\pi}x}\left(1-\frac{1}{x^2}\right)$ the second line is obtained after some algebraic manipulation. For $\nu <\frac{1+\sqrt{2}}{2\sin^2\phi}$, an upper bound can be expressed as
\begingroup
\allowdisplaybreaks
\begin{align*}
      \frac{\partial^2 f_{\Phi|N}(\phi|\nu)}{\partial \nu^2}
        <& -\frac{\cos\phi e^{-\nu\sin^2\phi}\left[2-(2\nu\sin^2\phi-1)^2\right]}{4\sqrt{\pi}\nu^{\frac{3}{2}}}\\
        &\quad\cdot\left[1-\frac{e^{-\nu\cos^2\phi}}{\sqrt{2\pi}\sqrt{2\nu}\cos\phi}\right]\\
        &+\frac{e^{-\nu}\cos^2\phi}{4\nu\pi}+\frac{e^{-\nu}\sin^4\phi}{2\pi}\\
         =& \frac{\cos\phi e^{-\nu\sin^2\phi}\left[(2\nu\sin^2\phi-1)^2-2\right]}{4\sqrt{\pi}\nu^{\frac{3}{2}}}\\
         &+\frac{e^{-\nu}[1+\sin^2\phi]}{4\nu\pi}+\frac{e^{-\nu}}{8\pi\nu^2}
\end{align*}%
\endgroup
where the inequality is obtained using the Q-function upper bound $Q(x) < \frac{\exp(-x^2/2)}{\sqrt{2\pi}x}$. Suppose we define $\theta_0 = \sin^{-1}\left(\sqrt{\frac{1+\sqrt{2}}{2\nu}}\right) > 0$. Then, we have
\begingroup
\allowdisplaybreaks
\begin{align*}
      &\int_{0}^{\theta}\frac{\partial^2 f_{\Phi|N}(\phi|\nu)}{\partial \nu^2}d\phi\\
        &\qquad<\;\int_{0}^{\theta}\Bigg[\frac{\cos\phi e^{-\nu\sin^2\phi}\left[(2\nu\sin^2\phi-1)^2-2\right]}{4\sqrt{\pi}\nu^{\frac{3}{2}}}\\
        &\qquad\qquad+\;\frac{e^{-\nu}[1+\sin^2\phi]}{4\nu\pi}+\frac{e^{-\nu}}{8\pi\nu^2}\Bigg]d\phi\\
        &\qquad\qquad+\;\int_{\theta_0}^{\theta}\frac{ e^{-\nu}\left[(2\nu\sin^2\phi-1)^2-2\right]}{16\pi\nu^{3}\cos^2\phi}\;d\phi\\
        &\qquad=\;\frac{e^{-\nu}}{8\pi\nu^2}\Bigg[(3\nu+1) \theta-4\sqrt{\pi}\nu^{\frac{3}{2}}\sin^3\theta e^{\nu\cos^2\theta}   \\
        &\qquad\qquad\qquad-2\sqrt{\pi}\sqrt{\nu}\sin\theta e^{\nu\cos^2\theta}
        -\frac{\nu \sin(2\theta)}{2}\Bigg]\\
        &\qquad\quad+\; \frac{e^{-\nu}}{16\pi\nu^3}\Bigg[(4\nu^2-4\nu-1)\tan\theta+\nu^2\sin(2\theta)\\
        &\qquad\qquad\qquad+\; 2(2-3\nu)\nu\theta\Bigg]\\
         &\qquad\quad- \frac{e^{-\nu}}{16\pi\nu^3}\Bigg[(4\nu^2-4\nu-1)\tan\theta_0\\
         &\qquad\qquad\qquad+\;\nu^2\sin(2\theta_0)+ 2(2-3\nu)\nu\theta_0\Bigg].
\end{align*}
\endgroup
We want to show that the above expression is negative for $\theta_0\in[0,\theta]$ and $\theta\in (0,\frac{\pi}{4}]$. Equivalently, by some algebraic manipulation and by expressing the $\nu$'s in terms of $\theta_0$, the claim is proven for Region 1 by showing that
\begingroup
\allowdisplaybreaks
\begin{align*}
      &3\theta-4\sqrt{\pi}\left(\frac{1+\sqrt{2}}{2}\right)^{\frac{3}{2}}\left(\frac{\sin\theta}{\sin\theta_0}\right)^3 e^{\frac{1+\sqrt{2}}{2}\cot^2\theta}   \\
      &\qquad-2\sqrt{\pi}\left(\frac{1+\sqrt{2}}{2}\right)^{\frac{1}{2}}\left(\frac{\sin\theta}{\sin\theta_0}\right) e^{\frac{1+\sqrt{2}}{2}\cot^2\theta}   \\
          &\qquad\qquad < \left[\frac{1+\sqrt{2}}{\sin^2\theta_0}-\frac{\sin^2\theta_0}{1+\sqrt{2}}-2\right]\left[\tan\theta_0 - \tan\theta\right]\\
          &\qquad\qquad\qquad+ \frac{1+\sqrt{2}}{2}\cot\theta_0 + \left[2-\frac{3(1+\sqrt{2})}{2\sin^2\theta_0}\right]\theta_0
\end{align*}%
\endgroup
A (stricter) inequality can be achieved by using the following lower bound for the first term of the RHS:
\begingroup
\allowdisplaybreaks
\begin{align*}
    &\left[\frac{1+\sqrt{2}}{\sin^2\theta_0}-\frac{\sin^2\theta_0}{1+\sqrt{2}}-2\right]\left[\tan\theta_0 - \tan\theta\right]\\
    &\qquad\qquad\qquad\qquad\qquad> \frac{1+\sqrt{2}}{\sin^2\theta_0}\left[\tan\theta_0 - \tan\theta\right],
\end{align*}
\endgroup
dropping the third term in the first $[\cdot]$ brackets of the LHS (which is negative), and the second term of the RHS (which is positive). Multiplying boths sides by $\sin^3(\theta_0)$ and rearranging the terms give us
\begin{equation}\label{eq:check_second_diff_reg1}
    \begin{split}
     &3\theta\sin^3\theta_0-\left[2\sin^2\theta_0-\frac{3(1+\sqrt{2})}{2}\right]\theta_0\sin\theta_0  
     \\
     &\quad\qquad < \sin\theta_0(1+\sqrt{2})\left[\tan\theta_0 - \tan\theta\right] \\
     &\qquad\qquad\qquad+ 4\sqrt{\pi}\left(\frac{1+\sqrt{2}}{2}\right)^{\frac{3}{2}}\sin^3\theta e^{\frac{1+\sqrt{2}}{2}\cot^2\theta}.    
    \end{split}
\end{equation}
For $\theta_0 = \theta$, the expression becomes
\begin{equation*}
    \begin{split}
     &\theta\sin^3\theta+\frac{3(1+\sqrt{2})\theta\sin\theta }{2}
 \\
 &\qquad\qquad\qquad< 4\sqrt{\pi}\left(\frac{1+\sqrt{2}}{2}\right)^{\frac{3}{2}}\sin^3\theta e^{\frac{1+\sqrt{2}}{2}\cot^2\theta} 
    \end{split}
\end{equation*}
which holds for $\theta\in(0,\frac{\pi}{4}]$. Moreover, for a fixed value of $\theta \in(0,\frac{\pi}{4}]$, it can be verified that the LHS of (\ref{eq:check_second_diff_reg1}) decreases and the RHS of (\ref{eq:check_second_diff_reg1}) increases as $\theta_0$ moves towards 0. As such, the claim holds for region 1.

\noindent\textbf{Region 2} ( $\nu \in [0,\frac{1+\sqrt{2}}{2\sin^2\theta)}$ ): 
In this region, we first give an upper bound of $\frac{\partial^2 f_{\Phi|N}\left(\phi|\nu\right)}{\partial\nu^2}$ for $\nu <\frac{1+\sqrt{2}}{2\sin^2\phi}$. This upper bound can be expressed as 
\begingroup
\allowdisplaybreaks
\begin{align*}
       &\frac{\partial^2 f_{\Phi|N}\left(\phi|\nu\right)}{\partial\nu^2}\\
        &\;< -\frac{\cos\phi e^{-\nu\sin^2\phi}\left[2-(2\nu\sin^2\phi-1)^2\right]}{4\sqrt{\pi}\nu^{\frac{3}{2}}}\left[1-\frac{e^{-\nu\cos^2\phi}}{2}\right]\\
        &\;\quad+\frac{e^{-\nu}\cos^2\phi}{4\nu\pi}+\frac{e^{-\nu}\sin^4\phi}{2\pi}\\
        &\;= -\frac{\cos\phi e^{-\nu\sin^2\phi}\left[2-(2\nu\sin^2\phi-1)^2\right]}{4\sqrt{\pi}\nu^{\frac{3}{2}}}+\frac{e^{-\nu}\cos^2\phi}{4\nu\pi}\\
        &\;\quad+\frac{e^{-\nu}\sin^4\phi}{2\pi}+\frac{\cos\phi e^{-\nu}[2-(2\nu\sin^2\phi-1)^2]}{8\sqrt{\pi}\nu^{\frac{3}{2}}},
\end{align*}%
\endgroup
where the second inequality is obtained using the Q-function upper bound $Q(x) < \frac{\exp(-x^2/2)}{2}$. Consequently, we have
\begingroup
\allowdisplaybreaks
\begin{align*}
        &\int_{0}^{\theta}\frac{\partial^2 f_{\Phi|N}(\phi|\nu)}{\partial \nu^2}d\phi\\
        &\qquad\quad< \frac{e^{-\nu}}{8\sqrt{\pi}\nu^{\frac{3}{2}}}\Bigg[\sin\theta-2\sin\theta e^{\nu\cos^2\theta} - 4\nu\sin^3\theta e^{\nu\cos^2\theta} \\
        &\qquad\qquad+ \frac{4\nu\sin^3\theta}{3} -\frac{4\nu^2\sin^5\theta}{5}\Bigg]+\frac{e^{-\nu}\left[\theta+\sin\theta\cos\theta\right]}{8\pi\nu}\\
        &\qquad\qquad+\frac{e^{-\nu}(12\theta - 8\sin(2\theta) + \sin(4\theta))}{64\pi}.
\end{align*}%
\endgroup
We want to show that the above expression is positive. Equivalently, after some algebraic manipulation, the claim is proven for Region 2 by showing that
\begin{align}\label{eq:check_second_diff_reg2}
        &\frac{\nu^{\frac{3}{2}}}{\sqrt{\pi}}\left[\frac{3\theta}{2}-\sin(2\theta) +\frac{\cos(2\theta)\sin(2\theta)}{4}\right] \nonumber\\
        &\qquad + 4\nu\sin^3\theta\left[\frac{1}{3}-e^{\nu\cos^2\theta}\right] + \sin\theta[1-2e^{\nu\cos^2\theta}]\nonumber\\
        &\qquad\qquad+ \sqrt{\frac{\nu}{\pi}}\left[\theta + \sin\theta\cos\theta\right] < \frac{4\nu^2\sin^5\theta}{5}
\end{align}
holds for all $\theta\in(0,\frac{\pi}{4}]$ and $\nu$ in Region 2. At $\nu = 0$, we have
\begin{equation*}
    -\sin\theta < 0
\end{equation*}
which is satisfied for all $\theta\in(0,\frac{\pi}{4}]$. Moreover, for any fixed $\theta\in(0,\frac{\pi}{4}]$, the RHS of (\ref{eq:check_second_diff_reg2}) increases as $\nu$ increases. Only the last term in the LHS of (\ref{eq:check_second_diff_reg2}) is nonnegative and this term grows at a slower rate than the other terms so the LHS of (\ref{eq:check_second_diff_reg2}) decreases as $\nu$ increases. As such, the claim holds for Region 2. Combining the results for both Region 2 and Region 1 shows that the first term of (\ref{eq:second_diff_W_2b-1}) is negative. Going back to (\ref{eq:second_diff_W_2b-1}), we have
\begingroup
\allowdisplaybreaks
\begin{align*}
         &\frac{\partial^2 W_{2^{b-1}}^{(b)}(\nu,\theta)}{\partial\nu^2}
             \\
         &\qquad\;=2\underbrace{\int_{0}^{\theta }\; \frac{\partial^2 f_{\Phi|N}(\phi|\nu)}{\partial \nu^2}\;d\phi}_{\leq 0} + \int_{ \theta}^{\frac{2\pi}{2^b}-\theta }\; \frac{\partial^2 f_{\Phi|N}(\phi|\nu)}{\partial \nu^2}\;d\phi\\
             &\qquad\;\leq \;\underbrace{\int_{0}^{\theta }\; \frac{\partial^2 f_{\Phi|N}(\phi|\nu)}{\partial \nu^2}\;d\phi}_{\leq 0} + \int_{ \theta}^{\frac{2\pi}{2^b}-\theta }\; \frac{\partial^2 f_{\Phi|N}(\phi|\nu)}{\partial \nu^2}\;d\phi\\
             &\qquad\;=\; \int_{ 0}^{\frac{2\pi}{2^b}-\theta }\; \frac{\partial^2 f_{\Phi|N}(\phi|\nu)}{\partial \nu^2}\;d\phi,
\end{align*}
\endgroup
which is negative when $ 0 < |\frac{2\pi}{2^b}-\theta| \leq \frac{\pi}{4}$. This is satisfied for $\theta\in[0,\frac{2\pi}{2^b})$ when $b > 3$. Note also that if $\theta = \frac{2\pi}{2^b}$, (\ref{eq:second_diff_W_2b-1}) becomes
\begingroup
\allowdisplaybreaks
\begin{align*}
     &\frac{\partial^2 W_{2^{b-1}}^{(b)}(\nu,\theta)}{\partial\nu^2}\Big|_{\theta = \frac{2\pi}{2^b}}\\
     &\qquad=\;2\int_{0}^{\frac{2\pi}{2^b} }\; \frac{\partial^2 f_{\Phi|N}(\phi|\nu)}{\partial \nu^2}\;d\phi + \int_{ \frac{2\pi}{2^b}}^{0}\; \frac{\partial^2 f_{\Phi|N}(\phi|\nu)}{\partial \nu^2}\;d\phi\\
     &\qquad=\underbrace{\int_{0}^{\frac{2\pi}{2^b} }\; \frac{\partial^2 f_{\Phi|N}(\phi|\nu)}{\partial \nu^2}\;d\phi}_{< 0},
\end{align*}
\endgroup
which is also negative for $b = 3$. Thus, the claim holds for $b \geq 3$.

\section{Sum of circular uniform distribution and an arbitrary circular distribution}\label{appendix_L}

Let $g(\cdot)$ be some arbitrary function that maps $[-\pi,\pi]$ to some real value, $A_1$ be a circular uniform distribution, and $A_2$ be some arbitrary circular distribution that is independent of $A_1$. Then, $\mathbb{E}_{A_1,A_2}\left[g(a_1+a_2\;\mathrm{mod}\; 2\pi)\right]$ becomes
\begingroup
\allowdisplaybreaks
\begin{align*}
     =& \int_{\mathrm{supp}\{A_2\}}\int_{-\pi}^{\pi}g(a_1+a_2\;\mathrm{mod}\; 2\pi)f_{A_1}(a_1)f_{A_2}(a_2)\;d{a_1}d{a_2}
\end{align*}
Let $A_3 = A_1 + A_2\;\mathrm{mod}\; 2\pi$. Then, $A_1 = A_3 - A_2 \;\mathrm{mod}\; 2\pi$ and $\mathbb{E}_{A_1,A_2}\left[g(a_1+a_2\;\mathrm{mod}\; 2\pi)\right]$ becomes 
\begin{align*}
    =& \int_{\mathrm{supp}\{A_2\}}\int_{-\pi+a_2\;\mathrm{mod}\;2\pi}^{\pi+a_2\;\mathrm{mod}\;2\pi}g(a_3)f_{A_1}(a_3-a_2\;\mathrm{mod}\;2\pi)\\
    &\qquad\qquad\qquad\qquad\qquad\qquad\cdot f_{A_2}(a_2)\;d{a_3}d{a_2}\\
     =&\int_{\mathrm{supp}\{A_2\}}\int_{-\pi+a_2\;\mathrm{mod}\;2\pi}^{\pi+a_2\;\mathrm{mod}\;2\pi}\frac{g(a_3)}{2\pi}f_{A_2}(a_2)\;d{a_3}d{a_2}\\
     =&\int_{\mathrm{supp}\{A_2\}}\int_{-\pi}^{\pi}\frac{g(a_3)}{2\pi}f_{A_2}(a_2)\;d{a_3}d{a_2}\\
     =&\underbrace{\int_{\mathrm{supp}\{A_2\}}f_{A_2}(a_2)\;d{a_2}}_{ = 1}\int_{-\pi}^{\pi}\frac{g(a_3)}{2\pi}\;d{a_3}\\
     =& \mathbb{E}_{A_3}\left[g(a_3)\right]\qquad \text{ where }A_3 \sim \mathrm{Unif}(-\pi,\pi),
\end{align*}
\endgroup
The first line is an expansion of the expectation using the random variable $A_3$. The second line follows from the distribution of $A_1$. The third line follows from the fact that the inner integration is over the whole circular domain $[-\pi,\pi]$ and is invariant of the offset $a_2$.

\end{appendices}

\ifCLASSOPTIONcaptionsoff
  \newpage
\fi

\bibliographystyle{ieeetr}
\bibliography{references}

\vskip 0pt plus -1fil
\begin{IEEEbiography}{Neil Irwin Bernardo} received the B.S. degree in Electronics and Communications Engineering from the University of the Philippines Diliman in 2014 and the M.S. degree in Electrical Engineering from the same university in 2016. He has been a faculty member of the University of the Philippines Diliman since 2014, and is currently on study leave to pursue the Ph.D. degree in Engineering at the University of Melbourne, Australia. His research interests include wireless communications, signal processing, and information theory.
\end{IEEEbiography}
\vskip 0pt plus -1fil
\begin{IEEEbiography}{Jingge Zhu} received the B.S. degree and M.S. degree in electrical engineering from Shanghai Jiao Tong University, Shanghai, China, in 2008 and 2011, respectively, the Dipl.-Ing. degree in technische Informatik from Technische Universit\"{a}t Berlin, Berlin, Germany in 2011 and the Doctorat \`{e}s Sciences degree from the Ecole Polytechnique F\'{e}d\'{e}rale (EPFL), Lausanne, Switzerland, in 2016. He was a post-doctoral researcher at the University of California, Berkeley from 2016 to 2018.  He is now a lecturer at the University of Melbourne, Australia. His research interests include information theory with applications in communication systems and machine learning. 

Dr. Zhu received the Discovery Early Career Research Award (DECRA) from the Australian Research Council in 2021, the IEEE Heinrich Hertz Award for Best Communications Letters in 2013, the Early Postdoc. Mobility Fellowship from Swiss National Science Foundation in 2015, and the Chinese Government Award for Outstanding Students Abroad in 2016.
\end{IEEEbiography}
\vskip 0pt plus -1fil
\begin{IEEEbiography}{Jamie Evans} was born in Newcastle, Australia, in 1970. He received the B.S. degree in physics and the B.E. degree in computer engineering from the University of Newcastle, in 1992 and 1993, respectively, where he received the University Medal upon graduation. He received the M.S. and the Ph.D. degrees from the University of Melbourne, Australia, in 1996 and 1998, respectively, both in electrical engineering, and was awarded the Chancellor's Prize for excellence for his Ph.D. thesis. From March 1998 to June 1999, he was a Visiting Researcher in the Department of Electrical Engineering and Computer Science, University of California, Berkeley. Since returning to Australia in July 1999 he has held academic positions at the University of Sydney, the University of Melbourne and Monash University. He is currently a Professor of Electrical and Electronic Engineering and Pro Vice-Chancellor (Education) at the University of Melbourne. His research interests are in communications theory, information theory, and statistical signal processing with a focus on wireless communications networks.
\end{IEEEbiography}

\end{document}